\documentclass[12pt]{article}
\pdfoutput=1
\usepackage{jheppub}
\usepackage{amsmath,bbm,array,amsfonts,graphicx,wrapfig,lscape,float,mathtools,multirow,longtable,rotating,subfigure,ulem,cancel}
\usepackage[section]{placeins}
\usepackage[all]{xy}

\newcommand{\be}{\begin{equation}}
\newcommand{\ee}{\end{equation}}
\newcommand{\beq}{\begin{equation}}
\newcommand{\beql}[1]{\begin{equation}\label{#1}}
\newcommand{\eeq}{\end{equation}}
\newcommand{\ba}{\begin{array}}
\newcommand{\ea}{\end{array}}
\newcommand{\bea}{\begin{eqnarray}}
\newcommand{\beal}[1]{\begin{eqnarray}\label{#1}}
\newcommand{\eea}{\end{eqnarray}}
\newcommand{\ben}{\begin{enumerate}}
\newcommand{\een}{\end{enumerate}}
\newcommand{\bean}{\begin{eqnarray*}}
\newcommand{\eean}{\end{eqnarray*}}

\newcommand{\btab}[1]{\begin{tabular}{#1}}
\newcommand{\etab}{\end{tabular}}

\newcommand{\comment}[1]{}

        \let\m=\mu

\newcommand{\qed}{\nobreak \ifvmode \relax \else
      \ifdim\lastskip<1.5em \hskip-\lastskip
      \hskip1.5em plus0em minus0.5em \fi \nobreak
      \vrule height0.75em width0.5em depth0.25em\fi}

\numberwithin{equation}{section}

\newcommand{\nc}{$ \ldots $}
\newcommand{\pal}{\ldots \text{pal}\ldots}
\newcommand{\slice}[1]{\mathcal{S}_{\mathcal N,#1} }
\newcommand{\svar}[2]{\mathcal{S}_{#1,#2} }
\newcommand{\bv}[1]{d_{BV}(#1)}
\newcommand{\orbit}[1]{\overline{\mathcal{O}}_{#1} }
\newcommand{\hs}[1]{g_{HS}^{#1}}
\newcommand{\higgs}{{\cal H}}
\newcommand{\coulomb}{{\cal C}}

\title{Quiver Theories and Hilbert Series of Classical Slodowy Intersections}

\author{Amihay Hanany,}
\author{Rudolph Kalveks}

\affiliation{
Theoretical Physics Group, The Blackett Laboratory,
Imperial College London, \\
Prince Consort Road, London SW7 2AZ, United Kingdom
}

\emailAdd{a.hanany@imperial.ac.uk, rudolph.kalveks09@imperial.ac.uk }

\preprint{Imperial/TP/19/AH/04}

\abstract{We build on previous studies of the Higgs and Coulomb branches of SUSY quiver theories having 8 supercharges, including $3d~{\cal N}=4$, and Classical gauge groups. The vacuum moduli spaces of many such theories can be parameterised by pairs of nilpotent orbits of Classical Lie algebras; they are transverse to one orbit and intersect the closure of the second. We refer to these transverse spaces as ``Slodowy intersections". They embrace reduced single instanton moduli spaces, nilpotent orbits, Kraft-Procesi transitions and Slodowy slices, as well as other types. We show how quiver subtractions, between multi-flavoured unitary or ortho-symplectic quivers, can be used to find a complete set of Higgs branch constructions for the Slodowy intersections of any Classical group. We discuss the relationships between the Higgs and Coulomb branches of these quivers and $T_{\sigma}^{\rho}$ theories in the context of $3d$ mirror symmetry, including problematic aspects of Coulomb branch constructions from ortho-symplectic quivers. We review Coulomb and Higgs branch constructions for a subset of Slodowy intersections from multi-flavoured Dynkin diagram quivers. We tabulate Hilbert series and Highest Weight Generating functions for Slodowy intersections of Classical algebras up to rank 4. The results are confirmed by direct calculation of Hilbert series from a localisation formula for normal Slodowy intersections that is related to the Hall Littlewood polynomials.

~\today}

\begin{document}

\maketitle

\listoftables

\listoffigures

\section{Introduction}
\label{sec:intro}

This paper builds on recent studies of SUSY quiver theories having 8 supercharges and Classical flavour and gauge groups, including both the Higgs and Coulomb branches of $3d~{\cal N}=4$ and the Higgs branches of $4d~{\cal N}=2$ theories. The vacuum moduli spaces of many such theories are related to the nilpotent orbits of Classical Lie algebras.

In essence, the nilpotent orbits ${\cal O}_{\rho}$ of the algebra $\mathfrak g$ of a Lie group $G$ are defined by equivalence classes of nilpotence conditions on representation matrices \cite{Collingwood:1993fk}. Such nilpotence conditions also describe the manner in which (combinations of) scalar fields in the F-term equations (derived from a superpotential) vanish at the SUSY vacuum, and can be specified (indirectly) by a quiver gauge theory. A ÒSlodowy sliceÓ ${\cal S}_{\rho} \equiv {\cal O}_{\rho}^{ \bot }$ is a space transverse to (or commuting with) a nilpotent orbit, yet lying within the adjoint orbit of the ambient group $G$ \cite{slodowy_1980}. These transverse spaces may be restricted to their intersections with the closure of any enclosing nilpotent orbit $\orbit{\sigma}$, leading to a wide variety of spaces, each parameterised by a pair of nilpotent orbits \cite{Kronheimer:1990ay}. We refer to these spaces, $\svar{\sigma}{\rho} \equiv \orbit{\sigma} \cap {\cal S}_{\rho}$ as ``Slodowy intersections".\footnote{There is no settled terminology: in \cite{maffei_2005} these are termed ``Slodowy varieties"; in \cite{Kronheimer:1990ay} and \cite{Gaiotto:2008ak} ``intersections"; in \cite{Baohua-Fu:2015nr} ``nilpotent Slodowy slices"; in \cite{proudfoot_schedler_2016} ``S3-varieties". Since the spaces are not generally nilpotent with respect to their symmetry group, $F \subseteq G$, and are always intersections (labelled by a pair of orbits), we prefer ``Slodowy intersections".} They embrace, as limiting cases, nilpotent orbits, Slodowy slices (intersected with the nilcone) and Kraft-Procesi transitions \cite{Kraft:1982fk}; other types appear for $G$ Classical of rank 3 upwards.

The connection was made in \cite{Gaiotto:2008sa} between (the $3d$ boundary conditions on) Type II brane systems \cite{Hanany:1996ie} in $4d~{\cal N}=4$ CFTs and Slodowy intersections. These $3d$ boundary conditions correspond to brane configurations that determine the Higgs branch vacuum moduli spaces of these theories. It was shown how the pole structure of these moduli spaces leads to their description as Slodowy intersections. This connection was deployed in \cite{Gaiotto:2008ak}, using S-duality and $3d$ mirror symmetry, to relate the Higgs and Coulomb branches of SUSY field theories characterised by D-brane configurations, with the latter encoded in quiver diagrams. These seminal papers have precipitated many studies.

More recently, in \cite{Hanany:2016gbz}, systematic methods were distilled for identifying SUSY quiver gauge theories whose Higgs or Coulomb branches correspond to the (closures of) nilpotent orbits of Classical algebras. In \cite{Cabrera:2018ldc}, this approach was extended to identify certain dual quiver theories whose Coulomb or Higgs branches are the Slodowy slices to these orbits.

This paper extends this systematic approach across the whole family of Slodowy intersections. In the interests of brevity, we draw extensively on \cite{Hanany:2016gbz} and \cite{Cabrera:2018ldc}, with cross-references to tables and formulae therein.

The language of quivers provides a rich description of the global, flavour and gauge symmetries underlying the quiver constructions for Slodowy intersections. In the case of the Higgs branch, there is a class of quivers, based on Nakajima (unitary) quiver varieties introduced in \cite{Nakajima:1994nu}. These were extended to ortho-symplectic types in \cite{Kraft:1982fk}. There is an overlapping set of unitary quivers based on Dynkin diagrams \cite{Dynkin:1957um}. Many of these quivers can be interpreted as intersections both on their Higgs and Coulomb branches. For $G$ Classical, the Higgs and Coulomb branches of certain dual quiver pairs can be related using the concept of $3d$ mirror symmetry \cite{Intriligator:1996ex, Nakajima:2015txa}.

Alternative descriptions exist for many Slodowy intersections in terms of $T_{\sigma} ^ {\rho}(G)$ theories \cite{Cremonesi:2014uva}. It is argued herein that an analysis in terms of pairs of nilpotent orbits of $G$ is (i) more complete for the intersections of $BCD$ groups, by virtue of dealing with special and non-special orbits on the same footing, and (ii) provides a clearer view of the mechanisms behind $3d$ mirror symmetry.

We deploy the technology of the Plethystics Program \cite{Feng:2007ur}, and its subsequent developments, to characterise Slodowy intersections by Hilbert series (``HS"), both in refined and unrefined form, and by their transformations to Highest Weight Generating (``HWG") functions \cite{Hanany:2014dia}. These provide a precise way of describing both the representation content and grading of theories (for example, by R-charges). In particular, refined Hilbert series and HWGs provide a means of testing whether (branches of) different quiver theories represent the same moduli spaces.

In section \ref{sec:Slodowy}, we summarise key aspects of the theory surrounding the nilpotent orbits and Slodowy slices of a Classical algebra $\mathfrak g$. Each Slodowy intersection $\svar{\sigma}{\rho}$ is defined by a pair of nilpotent orbits, where $\orbit{\sigma}$ contains ${{\mathcal{O}}_{\rho} }$. In cases where neither orbit contains the other, no intersection exists.

Matrices from (or functions on) an intersection $\svar{\sigma}{\rho}$ transform in a global symmetry, $F(\rho) \subseteq G$, which is some subgroup of the ambient symmetry group. The dimensions of the $\svar{\sigma}{\rho}$ are at most those of the nilcone (or maximal nilpotent orbit) of $G$, and their number increases rapidly with the rank of $G$.

Slodowy intersections can be analysed, either in terms of equivalence classes of sub-spaces of matrix representations, or, by virtue of the Jacobson Morozov Theorem, in terms of the embeddings of $\mathfrak{su(2)}$ into $\mathfrak g$ that define nilpotent orbits, along with the commuting sub-algebras ${\mathfrak f} \subseteq {\mathfrak g} $ that define Slodowy slices. In principle, either framework can be used to calculate Hilbert series. The latter approach, which we follow, fits naturally with the methods of quiver subtractions and Higgs or Coulomb branch HS constructions, developed in sections \ref{sec:ASeries} and \ref{sec:BCDSeries}.

Hilbert series for Slodowy intersections can also be constructed by purely group theoretic methods, using localisation formulae related to the Hall Littlewood polynomials. Several such formulae appear in the Literature \cite{Gadde:2011uv, Cremonesi:2014uva, Hanany:2017ooe}; their use requires careful attention to notational and other conventions. Building on the Nilpotent Orbit Normalisation formula \cite{Hanany:2017ooe}, we set out a Slodowy intersection formula (``SI formula") for calculating the HS of any $\svar{\sigma}{\rho}$ that intersects a normal nilpotent orbit $\orbit{\sigma}$.

In section \ref{sec:ASeries}, we build on \cite{Rogers:2018dez} and map out the quiver theories for Higgs and Coulomb branch constructions of $A$ series Slodowy intersections. These multi-flavoured linear unitary quivers can equally well be described as Nakajima or Dynkin type. We develop the methods of \cite{Cabrera:2018ann} to give a precise way of relating these quivers by an algebra of \emph{quiver subtractions} that draws on the concept of \emph{balance} \cite{Gaiotto:2008ak}, and is faithful to both Higgs and Coulomb branch dimensions. This systematises previous analyses in the Literature of $A$ series intersections, in preparation for the subsequent $BCD$ series analysis.

We use Higgs branch Weyl integration methods, the Coulomb branch unitary monopole formula and the SI formula to calculate the HS and HWGs of $A$ series quiver theories up to rank 4. As is known, these obey the rules of $3d$ mirror symmetry.

In section \ref{sec:BCDSeries}, we map out quiver theories for Higgs branch and certain Coulomb branch constructions of $BCD$ series Slodowy intersections. Both multi-flavoured linear ortho-symplectic and Dynkin quiver types are relevant. We show how the algebra of quiver subtractions extends to ortho-symplectic quivers, and is faithful to Higgs branch dimensions. 

We use Higgs branch Weyl integration methods and the SI formula to calculate the HS and HWGs of $BCD$ series quiver theories up to rank 4. The results from the Higgs branch and localisation methods are identical for normal Slodowy intersections. For non-normal Slodowy intersections $\svar{\sigma}{\rho}$, where $\orbit{\sigma}$ is a ``very even" orbit of $D_{2n}$, we show how the union of the normal components generated by the SI formula matches the non-normal Slodowy intersection given by Higgs branch methods.

We discuss the Coulomb branches of ortho-symplectic quivers for Slodowy intersections. These encounter similar problematic issues to those discussed in \cite{Cremonesi:2014kwa, Cabrera:2018ldc}.  As a consequence, the patterns of $3d$ mirror symmetry do not extend faithfully to $T_{\sigma}^{\rho}(G)$ theories for the Slodowy intersections of $BCD$ groups.

In section \ref{sec:Conclusions}, we summarise key findings and identify avenues for further work, for example by treating Slodowy intersections as building blocks for larger theories.

Appendix \ref{apx:1} summarises the key notation and conventions used by the Plethystics Program. Appendix \ref{apx:2} gives the details of the SI formula and its relationship to Hall Littlewood polynomials. Context permitting, we may refer to the closures of nilpotent orbits simply as ``nilpotent orbits", or ``orbits", to Slodowy slices as ``slices", and to Slodowy intersections as ``intersections".

\section{Slodowy Intersections}
\label{sec:Slodowy}

\subsection{Relationship to Slodowy Slices and Nilpotent Orbits}
\label{sec:SV_NO}

Throughout this text we assume a degree of familiarity with the concepts of nilpotent orbit and Slodowy slice. The reader is referred to \cite{Collingwood:1993fk} for a general grounding, or to \cite{Hanany:2016gbz, Cabrera:2018ldc} for a more specific introduction to our approach. The key properties of nilpotent orbits of Classical algebras up to rank 5, including their Characteristics \cite{Dynkin:1957um}, dimensions and the partitions of key irreps, are tabulated in \cite{Hanany:2016gbz}\footnote{In these tables Characteristics are referred to as ``root maps"}.

A nilpotent orbit $\cal O_{\rho}$ of the Lie algebra $\mathfrak g$ of a group $G$, is an equivalence class of nilpotent matrices that are conjugate under the action of $G$. By the Jacobson Morozov theorem, these nilpotent orbits are in one to one correspondence with equivalence classes of embeddings of $\mathfrak {su(2)}$ into $\mathfrak g$. Each such embedding is described by a homomorphism $\rho$, and this can be labelled either by the partition of a representation of $G$ (often taken as the fundamental or vector), or by a Characteristic, which uses Dynkin labels to specify the mapping of the roots (and hence weights) of $\mathfrak g$ onto $\mathfrak {su(2)}$. An orbit $\cal O_{\rho}$ has a closure ${\overline {\cal O}}_{\rho}$, defined as its union with (relevant) lower dimensioned orbits.

The closure of the maximal nilpotent orbit ${\overline {\cal O}}_{max}$, or nilcone $\cal N$, has the Hilbert series generated by symmetrising the character of the adjoint representation, modulo Casimir relations, graded by an R-charge (or $\mathfrak {su(2)}$ highest weight) fugacity $t^2$:
\begin{equation}
\label{eq:SV1}
{\cal N} \equiv {\overline {\cal O}}_{max} = PE\left[ {\chi _{adjoint}^G,{t^2}} \right]/PE \left[ {\sum\nolimits_{i = 1}^r {{t^{2{d_i}}}} } \right],
\end{equation}
where $\{d_1, \ldots,d_r \}$ are the degrees of symmetric Casimirs of $G$. The nilcone has dimension $|{\cal N}| = |{\mathfrak g}| - rank[\mathfrak g]$. Other key orbits include the sub-regular, minimal and trivial, the latter having dimension zero.

Nilpotent orbits can be arranged as a poset, or Hasse diagram, according to the inclusion relations of their \emph{closures}: 
\begin{equation}
\label{eq:SV2}
{{\cal N}}{ \equiv}{\overline {\cal O}}_{max} \supset {\overline {\cal O}}_{sub - reg} \ldots \supset {\overline {\cal O}}_{min} \supset {\overline {\cal O}}_{trivial} \equiv \{0\}.
\end{equation}

Each orbit ${\cal  O}_{\rho}$ has a transverse space termed a Slodowy slice, ${\cal S}_{\rho} \equiv {\cal  O}_{\rho}^{\bot}$. This transverse space intersects with the closures ${\cal \overline O}_{\sigma}$ of those orbits ranking higher in the Hasse diagram, generating a set of Slodowy \emph{intersections} ${{{\cal S}}_{\sigma ,\rho }}$, each defined by a pair of orbits of $G$:
\begin{equation}
\label{eq:SV3}
{{{\cal S}}_{\sigma ,\rho }} \equiv {\overline {\cal O}}_\sigma \cap {{\cal S}_\rho }.
\end{equation}
Necessarily, ${{{\cal S}}_{{\cal N} ,trivial }}= {\cal N}={\overline {\cal O}}_{max}$. The intersection with the nilcone ${{{\cal S}}_{{\cal N} ,\rho }}$ differs from the slice ${{{\cal S}}_{\rho }}$ only by the Casimir relations, and so, context permitting, we also refer to ${{{\cal S}}_{{\cal N} ,\rho }}$ as a Slodowy \emph{slice}. 

The number of non-trivial intersections rises rapidly with the number $n$ of nilpotent orbits of $G$, being bounded by the binomial coefficient $_n C_2$ (with saturation if the Hasse diagram of orbits of $G$ is linear).

\subsection{Symmetry Groups and Dimensions}
\label{sec:SV_Dim}

The dimension of each Slodowy intersection follows from the difference in dimensions of its defining pair of orbits:
\begin{equation}
\label{eq:SV4}
\left| {\cal S}_{\sigma ,\rho } \right| = \left| {{{ \cal O }_\sigma }} \right| - \left| {{{ \cal O }_\rho }} \right|.
\end{equation}

The dimension of an orbit  ${{{\cal O}_\rho }}$ is related to the partition of the adjoint of $G$ via its Characteristic ${\bf q} \equiv [q_1,\ldots,q_r]$, where $q_i \in  \{0,1,2\}$. A Characteristic provides a map between the simple root and weight fugacities ${\bf z} \equiv \{z_1,\ldots,z_r\}$ and ${\bf x} \equiv \{x_1,\ldots,x_r\}$ of $G$ and those of $SU(2)$, say $\{ {z}\}$ and $\{t\}$, respectively:
 \begin{equation}
\label{eq:SV5}
\begin{aligned}
{\rho:\{z_1,\ldots,z_r\} }&\to { \{ { z}^{q_1/2}, \ldots, { z}^{q_r/2} \}},\\
{\rho:\{x_1,\ldots,x_r\} }&\to {\{t^{\omega_1},\ldots,t^{\omega_r}\}}.
\end{aligned}
\end{equation}
The charges are related by the Cartan matrix $\bf A$ of $G$: ${{\bf q = A} \cdot {( \omega_1,\ldots, \omega_r )}}$. Under the map $\rho$, the adjoint of $G$ decomposes to irreps of $SU(2)$ with multiplicities $a_n$:
\begin{equation}
\label{eq:SV6}
\rho: \chi _{[adjoint]}^G({\bf x}) \to \mathop  \bigoplus \limits_{[n]} {a_n}(\rho)  \chi _{[n]}^{SU(2)}(t).
\end{equation}
Conventionally, these decompositions are expressed in condensed partition notation $\left( \ldots ,{{(n+1)}^{{a_n}}},\ldots,{1^{{a_0}}} \right)$, replacing $SU(2)$ Dynkin labels by irrep dimensions and (non-zero) multiplicities by exponents. The dimension of ${{{\cal O}_\rho }}$ is then found by subtracting the number of $SU(2)$ irreps in the adjoint partition from the dimension of $G$:
%
\begin{equation}
\label{eq:SV7}
|{{{\cal O}_\rho }}|=|G|-\sum\limits_n {{a_n}(\rho)}.
\end{equation}

The dimension of an intersection ${\cal S}_{\sigma ,\rho }$ follows via \ref{eq:SV4} and \ref{eq:SV7} from the multiplicities in the adjoint partitions of the relevant pair of orbits:
\begin{equation}
\label{eq:SV8}
\left|{{{\cal S}_{\sigma ,\rho }}} \right|=\sum\limits_n {{a_n}(\rho)} - \sum\limits_m {{a_m}(\sigma)}.
\end{equation}

Whereas nilpotent orbits have a symmetry group $G$, the intersections ${\cal S}_{\sigma ,\rho }$ have a symmetry group $F(\rho) \subseteq G$, where $F(\rho)$ need not be semi-simple, and may contain Abelian or finite subgroups. Note that intersections ${\cal S}_{\sigma ,\rho }$ are not generally nilpotent.

Now, refine the branching \ref{eq:SV6}, by introducing irreps $\chi _{\bf [m]}^F$ of $F(\rho)$ with dimensions $ |\chi _{\bf [m]} ^F |$ and multiplicities ${a_{[n],{\bf [m]}}}$, such that:
\begin{equation}
\label{eq:SV9}
\begin{aligned}
{a_n} &= \sum\limits_{[\bf m]} {{a_{[n],{\bf [m]}}} |\chi _{\bf [m]} ^F |},\\
\chi _{[adjoint]} ^G &= \mathop  \bigoplus \limits_{[n],{\bf [m]}} {a_{[n],{\bf [m]}}}\left( {\chi _{[n]}^{SU(2)} \otimes \chi _{\bf [m]}^F} \right).
\end{aligned}
\end{equation}
Due to the \emph{regular} nature of the branching $G \to SU(2)  \otimes F(\rho) $, the algebra $ {\mathfrak g}$ contains both $ {\mathfrak {su(2)}}$ and ${\mathfrak f}$ as sub-algebras.\footnote{${\mathfrak f}$ is the commutant, or centraliser, of ${\mathfrak {su(2)}}$ inside ${\mathfrak g}$ \cite{liebeck_seitz_2012}.} Hence, ${a_{[2],[singlet]}} =1$ and ${a_{[0],[adjoint]}}=1$. Consequently, the identity of $F(\rho)$ can often be determined by matching $|F|$ to the coefficient $a_0$, and by applying the constraint that $rank[F] \le rank[G]$. The structure of $F(\rho)$ is also provided by quiver theories, as explained in sections \ref{sec:ASeries} and \ref{sec:BCDSeries}.

It is useful to introduce weight space fugacities ${\bf y} \equiv \left\{ {{y_1}, \ldots ,{y_{rank[F]}}} \right\}$ and to refine \ref{eq:SV5}, by mapping the fugacities of $G$ to (monomials in) fugacities for $SU(2) \otimes  F $:
\begin{equation}
\label{eq:SV10}
\rho: {x_i} \to        {x_i ({\bf y}, t)}   =      {t^{{\omega _i}}}           \prod\limits_{j = 1}^{rank[F]} {y_j^{{\omega_{ij}}}},
\end{equation}
where the $\omega _{ij}$ are integers. Often, alternative sets of charges $\omega _{ij}$ are equivalent under conjugation by the Weyl group of $G$. All the basic irreps of $G$ (those whose Dynkin labels contain a single $1$) project to irreps of $SU(2) \otimes F $ under a valid fugacity map.

For example, the homomorphism $\rho$ with Characteristic $[020]$, which generates the 8 dimensional nilpotent orbit of $A_3$, also induces the following maps:
 \begin{equation}
\label{eq:SV11}
\begin{aligned}
&{\rho:\{z_1, z_2 ,z_3 \} }\to { \{ 1,{ z}, 1 \}},\\
&{\rho:\{x_1,x_2,x_3\} }\to {\{t ,t^2,t \}},\\
&{\rho: [1,0,0] }  \to{ 2 [1]_{SU(2)}}  \Leftrightarrow \chi _{fund.} ^{A_3} \to (2^2),\\
&{\rho: [1,0,1] }  \to{ 4 [2]_{SU(2)} \oplus 3 [0]_{SU(2)}}  \Leftrightarrow \chi _{[adjoint]} ^{A_3} \to (3^4,1^3).
\end{aligned}
\end{equation}
Noting the $F(\rho)=A_1$ symmetry implied by the multiplicity 3 of the  $ [0]_{SU(2)}$ singlet in the adjoint map, the weight map can be refined by introducing the $A_1$ fugacity $x$:
\begin{equation}
\label{eq:SV12}
\begin{aligned}
&{\rho:\{x_1,x_2,x_3\} }\to {\{ x t , t^2 , t /x  \}},\\
&{\rho: [1,0,0] }  \to{ [1]_{SU(2)} [1]_A },\\
&{\rho :[1,0,1] \to {  {  [2]_{SU(2)}} {[2]_A} }  \oplus {[2]_{SU(2)}   }  \oplus [2]_A  }.
\end{aligned}
\end{equation}

A degree of trial and error may be required to find a valid weight map containing the fugacities for $F(\rho)$; this complication is a byproduct of the conjugation deployed by Dynkin \cite{Dynkin:1957um}, when defining a Characteristic to contain only the integers $\{0,1,2\}$.

\subsection{Hilbert Series}
\label{sec:SV_HS}

The Hilbert series for a Slodowy slice ${\cal S}_{{\cal N},\rho}$ can be found directly from the adjoint branching of $G$ in \ref{eq:SV6} or \ref{eq:SV9} \cite{Cabrera:2018ldc}. The HS of this transverse space is obtained by (i) replacing the $SU(2)$ characters by their highest weights, $\chi_{[n]}^{SU(2)} \to t^n$, (ii) symmetrising the representations of $F(\rho)$ under a grading by $t^2$, and (iii) taking a quotient by the Casimirs of $G$. This leads to the refined and unrefined HS:
\begin{equation}
\label{eq:SV13}
\begin{aligned}
g_{HS}^{{{\cal S}_{{\cal N}},\rho }}({\bf y},t) &= PE\left[ {\mathop  \bigoplus \limits_{[n], \bf [m]}  a_{n,\bf m} {~} \chi _{\bf [m]}^F({\bf y}){t^{n + 2}} - \sum\limits_{i = 1}^r {{t^{2{d_i}}}} } \right],\\
g_{HS}^{{{{\cal S}}_{{{\cal N}},\rho }}}(1,t) &= PE\left[ {\sum\limits_n {{a_n}{t^{n + 2}} - \sum\limits_{i = 1}^r {{t^{2{d_i}}}} } } \right].
\end{aligned}
\end{equation}
In order to generalise \ref{eq:SV13} to the Hilbert series corresponding to a Slodowy intersection ${\cal S}_{{\sigma},\rho}$, we need to restrict the representation content in $g_{HS}^{{\cal S}_{{\cal N},\rho }}$ by the orbit ${\overline {\cal O} }_{\sigma}$. This is achieved by applying a quotient
%
\begin{equation}
\label{eq:SV14}
\begin{aligned}
\hs{\svar{\sigma}{\rho}}({\bf y},t)= \hs{\slice{\rho}}({\bf y},t) {\left. {\frac{\hs{\orbit{\sigma}} ({\bf x},t)}{ {\hs{\cal N}} ({\bf x}, t)  }} \right|_{{\bf x} \to {\bf x}({\bf y},t)}}
\end{aligned}
\end{equation}
Note that the the fugacity map \ref{eq:SV10} associated with $\rho$ is applied after taking the quotient ${\overline{ \cal O} }_{\sigma}/{\cal N}$, to avoid divergences. We can motivate \ref{eq:SV14} by considering the limiting cases of nilpotent orbits and Slodowy slices:
\begin{equation}
\label{eq:SV15}
\begin{aligned}
g_{HS}^{{{\overline {\cal O} }_\sigma }} &\equiv \hs{\svar{\sigma}{trivial}}  &=\hs{\slice{trivial}}  {~}{\left. {\frac{\hs{\orbit{\sigma}}} { {\hs{\cal N}}   }} \right|} &=g_{HS}^{{{\overline {\cal O} }_\sigma }}, \\
g_{HS}^{{{\cal S}}_{{{\cal N}},\rho }} & &=\hs{\svar{\cal{N}}{\rho}} {~} {\left. {\frac{  \hs{\orbit{max}}    }   {\hs {{{\cal N}}}} } \right| }&= g_{HS}^{{{\cal S}}_{{{\cal N}},\rho }},
\end{aligned}
\end{equation}
where we have used the identities ${{{\cal S}}_{{\cal N} ,trivial }}= {\cal N}={\overline {\cal O}}_{max}$.

The refined Hilbert series $\hs{\orbit{\sigma}}$ required by \ref{eq:SV14} can be found by Higgs or Coulomb branch methods \cite{Hanany:2016gbz}, or, if $\orbit\sigma$ is normal, from the Nilpotent Orbit Normalisation formula \cite{Hanany:2017ooe}. The latter choice leads to a localisation formula (``SI formula") for Slodowy intersections, described further in Appendix \ref{apx:2}. This provides a purely group theoretic calculation of $\hs{\svar{\sigma}{\rho}}$, which has been used to validate the quiver constructions herein.

Once the refined Hilbert series for an intersection has been calculated, it can be transformed to a Highest Weight Generating function:
\begin{equation}
\label{eq:apxB12}
\begin{aligned}
g_{HWG}^{{\cal S}_{\sigma, \rho}}({\bf m}, t) = \oint\limits_F {d{\mu ^F}{(\bf y)}} {~}{g_\chi ^F \left( \bf y^*, m \right)}{~}g_{HS}^{{\cal S}_{\sigma, \rho}} \left( {\bf y},t \right),
\end{aligned}
\end{equation}
where ${d{\mu ^F} {(\bf y)}}$ is a Haar measure for $F$, and $ {g_\chi ^F \left( \bf y, m \right)}$ is a generating function for the characters of $F$ (see \cite{Hanany:2014dia}). Alternatively, the HS can be simplified to unrefined form $g_{HS}^{{\cal S}_{\sigma, \rho}}(t)$, by setting $\{\forall i: y_i \to 1 \}$.

\subsection{Kraft Procesi Transitions}
\label{sec:SV_KP}

An intersection ${\cal S}_{\rho' , \rho }$ between a pair of orbits $(\rho' , \rho)$ that are adjacent in a Hasse diagram is often termed a Kraft-Procesi transition \cite{Kraft:1982fk}. For Classical groups, such transitions are either minimal nilpotent orbits, or two (complex) dimensional singularities ${{\mathbb C}^2}/ \Gamma$, where $\Gamma$ is a finite group of $ADE$ type. We follow convention and label the minimal nilpotent orbit of $G$ of rank $r$ by $g_r$, and a finite group singularity of $ADE$ type by $G_r$.

Thus, the set of non-trivial Slodowy intersections for $G$ is bounded by a set of nilpotent orbits, a set of Slodowy slices, and Kraft-Procesi transitions. These can conveniently be considered as an upper triangular schema: $\{{\cal S}_{\sigma, \rho}: {\cal N} \ge \sigma >  \rho \ge \{0\} \}$.

\subsection{Barbasch-Vogan Map and Duality}
\label{sec:SV_dual}

Importantly, while nilpotent orbits of $A_n$ correspond to partitions of $n+1$, denoted ${\cal P}(n+1)$, not all partitions $ {\cal P}(2n+1), {\cal P}(2n)$ or ${\cal P}(2n)$ biject to orbits of $ B_n, C_n$ or $D_n$. Those that do are termed $B$, $C$ or $D$ partitions, denoted ${\cal P}_{B/C/D}$.\footnote{In a $B$ or $D$ (resp. $C$) partition, any even (resp. odd) number appears an even number of times.} Thus, while the Lusztig-Spaltenstein map (transposition of partitions), being an involutive automorphism, provides a basis for dualities between $A$ series orbits, the more sophisticated Barbasch-Vogan map \cite{barbasch_vogan_1985} is required to formulate $BCD$ dualities. The Barbasch-Vogan map, ${d_{BV}}(\rho )$, of a partition $\rho$ depends on $G$, as defined in table \ref{tab:slod1}.
\begin{table}[htp]
\small
\centering
\begin{tabular}{|c|c|c|c|}
\hline
$G$ & $\rho$ & $ \text{Transformation}$ & $d_{BV}(\rho)$ \\
\hline
$A_n$ & $ \rho \in {\cal P}(n+1) $ & $ \rho  \to {\rho ^T} $ & $ d_{BV}(\rho) \in {\cal P}(n+1) $\\
$B_n$ & $ \rho \in {\cal P}_B (2n+1) $ & $ \rho  \to { {{{\left( {{{\left( \rho^T  \right)}_{N -  > N - 1}}} \right)}_C}} } $ & $ d_{BV}(\rho) \in {\cal P}_C(2n) $\\
$C_n$ & $ \rho \in {\cal P}_C (2n) $ & $ \rho  \to { {{{\left( {{{\left( \rho^T  \right)}_{N -  > N + 1}}} \right)}_B}}} $ & $ d_{BV}(\rho) \in {\cal P}_B(2n+1) $\\
$D_n$ & $ \rho \in {\cal P}_D (2n) $ & $ \rho  \to {{{\left( {{\rho ^T}} \right)}_D}}$ & $ d_{BV}(\rho) \in {\cal P}_D(2n) $\\
\hline
\end{tabular}
\caption{Barbasch-Vogan Map}
\label{tab:slod1}
\end{table}

Here $N \to N +(-)1$ indicates incrementing (decrementing) a partition and $()_{B/C/D}$ indicates collapse (as necessary) to a $B$, $C$ or $D$ partition, respectively. A partition collapse loses information and, consequently, the Barbasch-Vogan map is many to one for some $BCD$ orbits. Those orbits/partitions for which $d_{BV}$ is an involution, $d_{BV}^2=1$, are termed ``special", including all $A$ series orbits.

We can use the Barbasch-Vogan map to define the ``Special dual" of a Slodowy intersection, by exchanging a pair of special orbits of $G$ and taking $d_{BV}$ duals of their fundamental/vector partitions. The Special dual intersection to ${{{\cal S}}_{\sigma ,\rho }}$ is thus ${{{\cal S}^\vee}_{\sigma ,\rho }} \equiv {{{\cal S}}_{{d_{BV}}(\rho ),{d_{BV}}(\sigma )}}$. Notably, the Special dual is a form of GNO duality \cite{Goddard:1976qe} and switches between $B$ and $C$ partitions.

\FloatBarrier

\section{$A$ Series Quivers and Hilbert Series}
\label{sec:ASeries}

\subsection{Quiver Types}
\label{subsec:AQuivers}

Quivers for the closures of $A$ series nilpotent orbits ${\overline {\cal O}}_{\rho}$, or Slodowy slices ${{{\cal S}}_{{\cal N} ,\rho }}$, whether constructed on the Higgs or Coulomb branch, are either \emph{single-flavour linear} quivers ${\cal L}_A$\footnote{Quivers ${\cal L}_A (\lambda)$ consist of a single $SU(N_0)$ flavour node connected to a linear chain of $U(N_{i})$ gauge nodes, where the decrements between nodes, $\lambda_i = N_{i-1} - N_{i}$, constitute a partition of $N_0$, $\lambda \equiv \{\lambda_1,\ldots,\lambda_{{k}}\}$, where $\lambda_i \ge \lambda_{i+1}$ and $\sum \nolimits_{i = 1}^{k} {\lambda _i} = N_0$.}, or \emph{multi-flavour balanced} quivers ${\cal B}_A$ \cite{Cabrera:2018ldc}. In both cases, the unitary gauge nodes form an $A$ series Dynkin diagram. The two types are related by $3d$ mirror symmetry.

\begin{figure}[htbp]
\centering
\includegraphics[scale=0.5]{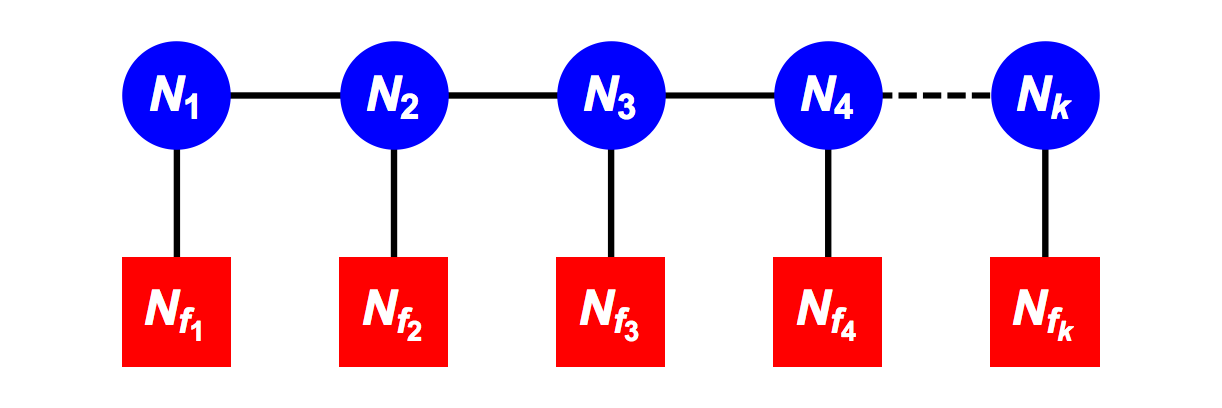}\\
\caption[$A$ Series Multi-flavour Dynkin Quiver.]{$A$ Series Multi-flavour Dynkin Quiver. The unitary gauge nodes (blue/round) with fundamental dimension $N_i$ are linked by conjugate pairs of bifundamental fields to adjacent gauge nodes and to unitary flavour nodes (red/square) with fundamental dimension $N_{f_i}$. The ${N_{{i}}}$ are typically non-zero and at least one of the ${N_{{f_i}}}$ should be non-zero.}
\label{fig:A1}
\end{figure}

The constructions for Slodowy intersections ${{{\cal S}}_{{\sigma},\rho }}$ draw upon the more general multi-flavour $A$ series Dynkin quiver \cite{nakajima_1994}, which subsumes both these types. These quivers ${\cal M}_A ({\bf N},{\bf N_f})$ can be drawn, as in figure \ref{fig:A1}, as a sequence of $k$ unitary gauge nodes, with ranks ${\bf N} \equiv \{N_1,\ldots, N_k \}$, with each gauge node connected to a unitary flavour node, with ranks ${\bf N_f} \equiv \{N_{f_1},\ldots, N_{f_k} \}$, where $\forall (i) : {N_f}_i \ge 0$.

Figure \ref{fig:A1} embraces many different quivers. In order to delineate those associated with Slodowy intersections it is helpful to deploy the concept of balance \cite{Gaiotto:2008ak}. A balance vector ${\bf B} \equiv \{B_1,\ldots, B_k \}$ for a quiver of type ${\cal M}_A ({\bf N},{\bf N_f})$ can be defined as:
\begin{equation} 
\label{eq:A1}
\begin{aligned}
{\bf B} ={\bf {N_f} - A \cdot N},
 \end{aligned}
\end{equation}
where $\bf A$ is the Cartan matrix for $A_k$. Following \cite{Gaiotto:2008ak}, we work here with quivers that are ``good" and do not have nodes with negative balance, so ${\bf B \ge 0}$ (defined as $ \forall (i):{B_i} \ge 0$). We shall show how any good quiver ${\cal M}_A ({\bf N},{\bf N_f})$ can be defined by a pair of partitions $\left( {\sigma, \rho} \right)$ and denote such a quiver ${\cal M}_A \left( {\sigma, \rho} \right)$.

We define ${\cal M}_A \left( {\rho, 0 } \right) \equiv {\cal L}_A (\rho ^T) $, where $\rho$ is a partition. It follows from $\rho^T_i \ge \rho^T_{i+1}$ that the quiver $ {\cal M}_A \left( {\rho, 0 } \right) $ has a balance vector that has no negative components. The partition data fixes the flavour and gauge nodes of this quiver $ {\cal L}_A (\rho ^T) \to  {\cal M}_A ({\bf N},{\bf N_f})$. Its Higgs branch is the closure of an $A$ series nilpotent orbit ${\overline {\cal O}}_{\rho}$, while its Coulomb branch is a Slodowy slice ${{{\cal S}}_{{\cal N}, \rho^T }}$ \cite{Cabrera:2018ldc}:
\begin{equation} 
\label{eq:A2a}
\begin{aligned}
{{{{\bar {\cal O}}}_\rho }} & =  \higgs[  {\cal M}_A \left( {\rho,0 } \right) ],\\
{{{\cal S}}_{{\cal N}, \rho^T }} & =  \coulomb[  {\cal M}_A \left( {\rho, 0 } \right) ].
 \end{aligned}
\end{equation}

A complete set of quivers, whose Higgs branches are $A$ series Slodowy \emph{intersections}, ${{{\cal S}}_{{\sigma} ,\rho }}$ can be found by carrying out \emph{quiver subtractions}, as suggested by the dimensional relations \ref{eq:SV4}. Various approaches to $A$ series quiver subtractions have been elaborated  \cite{Rogers:2018dez, Cabrera:2018ann}. The method described here draws explicitly on the concept of balance, which also serves to organise the resulting quivers. Thus, we claim:
\begin{equation} 
\label{eq:A2b}
\begin{aligned}
{{{{\cal S}}_{\sigma ,\rho }}} & =\higgs[  {\cal M}_A \left( {\sigma ,\rho } \right)],
 \end{aligned}
\end{equation}
where
\begin{equation} 
\label{eq:A2c}
\begin{aligned}
{\cal M}_A \left( {\sigma ,\rho } \right) & \equiv {\cal M}_A \left( {\sigma,0 } \right) \ominus {\cal M}_A \left( {\rho,0 } \right),
 \end{aligned}
\end{equation}
and the operation $\ominus$ of quiver subtraction is as defined below. 

Recall, the (complex) dimension of the HS of the Higgs branch of a quiver ${\cal M}_A \left( {\bf N,{N_f}} \right)$, is found by summing the dimensions of the conjugate pairs of bifundamental fields, and subtracting the gauge group dimensions twice (once for the Weyl integration and once for the adjoint relations)\footnote{This dimension formula, expressed in \cite{nakajima_1994} in terms of real dimensions, is valid for "good quivers", since these do not suffer from ``incomplete Higgsing'', (which would otherwise invalidate the HyperK\"ahler quotient).}:
\begin{equation} 
\label{eq:A3}
\begin{aligned}
\left | \higgs[  {\cal M}_A \left( \bf {N,{N_f}} \right) ] \right|=2 {\bf N \cdot N_f} - {\bf N \cdot A \cdot N}.
 \end{aligned}
\end{equation}
Now consider two quivers ${\cal M}_A \left( {\bf N_a,{N_f}_a} \right)$ and ${\cal M}_A \left( {\bf N_b,{N_f}_b} \right)$, with Higgs branches of dimension $|{\higgs}_{\bf a}|$ and $|{\higgs}_{\bf b}|$, respectively. We have:
\begin{equation} 
\label{eq:A4}
\begin{aligned}
\left| {\higgs}_{\bf a} \right| & = 2{\bf N_a} \cdot {\bf{N_f}_a}. - {\bf N_a} \cdot {\bf{A}} \cdot {\bf N_a}\\
\left| {\higgs}_{\bf b} \right| & = 2{\bf N_b}\cdot {\bf {N_f}_b} - {\bf N_b} \cdot {\bf{A}} \cdot {\bf N_b}
 \end{aligned}
\end{equation}
If we make the assumption that the two quivers have the same balance, $\bf B_a = B_b$, then ${\bf {N_f}_b}$ can be eliminated using \ref{eq:A1}, and \ref{eq:A4} yields:
\begin{equation} 
\label{eq:A5}
\begin{aligned}
\Delta \equiv \left| {\higgs}_{\bf a} \right| - \left| {\higgs}_{\bf b} \right|   = 2\left( {\bf N_a - N_b} \right) \cdot {\bf{N_f}_a} - \left( \bf N_a - N_{\bf b} \right) \cdot {\bf A} \cdot \left( {\bf N_a - N_b} \right).
 \end{aligned}
\end{equation}
Thus, $\Delta$ matches the dimension of a third quiver ${\cal M}_A ({\bf N_a} - {\bf N_b}, {\bf{N_f}_a})$, and this suggests a rule for subtracting two quivers with the same flavours $ {\bf{N_f}_a}$:
\begin{equation} 
\label{eq:A6}
\begin{aligned}
 {\cal M}_A \left( {\bf N_a, {N_f}_a} \right) & \ominus {\cal M}_A ({\bf N_a} - {\bf N_b}, {\bf{N_f}_a}) =  {\cal M}_A \left( {\bf N_b, {N_f}_b} \right), \\
 & \text{where}\\
 {\bf {N_f}_b} & = { \bf {N_f}_a - {\bf A} \cdot ({N_a - N_b}}).
 \end{aligned}
\end{equation}
Redefining the gauge vector, $\bf N_b \to N_a - N_b$, \ref{eq:A6} transforms to:
\begin{equation} 
\label{eq:A7}
\begin{aligned}
{\cal M}_A \left( {\bf N_a, {N_f}_a} \right) & \ominus {\cal M}_A \left( {\bf N_b, {N_f}_a} \right) ={\cal M}_A \left({\bf N_a - N_b,  {N_f}_b} \right),\\
 &\text{where}\\
 {\bf {N_f}_b} &= { \bf {N_f}_a - A \cdot N_b }. 
 \end{aligned}
\end{equation}
Naturally, the gauge ranks in the vector $\bf N_a - N_b$ must be non-negative for the quiver subtraction to be valid. Note that nodes of zero gauge rank do not contribute to the dimension formula \ref{eq:A3}.

Now, the quivers ${\cal M}_A \left( {\lambda,0 } \right)$, where $\lambda$ is a partition of the fundamental of $A_n$, all share the same flavour node vector $ {\bf N_f}=\{n+1,0,\dots,0\}$. Consequently, by allowing $\sigma$ and $\rho$ to range over the partitions of $n+1$, and by using \ref{eq:A2b}, \ref{eq:A2c} and \ref{eq:A7}, we can obtain a full set of quiver candidates ${\cal M}_A (\sigma,\rho)$, that have dimensions consistent with Higgs branch constructions of $A_n$ intersections ${\cal S}_{\sigma ,\rho }$.

\begin{figure}[htbp]
\centering
\includegraphics[scale=0.45]{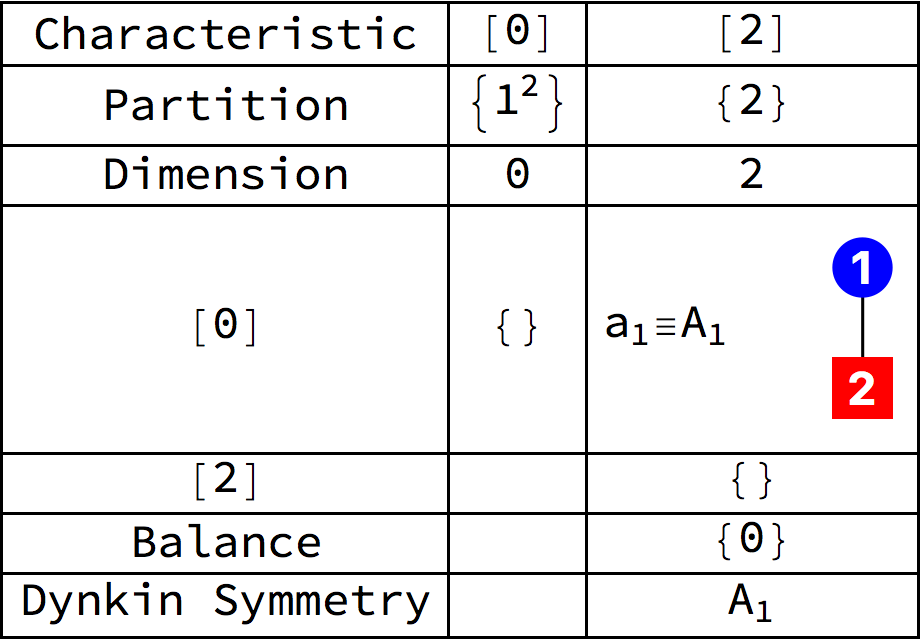}\\
\caption[Unitary Quivers for $A_1$ Slodowy Intersections.]{Quivers for $A_1$ Slodowy Intersections. The Higgs branch of this unitary quiver is the nilcone ${\cal S}_{A[2] ,A[0]}$. Rows (columns) are labelled by Characteristics of ${ {\cal O}}_\rho$ (${ {\cal O}}_\sigma$).}
\label{fig:A2a}
\end{figure}

\begin{figure}[htbp]
\centering
\includegraphics[scale=0.45]{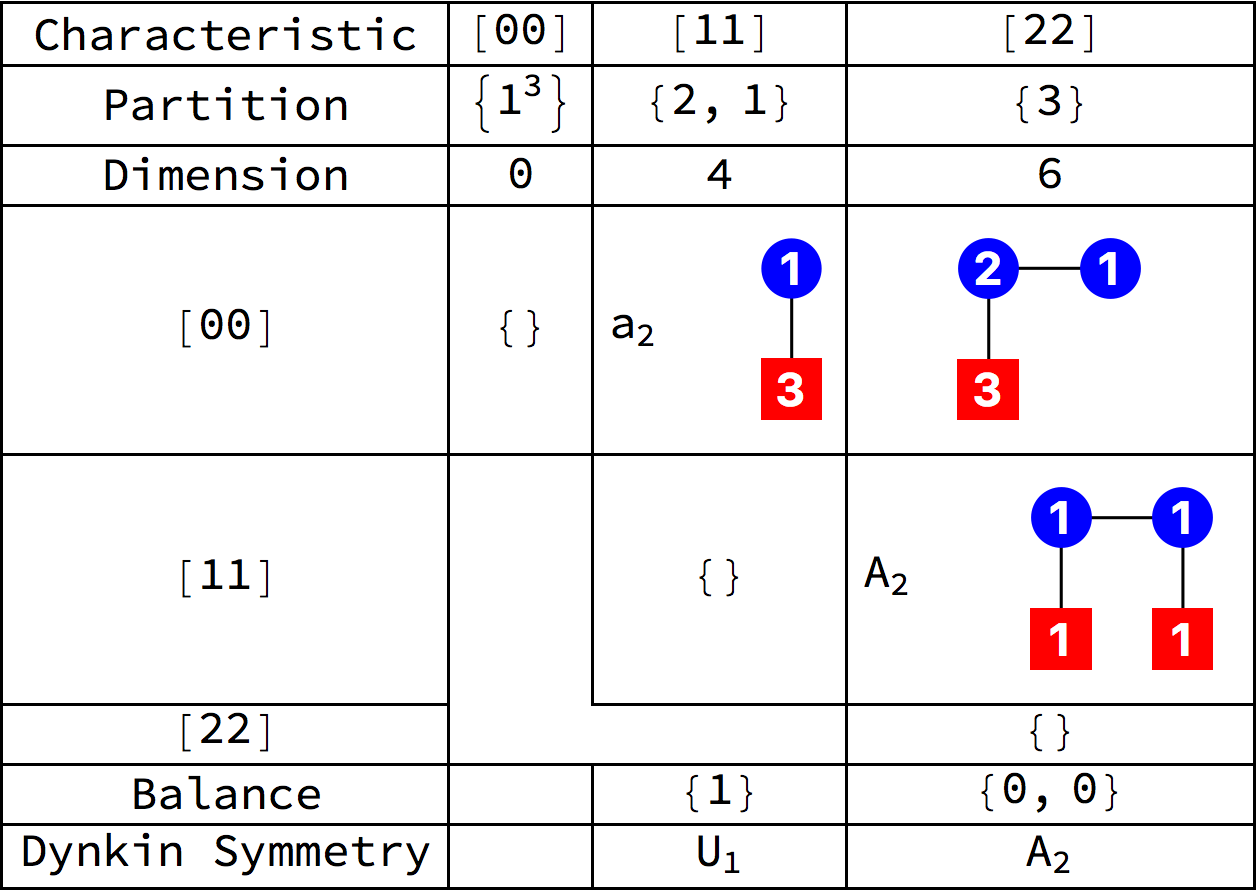}\\
\caption[Unitary Quivers for $A_2$ Slodowy Intersections.]{Quivers for $A_2$ Slodowy Intersections. The Higgs branches of these unitary quivers are Slodowy intersections ${\cal S}_{\sigma ,\rho }$. Rows (columns) are labelled by Characteristics of ${ {\cal O}}_\rho$ (${ {\cal O}}_\sigma$). }
\label{fig:A2b}
\end{figure}

\begin{figure}[htbp]
\centering
\includegraphics[scale=0.45]{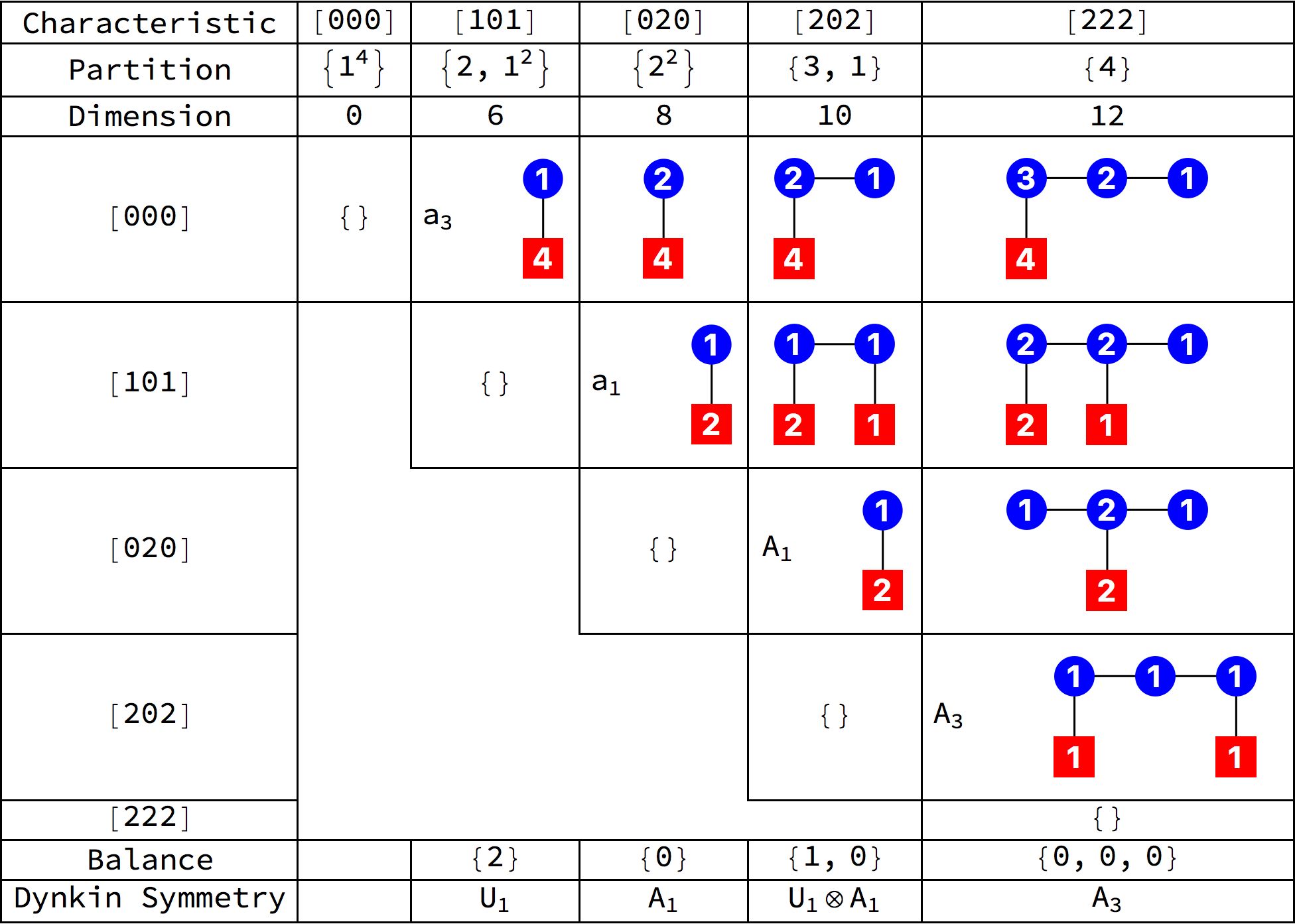}\\
\caption[Unitary Quivers for $A_3$ Slodowy Intersections.]{Quivers for $A_3$ Slodowy Intersections. The Higgs branches of these unitary quivers are Slodowy intersections ${\cal S}_{\sigma ,\rho }$. Rows (columns) are labelled by Characteristics of ${ {\cal O}}_\rho$ (${ {\cal O}}_\sigma$).}
\label{fig:A2c}
\end{figure}

\begin{figure}[htbp]
\centering
\includegraphics[scale=0.45]{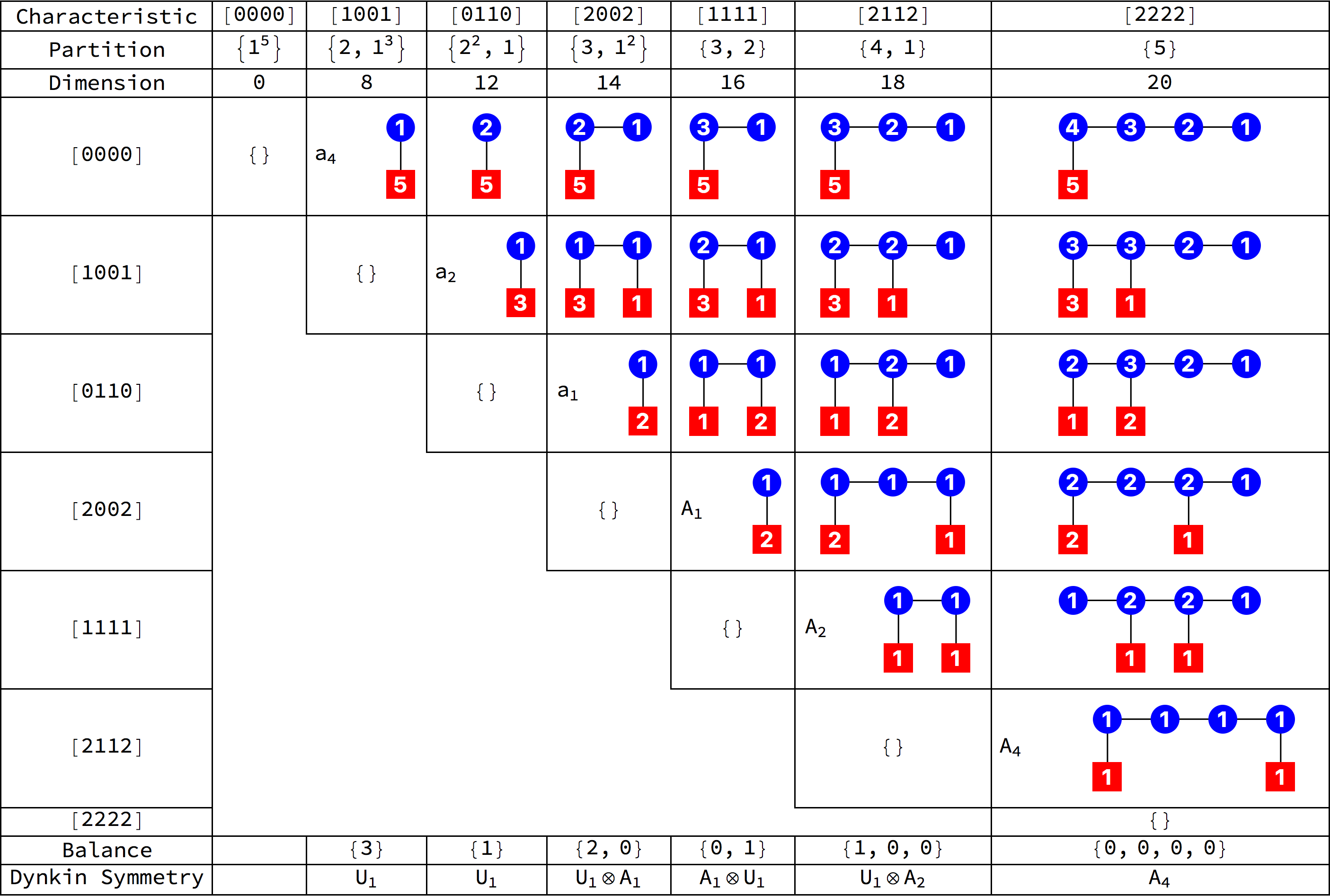}\\
\caption[Unitary Quivers for $A_4$ Slodowy Intersections.]{Quivers for $A_4$ Slodowy Intersections. The Higgs branches of these unitary quivers are Slodowy intersections ${\cal S}_{\sigma ,\rho }$. Rows (columns) are labelled by Characteristics of ${ {\cal O}}_\rho$ (${ {\cal O}}_\sigma$).}
\label{fig:A2d}
\end{figure}

These sets of quivers for the ${\cal S}_{\sigma ,\rho }$ of $A_n$ up to rank $4$ are shown in figures \ref{fig:A2a} through \ref{fig:A2d}. They are arranged as matrices, with rows and columns labelled by the Characteristics of ${ {\cal O}}_\rho$ and ${ {\cal O}}_\sigma$, respectively. Fundamental partitions and dimensions of ${ {\cal O}}_\sigma$ are also shown, as well as the balance vector $\bf B$, which is constant (by construction) for each column, and the Dynkin diagram symmetry - see below. Trivial self-intersections, with $g_{HS}^{{{{\cal S}}_{\rho, \rho }}} = 1$, are denoted $\{ \}$. The Kraft-Procesi transition for each row is labelled by its minimal singularity, as described in \ref{sec:SV_KP}. Empty entries indicate the absence of any intersection.\footnote{These appear above the diagonal from $A_5$ upwards, due to non-linear Hasse diagrams.} Gauge nodes of zero rank are truncated.

The matrices are all of upper triangular form. Each top row contains quivers ${\cal L}_A(\sigma^T)$ whose Higgs branches are closures of nilpotent orbits ${\overline {\cal O}}_\sigma$. Each rightmost column contains balanced quivers ${\cal B}_A (\rho)$, with $\bf B=0$, whose Higgs branches are Slodowy slices ${\cal S}_{{\cal N} ,\rho }$. The first non-empty entries above the diagonal are Kraft-Procesi transitions. More general intersections appear from $A_3$ upwards. Any two quivers ${\cal S}_{\sigma_1 ,\rho }$ and ${\cal S}_{\sigma_2 ,\rho }$ in the same row, where $\sigma_1 > \sigma_2$, are related by quiver subtraction to a third quiver ${\cal S}_{\sigma_1 ,{\sigma_2}}$ in a row below. All the quivers have non-negative balance, ${\bf B \ge 0}$.



The intersections ${\cal S}_{\sigma ,\rho }$ in each row transform in the same group $F(\rho)$ as the slice ${\cal S}_{{\cal N} ,\rho }$, although when $F(\rho)$ is a product group, lower dimensioned intersections (such as Kraft-Procesi transitions), may transform trivially under some component(s) of $F(\rho)$.

While each Slodowy intersection of $A_n$ is constructed from a pair of partitions of $n+1$, it can also be identified as a partition of $n'+1$ through the summation, $n'+1 =\sum\nolimits_{i = 1}^k {i {N_{f_i}}} $, and hence constructed from a pair of partitions of $n'+1$.\footnote{The partitions can be found from any given ${\cal M}_A \left( {\bf N,{N_f}} \right)$ with $\bf B>0$ by considering two linear quivers ${\cal L}_A$ with $N'_0 =n'+1$ and using \ref{eq:A7}.} If the intersection transforms trivially under some component of $F(\rho)$, then $n' < n$ and the intersection of $A_n$ also appears amongst the intersections of $A_{n'} \subset A_n$.

Significantly, for any intersection, the gauge nodes with $B_i=0$ form the Dynkin diagram of a semi-simple group, while gauge nodes with $B_i>0$ contribute Abelian $U(1)$ factors. These Dynkin diagrams determine the global symmetry $F(\rho)$ that appears on the \emph{Coulomb} branch of a quiver.

Corresponding tables can easily be constructed of the $3d$ mirror quivers whose \emph{Coulomb} branches are $A$ series Slodowy intersections, by using the Special duality map from table \ref{tab:slod1}, $(\sigma,\rho) \mathop \to \limits_{} (\rho^T, \sigma^T)$. In figures \ref{fig:A2a} through \ref{fig:A2d}, the nilpotent orbits in the top row have been ordered such that transposition acts as an order reversing involution on the set of fundamental partitions. (This is always possible for the $A$ series due to the bijection between partitions of $n+1$ and orbits of $A_n$.) Under this convention, Special duality is realised by matrix reflection in the (lower-left to top-right) diagonal of pairs of partitions that are BV self-dual: $ (\rho^T, \sigma^T)=(\sigma,\rho)$.

Recall, the complex dimension of the Coulomb branch of a unitary quiver is given by twice the sum of the ranks of the gauge nodes:

\begin{equation} 
\label{eq:A7a}
\begin{aligned}
 \left | \coulomb[  {\cal M}_A \left( \bf {N,{N_f}} \right) ] \right | =2 \sum\limits_{i = 1}^k {{N_i}} .
 \end{aligned}
\end{equation}
Since, under quiver subtraction \ref{eq:A7}, the ranks of gauge nodes are related by $\bf N_a \ominus N_b = N_a - N_b$, it follows from \ref{eq:A7a} that the dimensions of the Coulomb branches of the quivers $ {\cal M}_A \left( \bf {N,{N_f}} \right)$ form a self-consistent set. Moreover, \ref{eq:A2a} entails that $\coulomb[ {\cal M}_A (\rho^T,0)] = {{\cal S}_{{{\cal N}},\rho }}$, so it also follows that the Coulomb branch dimensions of the quivers $ {\cal M}_A (\rho^T,\sigma^T)$, obtained by quiver subtraction, match those of the ${\cal S}_{\sigma ,\rho }$ calculated on the Higgs branch.

\FloatBarrier


\subsection{Hilbert Series}
\label{subsec:AHS}

Hilbert series for $A$ type Slodowy intersections ${{\cal S}_{\sigma, \rho}}$ can be calculated from the quivers ${\cal M}_A \left( {\sigma,\rho } \right)$ using the Higgs branch formula described in \cite{Cabrera:2018ldc} (in section 3.2 thereof).

The results, labelled by pairs of Characteristics $(\sigma, \rho)$, are summarised in tables \ref{tab:AHS1} and \ref{tab:AHS2}. These set out, for each non-trivial Slodowy intersection, its dimension, its symmetry group $F(\rho) \subseteq A_n$, its unrefined Hilbert series, and the HWG (expressed as a PL) that decodes its HS into irreps of $F(\rho)$ \cite{Hanany:2014dia}. Refined HS and HWGs lacking finite PLs are not tabulated (due to space constraints). Trivial self-intersections, $g_{HS}^{{\cal S}_{\rho, \rho}}=1$, are omitted. 

\begin{sidewaystable}[h]
	\centering
	\begin{tabular}{|c|c|c|c|c|c|}
\hline
	$\rho$ & $\sigma$ & $\left | {{\cal S}_{\sigma, \rho}} \right |$ & $F(\rho)$ & Unrefined HS & PL[HWG] \\
 \hline
	$[0]$ & $[2]$ & 2 & $A_1$ & $\frac{1-t^4}{(1-t^2)^3}$ & $m^2 t^2$  \\
\hline
\hline
 	$[00]$ & $ {[11]}$ & $4$  & $A_2$ & ${ \frac{1+4 t^2+t^4}{(1-t^2)^4} }$ &${ m_1 m_2 t^2}$ \\
	${}$ & $ {[22]}$ & $6$  & ${}$ & ${\frac{(1-t^4) (1-t^6)}{(1-t^2)^8}} $ &$ {m_1 m_2 t^2 + m_1 m_2 t^4 + m_1^3 t^6 + m_2^3 t^6 - m_1^3 m_2^3 t^{12}} $ \\
 \hline
 	$[11]$ & $ {[22]}$ & $2$  & $U(1)$ & $\frac{1-t^6}{(1-t^2)(1-t^3)^2}$ &${ t^2 +(q+1/q)t^3}-t^6$ \\
 \hline
 \hline
 	$[000]$ & $ {[101]}$ & $6$  & $A_3$ & $\frac{(1+t^2) (1+8 t^2+t^4)}{(1-t^2)^6}$ &${m_1 m_3 t^2}$ \\
	${}$ & $ {[020]}$ & $8$  & ${}$ & $\frac{(1+t^2)^2 (1+5 t^2+t^4)}{(1-t^2)^8}$ &$ m_1 m_3 t^2 + m_2^2 t^4$\\
	${}$ & $ {[202]}$ & $10$  & ${}$ & $\frac{(1+t^2) (1+4 t^2+10 t^4+4 t^6 + t^8)}{(1-t^2)^{10}}$ &$ {m_1 m_3 t^2 + m_2^2 t^4 + m_1 m_3 t^4 + m_1^2 m_2 t^6 + m_2 m_3^2 t^6 - m_1^2 m_2^2 m_3^2 t^{12}}$ \\
	${}$ & $ {[222]}$ & $12$  & ${}$ & $\frac{(1-t^4) (1-t^6) (1-t^8)}{(1-t^2)^{15}}$ & \nc \\
    	 \hline
	${[101]}$ & $ {[020]}$ & $2$  & $A_1\otimes U(1)$ & $\frac{1-t^4}{(1-t^2)^3}$ &$ m^2 t^2$\\
	${}$ & $ {[202]}$ & $4$  & ${}$ & $\frac{1+2 t^2+2 t^3+2 t^4+t^6}{(1-t^2)^2 (1-t^3)^2}$ &$t^2 + m^2 t^2 + m ( q+1/q) t^3 - m^2 t^6$ \\
	${}$ & $ {[222]}$ & $6$  & ${}$ & $\frac{(1-t^6) (1-t^8)}{(1-t^2)^4 (1-t^3)^4}$ &\nc \\
	  \hline
	${[020]}$ & $ {[202]}$ & $2$  & $A_1$ & $\frac{1-t^4}{(1-t^2)^3}$ & $m^2 t^2$ \\
	${}$ & $ {[222]}$ & $4$  & ${}$ & $\frac{(1-t^6) (1-t^8)}{(1-t^2)^3 (1-t^4)^3}$ & $m^2 t^2 + t^4 + m^2 t^4 + m^2 t^6 - m^4 t^{12}$ \\
    	 \hline
	 ${[202]}$ & $ {[222]}$ & $2$  & $U(1)$ & $\frac{1-t^8}{(1-t^2) (1-t^4)^2}$ & $t^2 + (q+1/q) t^4 - t^8$ \\
    	 \hline
	\end{tabular}
	\caption[$A_1$, $A_2$ and $A_3$ Slodowy Intersections]{$A_1$, $A_2$ and $A_3$ Slodowy Intersections. Only HWGs that are complete intersections are shown.}
	\label{tab:AHS1}
\end{sidewaystable}

\begin{sidewaystable}[h]
	\centering
	\small
	\begin{tabular}{|c|c|c|c|c|c|}
\hline
	$\rho$ & $\sigma$ & $\left | {{\cal S}_{\sigma, \rho}} \right |$ & $F(\rho)$ & Unrefined HS & PL[HWG] \\
 \hline
	$[0000]$ & $[1001]$ & 8 & $A_4$ & $\frac{1+16 t^2+36 t^4+16 t^6+t^8}{(1-t^2)^8}$ & $m_1 m_4 t^2$  \\
	$ $ & $[0110]$ & 12 & ${}$ & $\frac{1+12 t^2+53 t^4+88 t^6+\pal+t^{12}}{(1-t^2)^{12}}$ & $m_1 m_4 t^2 + m_2 m_3 t^4$  \\
	$ $ & $[2002]$ & 14 & ${}$ & $\frac{ 1+9 t^2+45 t^4+65 t^6+\pal +t^{12}}{(1-t^4)^{-1}(1-t^2)^{15}}$ & 
 $ \begin{array}{c} m_1 m_4 t^2 +(m_2 m_3 + m_1 m_4) t^4 +\\ (m_1^2 m_3 + m_2 m_4^2) t^6 - m_1^2 m_2 m_3 m_4^2 t^{12} \end{array}$\\
	$ $ & $[1111]$ & 16 & ${}$ & $\frac{ 1+6 t^2+22 t^4+37 t^6+\pal +t^{12} }{(1-t^4)^{-2}(1-t^2)^{18}}$ & \nc  \\
	$ $ & $[2112]$ & 18 & ${}$ & $\frac{1+4 t^2+10 t^4+20 t^6+\pal +t^{12}}{(1-t^4)^{-1} (1-t^6) ^{-1}(1-t^2)^{20}}$ &\nc  \\
	$ $ & $[2222]$ & 20 & ${}$ & $\frac{(1-t^4) (1-t^6) (1-t^8) (1-t^{10})}{(1-t^2)^{24}}$ & \nc  \\
\hline
	$[1001] $ & $[0110]$ & 4 & $A_2 \otimes U(1)$ & $\frac{1+4 t^2+t^4}{(1-t^2)^4}$ & $m_1 m_2 t^2$  \\
	$ $ & $[2002]$ & 6 & ${}$ & $\frac{ 1+t+6 t^2+9 t^3+15 t^4 +12 t^5+\pal +t^{10}}{(1-t)^{-1}(1-t^2)^4 (1-t^3)^3}$ & $t^2 + m_1 m_2 t^2  + (m_1 q + m_2/q) t^3 - m_1 m_2 t^6$\\
	$ $ & $[1111]$ & 8 & ${}$ & $\frac{ 1+t+3 t^2+6 t^3+9 t^4+9 t^5+\pal+t^{10} }{(1-t)^{-1} (1+t^2)^{-1} (1-t^2)^6 (1-t^3)^3}$ & \nc  \\
	
	$ $ & $[2112]$ & 10 & ${}$ & $\frac{  (1+2 t+4 t^2+5 t^3+4 t^4+2 t^5+t^6) (1+t^2+t^3+4 t^4+t^5+t^6+t^8)}{(1-t)^{-2}(1-t^6)^{-1} (1-t^2)^7 (1-t^3)^6}$ &\nc  \\
	$ $ & $[2222]$ & 12 & ${}$ & $\frac{(1-t^6) (1-t^8) (1-t^{10})}{(1-t^2)^9 (1-t^3)^6}$ & \nc  \\
\hline
	$[0110] $ & $[2002]$ & 2 & $A_1 \otimes U(1)$ & $\frac{1-t^4}{(1-t^2)^3}$ & $m^2 t^2$\\
	$ $ & $[1111]$ & 4 & ${}$ & $\frac{1+2 t^2+2 t^3+2 t^4+t^6}{(1-t^2)^2 (1-t^3)^2}$ & $t^2 + m^2 t^2 + m ( q+1/q) t^3 - m^2 t^6$ \\
	$ $ & $[2112]$ & 6 & ${}$ & $\frac{1+t^2+2 t^3+2 t^4+2 t^5+t^6+t^8}{(1-t^6)^{-1} (1-t^2)^3 (1-t^3)^2 (1-t^4)^2}$ &\nc  \\
	$ $ & $[2222]$ & 8 & ${}$ & $\frac{(1-t^6) (1-t^8) (1-t^{10})}{(1-t^2)^4 (1-t^3)^4 (1-t^4)^3}$ & \nc  \\
\hline
	$[2002] $ & $[1111]$ & 2 & $A_1 \otimes U(1)$ & $\frac{1-t^4}{(1-t^2)^3}$ & $m^2 t^2$ \\
	$ $ & $[2112]$ & 4 & ${}$ & $\frac{1+2 t^2+4 t^4+2 t^6+t^8}{(1-t^2)^2 (1-t^4)^2}$ &$t^2 + m^2 t^2 + m ( q+1/q ) t^4 - m^2 t^8$  \\
	$ $ & $[2222]$ & 6 & ${}$ & $\frac{(1-t^8) (1-t^{10})}{(1-t^2)^4 (1-t^4)^4}$ & \nc  \\
\hline
	$ [1111]$ & $[2112]$ & 2 & $U(1)$ & $\frac{1-t^6}{(1-t^2) (1-t^3)^2}$ &$t^2 + (q+1/q ) t^3 - t^6$  \\
	$ $ & $[2222]$ & 4 & ${}$ & $\frac{(1-t^8) (1-t^{10})}{(1-t^2) (1-t^3)^2 (1-t^4) (1-t^5)^2}$ &$ t^2 + (q+1/q) t^3 + t^4 + (q+1/q) t^5 - t^8 - t^{10}$\\
\hline
	$[2112] $ & $[2222]$ & 2 & $U(1)$ & $\frac{1-t^{10}}{(1-t^2) (1-t^5)^2}$ & $t^2 + (q+1/q) t^5 - t^{10}$\\
\hline
	\end{tabular}
	\caption[$A_4$ Slodowy Intersections]{$A_4$ Slodowy Intersections. Only HWGs that are complete intersections are shown.}
	\label{tab:AHS2}
\end{sidewaystable}

The HS are consistent both with the dimension formulae given above, and with established results in the Literature for $A$ series orbits, Slodowy slices and KP transitions. Equivalent results are obtained on the Coulomb branch, applying the unitary monopole formula, described in \cite{Cremonesi:2013lqa} or \cite{Cabrera:2018ldc} (in section 3.3 thereof), to the quivers ${\cal M}_A \left( {\rho^T, \sigma^T} \right)$, or alternatively by using the SI formula \ref{eq:SV14}. As a further non-trivial check, the HS for the different intersections ${\svar{\sigma}{ \rho}}$ within each slice (fixed by $\rho$) obey inclusion relations that match those of the poset of orbits $\orbit{\sigma}$ in the parent group Hasse diagram \cite{Kraft:1982fk}.
\FloatBarrier

Several observations can be made about the Hilbert series of $A$ type Slodowy intersections ${{\cal S}_{\sigma, \rho}}$ and their HWGs:

\begin{enumerate}

\item{All the unrefined HS are normal and palindromic. If an intersection is a Slodowy slice, (or one which matches a Slodowy slice of a lower rank algebra), its unrefined HS is also a complete intersection}. 

\item{$F(\rho)$ is a product group with unitary and/or special unitary components. It always has a rank one below the sum of unitary flavours $\sum\nolimits_i {{N_{{f_i}}}}$ in the Higgs quiver, due to an overall SU condition on the flavour nodes.}

\item{The adjoint representation of $F(\rho)$ (or its relevant subgroup) always appears at counting order $t^2$. Other representations of $F(\rho)$ only appear at higher orders.}

\item{The Slodowy intersections are series of real representations of $F(\rho)$, so any complex irreps of $F(\rho)$ that appear are coupled with their conjugates at each counting order.}

\item{The same HS may recur for different pairs of partitions and for different ranks of ambient group $A_n$. Such recurrences can be identified directly from the quiver diagrams and their outer automorphisms.}

\end{enumerate}


\subsection{Relationship to $T_{\sigma}^{\rho}$ theories}
\label{sec:A_Tsr}

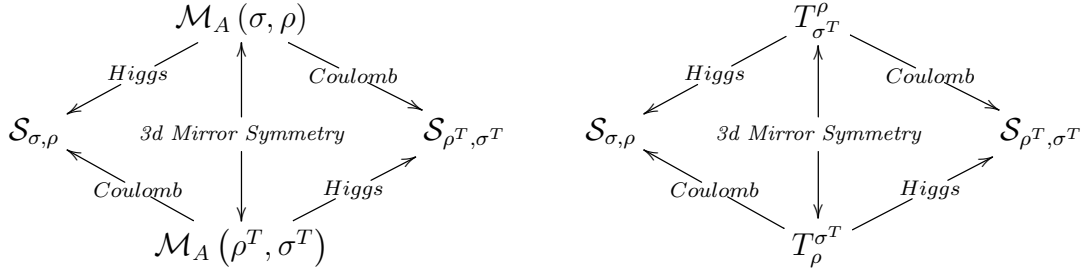
\begin{figure}[htp]
\centering
\begin{displaymath}
    \xymatrix{
  &  {\cal M}_A \left( {\sigma,\rho } \right) \ar[dl]|{Higgs} \ar[dr]|{Coulomb} & & & T_{\sigma^T}^{\rho}   \ar[dl]|{Higgs} \ar[dr]|{Coulomb} & \\
 {{\cal S}_{\sigma, \rho}}  &\text{\scriptsize \it 3d~Mirror Symmetry} \ar[d] \ar[u] &  {{\cal S}_{\rho^T, \sigma^T}}   & {{\cal S}_{\sigma, \rho}}  & \text{\scriptsize \it 3d~Mirror Symmetry} \ar[d] \ar[u] & {{\cal S}_{\rho^T, \sigma^T}} \\
   & {\cal M}_A \left( {\rho^T, \sigma^T} \right) \ar[ul]|{Coulomb} \ar[ur]|{Higgs} & & & T_{\rho}^{\sigma^T}   \ar[ul]|{Coulomb} \ar[ur]|{Higgs}  & }
\end{displaymath}
\caption[$A$ Series 3d Mirror Symmetry]{$A$ series $3d$ mirror symmetry.  Under Special duality, the nilpotent orbit partitions $\rho$ and $\sigma$ are dualised under the Lusztig-Spaltenstein map, to $\rho^T$ and $\sigma^T$, and then interchanged ${{{\cal S}^\vee}_{\sigma ,\rho }} \equiv {{{\cal S}}_{\rho^T, \sigma^T}}$. All the constructions yield refined Hilbert series.}
\label{fig:Ams}
\end{figure}

As discussed in \cite{Cremonesi:2014uva}, $A$ series Slodowy intersections are related to a class of quiver theories known as $T_{\sigma}^{\rho}$ theories:

\begin{equation} 
\label{eq:A8}
\begin{aligned}
{ \higgs[T_{\sigma}^{\rho}(SU(N))] } &= {\svar{\sigma^T}{ \rho}}= {\coulomb[T_{\rho}^{\sigma}(SU(N))]}.  
 \end{aligned}
\end{equation}
We therefore have the following correspondence between $A$ series multi-flavour Dynkin quivers and $T_{\sigma}^{\rho}$ theories:
\begin{equation} 
\label{eq:A9}
\begin{aligned}
{\cal M}_A \left( {\sigma,\rho } \right) \cong T_{\sigma^T}^{\rho}(SU(N)).
 \end{aligned}
\end{equation}
Under our approach, we label the quiver according to the partitions $\rho$ and $\sigma$ for orbits of the ambient group $G$. The phenomenon of $3d$ mirror symmetry can thus be understood, as in figure \ref{fig:Ams}, as a consequence of the composition of the interchange of a pair of nilpotent orbits with the Lusztig-Spaltenstein map (i.e. $\bv\rho$ for the $A$ type).


\section{$BCD$ Series Quiver Constructions}
\label{sec:BCDSeries}

\subsection{Quiver Types}
\label{subsec:BCDQuivers}

\begin{figure}[htbp]
\includegraphics[scale=0.5]{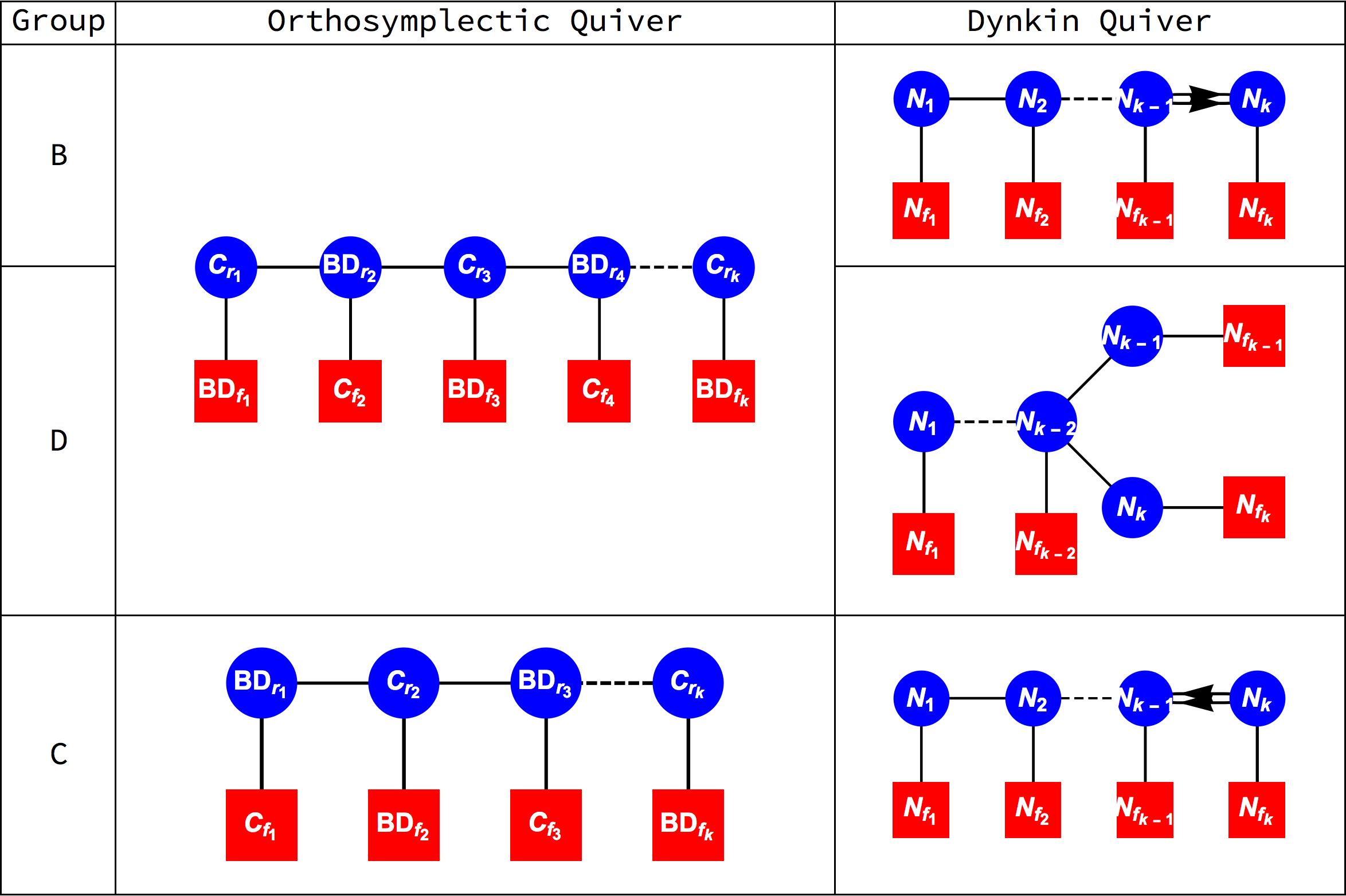}\\
\caption[$BCD$ Series Multi-flavoured Quiver Types]{$BCD$ multi-flavoured quiver types. Ortho-symplectic linear quivers ${\cal M}_{BCD} \left( {\bf N,{N_f}} \right)$ have gauge (blue/round) and flavour (red/square) nodes, with ranks ${r_i} \ge 0$ and ${f_i} \ge 0$, linked by bi-vector fields. $B_r$, $C_r$ or $D_r$ means $SO/O(2r+1)$, $USp(2r)$ or $SO/O(2r)$, and $BD$ indicates a node of one of the two series. Dynkin quivers ${\cal D}_{G} \left( {\bf N,{N_f}} \right)$ have unitary gauge and flavour nodes, with ranks $N_i  > 0$ and $N_{f_i}  \ge 0$, linked by conjugate pairs of bifundamental fields.}
\label{fig:BCD1}
\end{figure}

Numerous field theories generate subsets of Slodowy intersections ${\cal S}_{\sigma,\rho}$ of $BCD$ algebras, (although this is not always recognised). Their constructions include the Higgs and Coulomb branches of quivers \cite{Cremonesi:2014uva, Hanany:2016gbz, Cabrera:2018ldc}, as well as plethystic formulae related to Hall Littlewood polynomials \cite{Cremonesi:2014kwa}. The main quiver types fall into one of two categories, as shown in figure \ref{fig:BCD1}: ortho-symplectic linear quivers, ${\cal M}_{BCD} \left( {\bf N,{N_f}} \right)$, which have alternating orthogonal or symplectic gauge and flavour nodes, and quivers ${\cal D}_{G} \left( {\bf N,{N_f}} \right)$, with unitary gauge nodes arranged as a Dynkin diagram of $G$. Significantly, not all approaches provide a complete set of constructions and there are several subtleties, depending, for example, on whether or not the orbits are normal or special.

\subsubsection{Ortho-symplectic Quivers}
\label{subsec:BCDQuiverOrtho}

The ortho-symplectic quivers combine the multi-flavoured aspect of the $A$ series Dynkin quivers \cite{nakajima_1994}, with alternating orthogonal and symplectic gauge nodes \cite{Kraft:1982fk}. When working with these quivers, it is again helpful to use the concept of balance. Proceeding as before, we use partition data $\rho$ to construct vectors $({\bf N},{\bf N_f})$ for the vector irrep dimensions of the gauge and flavour nodes, and apply \ref{eq:A1} to calculate a balance vector $\bf B$. The linear quivers ${\cal M}_{BCD}(\rho,0) \to {\cal M}_{BCD} \left( {\bf N,{N_f}} \right)$, whose Higgs branches are $BCD$ nilpotent orbits ${\overline {\cal O}}_{\rho}$, each correspond to a ($B$, $C$ or $D$) partition of a vector irrep, and so $\bf B \ge 0$ \cite{Hanany:2016gbz}.

Remarkably, a full set of quivers whose Higgs branches are $BCD$ group Slodowy intersections ${{{\cal S}}_{{\sigma} ,\rho }}$ can be obtained by following the dimensional logic of \ref{eq:SV4}, thereby extending the method of $A$ series quiver subtractions discussed in section \ref{subsec:AQuivers}. Thus:
\begin{equation} 
\label{eq:BCD1}
\begin{aligned}
{{{{\cal S}}_{\sigma ,\rho }}} & =\higgs[  {\cal M}_{BCD}\left( {\sigma ,\rho } \right)],
 \end{aligned}
\end{equation}
where
\begin{equation} 
\label{eq:BCD2}
\begin{aligned}
{\cal M}_{BCD}\left( {\sigma ,\rho } \right) & \equiv {\cal M}_{BCD} \left( {\sigma,0 } \right) \ominus {\cal M}_{BCD} \left( {\rho,0 } \right),
 \end{aligned}
\end{equation}
and the operation $\ominus$ of quiver subtraction for the $BCD$ series is as defined below.

For ortho-symplectic quivers, the (complex) dimension of the Higgs branch of ${\cal M}_{BCD} \left( {\bf N,{N_f}} \right)$ is found by summing the dimensions of the bi-vector fields and subtracting the gauge group dimensions twice.\footnote{The dimension formula assumes that quivers do not suffer from ``incomplete Higgsing'', which would invalidate the assumed HyperK\"ahler quotient. Quivers from partitions do not have this problem.} This leads to the formula:
\begin{equation} 
\label{eq:BCD3}
\begin{aligned}
 \left | \higgs[ {\cal M}_{BCD} \left( \bf {N,{N_f}} \right) ] \right |= {\bf N \cdot (N_f + K)} -\frac{1}{2} {\bf N \cdot A \cdot N},
 \end{aligned}
\end{equation}
where ${\bf K} = \{K_1, \ldots, K_k  \}$, $K_i =+1$ for an orthogonal node or $-1$ for a symplectic node, and $k$ is the number of gauge nodes. Note again that nodes with $N_i=0$ do not contribute and can be dropped from (or added to) a quiver.

Now consider two quivers ${\cal M}_{BCD} \left( {\bf N_a,{N_f}_a} \right)$ and ${\cal M}_{BCD} \left( {\bf N_b,{N_f}_b} \right)$, with Higgs branches ${\higgs}_{\bf a}$ and ${\higgs}_{\bf b}$, respectively. We have:
\begin{equation} 
\label{eq:BCD4}
\begin{aligned}
\left |{\higgs}_{\bf a} \right | & = {\bf N_a} \cdot ( {\bf{N_f}_a}+{\bf K_a}) -\frac{1}{2} {\bf N_a} \cdot {\bf{A}} \cdot {\bf N_a}\\
\left | {\higgs}_{\bf b} \right | & = {\bf N_b}\cdot ( {\bf {N_f}_b}+{\bf K_b}) -\frac{1}{2} {\bf N_b} \cdot {\bf{A}} \cdot {\bf N_b}
 \end{aligned}
\end{equation}
If we make the assumption that the two quivers have the same balance, $\bf B_a = B_b$, and a compatible orthosymplectic node pattern, $\bf K_a = K_b \equiv K $, then ${\bf {N_f}_b}$ can be eliminated using \ref{eq:A1}, and \ref{eq:BCD4} yields:

\begin{equation} 
\label{eq:BCD5}
\begin{aligned}
\Delta &\equiv \left |{\higgs}_{\bf a} \right |- \left |{\higgs}_{\bf b} \right |\\
& = ( {\bf N_a - N_b} ) \cdot ( {\bf{N_f}_a + K} )- \frac{1}{2} \left( \bf N_a - N_b \right) \cdot {\bf A} \cdot \left( {\bf N_a - N_b} \right).
 \end{aligned}
\end{equation}
Like the $A$ series, $\Delta$ matches the dimension of a third quiver ${\cal M}_{BCD} ({\bf N_a} - {\bf N_b}, {\bf{N_f}_a})$, and this suggests a rule for subtracting two ortho-symplectic quivers with the same flavours $ {\bf{N_{f_a}}}$ and compatible node pattern $\bf K$:

\begin{equation} 
\label{eq:BCD6}
\begin{aligned}
 {\cal M}_{BCD} \left( {\bf N_a, {N_f}_a} \right) & \ominus {\cal M}_{BCD} ({\bf N_a} - {\bf N_b}, {\bf{N_f}_a}) = {\cal M}_{BCD} \left( {\bf N_b, {N_f}_b} \right), \\
 &\text{where}\\
 {\bf {N_f}_b} &= { \bf {N_f}_a - {\bf A} \cdot ({N_a - N_b}}).
 \end{aligned}
\end{equation}
Redefining the gauge vector, $\bf N_b \to N_a - N_b$, this transforms to:
\begin{equation} 
\label{eq:BCD7}
\begin{aligned}
 {\cal M}_{BCD} \left( {\bf N_a, {N_f}_a} \right) & \ominus {\cal M}_{BCD} \left( {\bf N_b, {N_f}_a} \right) = {\cal M}_{BCD} \left({\bf N_a - N_b,  {N_f}_b} \right),
 \\ &\text{where}\\
 {\bf {N_f}_b} &= { \bf {N_f}_a - A \cdot N_b }. 
 \end{aligned}
\end{equation}
Naturally, the gauge ranks in the vector $\bf N_a - N_b$ must be non-negative for the quiver subtraction to be valid. Significantly, the formulae for quiver subtraction, \ref{eq:A7} and \ref{eq:BCD7}, are similar for unitary and ortho-symplectic quivers.


Now, if $\lambda$ is a $B_n$ partition, the quivers ${\cal M}_{BCD}\left( {\lambda, 0} \right)$, have the same flavour node vector $ {\bf N_f}=\{2n+1,0,\dots,0\}$, and similarly, if $\lambda$ is a $C_n$ or $D_n$ partition, they share $ {\bf N_f}=\{2n,0,\dots,0\}$. Consequently, by allowing $\sigma$ and $\rho$ to range over each set of $B_n$, $C_n$ and $D_n$ partitions in turn, and by using \ref{eq:BCD1}, \ref{eq:BCD2} and \ref{eq:BCD7}, we can obtain a full set of quiver candidates ${\cal M}_{BCD}(\sigma,\rho)$ for the Higgs branch constructions of $B$, $C$ and $D$ series Slodowy intersections ${\cal S}_{\sigma,\rho}$.

\begin{figure}[htbp]
\centering
\includegraphics[scale=0.45]{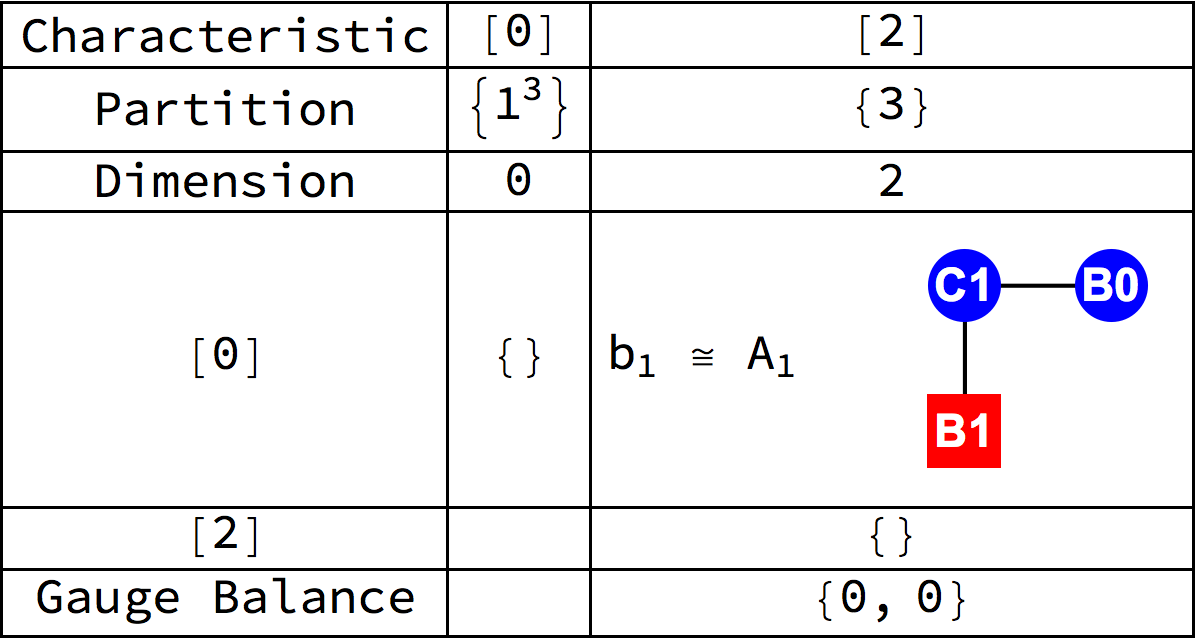}{~~~~~~~~~~}
\includegraphics[scale=0.45]{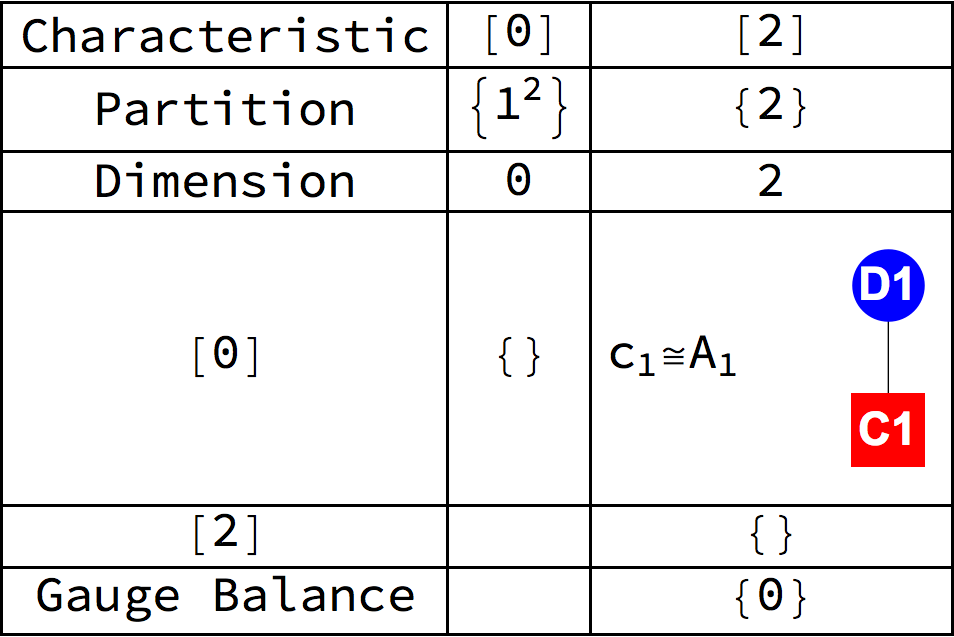}\\
\caption[O-USp Quivers for $B_1 \cong C_1$ Slodowy Intersections.]{O-USp quivers for $B_1 \cong C_1$ Slodowy Intersections. The Higgs branches are Slodowy intersections ${\cal S}_{\sigma ,\rho }$. Rows (columns) are labelled by Characteristics of ${ {\cal O}}_\rho$ (${ {\cal O}}_\sigma$).}
\label{fig:B1}
\end{figure}

\begin{figure}[htbp]
\centering
\includegraphics[scale=0.35]{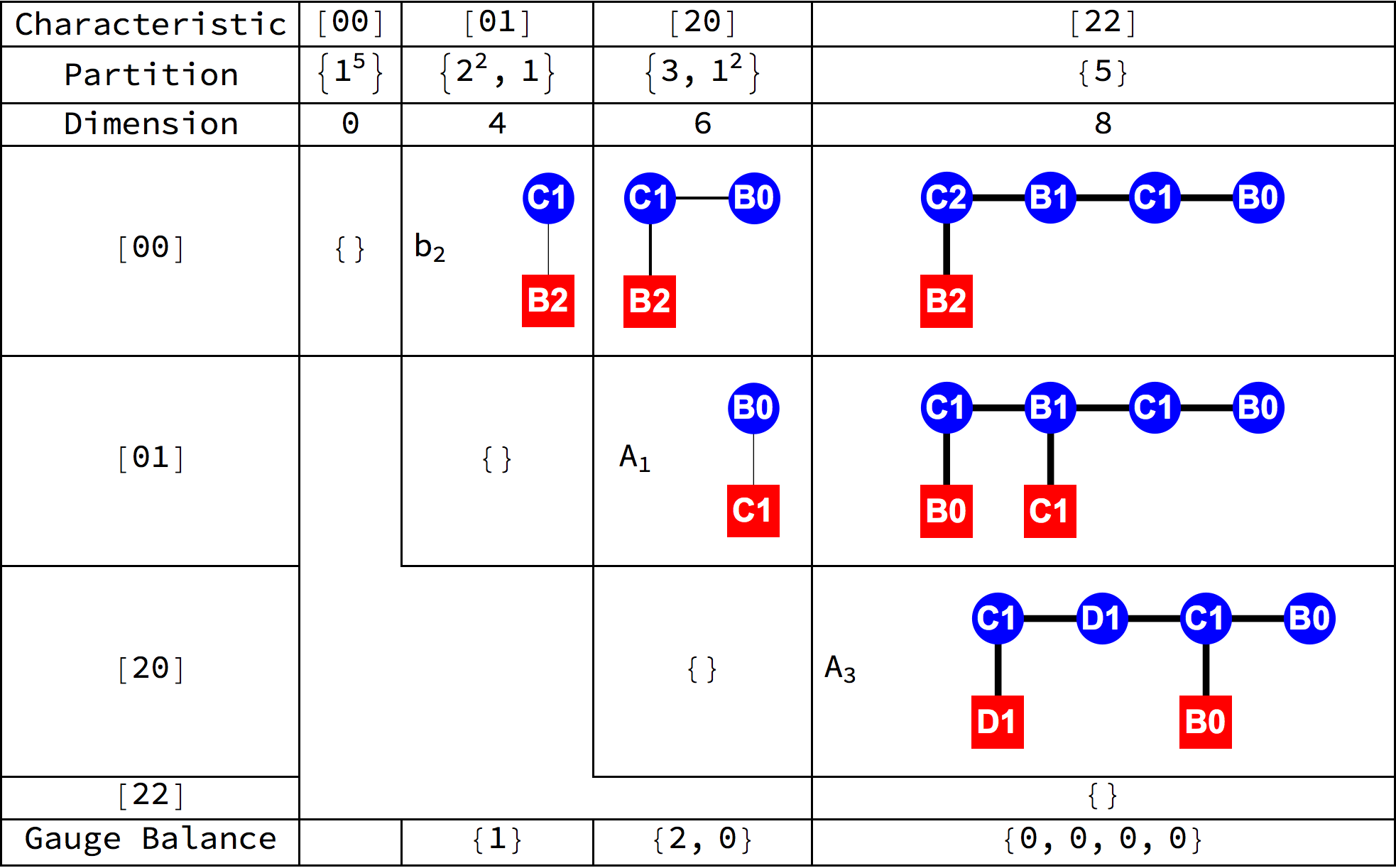}{~~~~~}
\includegraphics[scale=0.35]{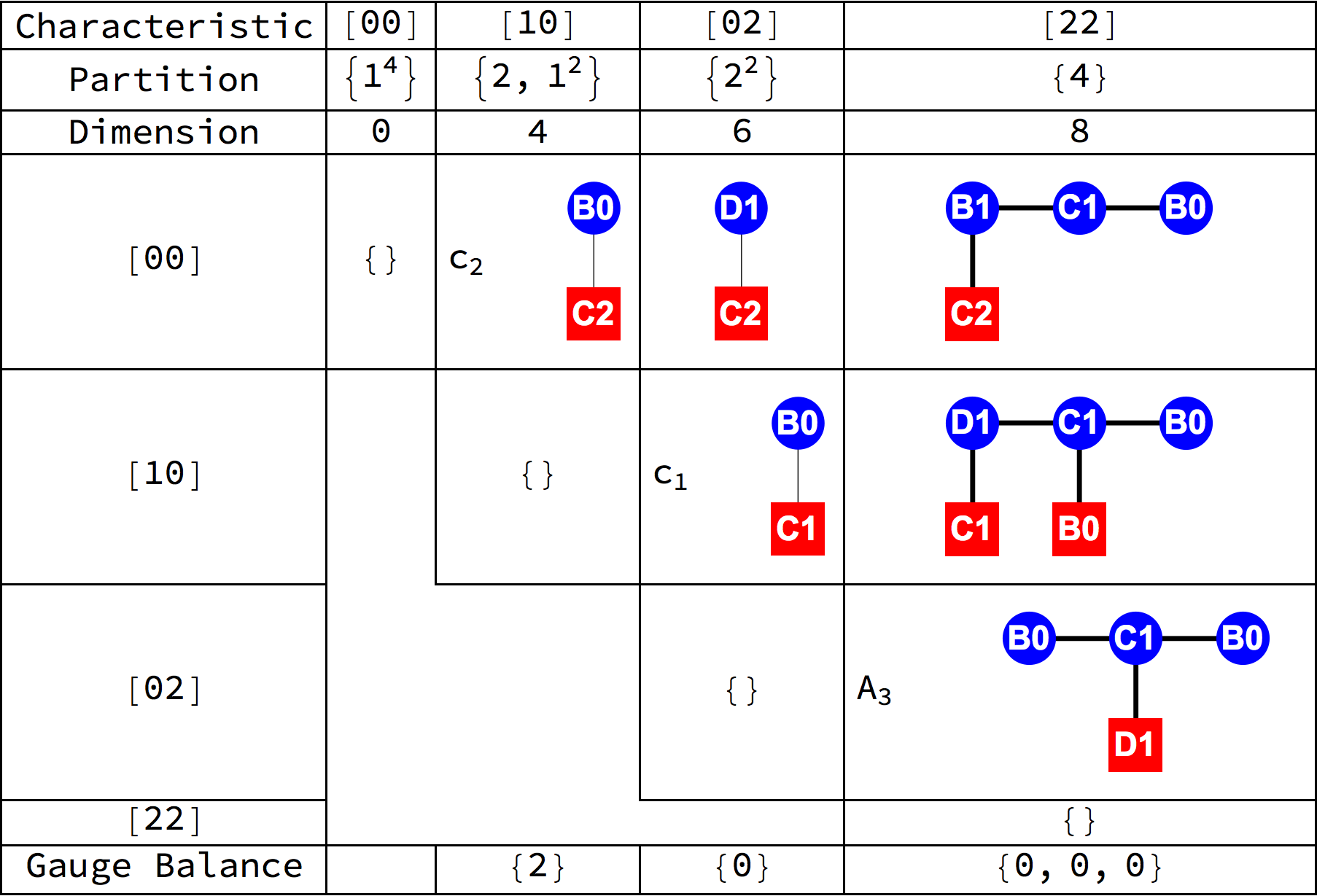}\\
\caption[O-USp Quivers for $B_2  \cong C_2$ Slodowy Intersections.]{O-USp quivers for $B_2  \cong C_2$ Slodowy Intersections. The Higgs branches are Slodowy intersections ${\cal S}_{\sigma ,\rho }$. Rows (columns) are labelled by Characteristics of ${ {\cal O}}_\rho$ (${ {\cal O}}_\sigma$).}
\label{fig:B2}
\end{figure}

\begin{figure}[htbp]
\centering
\includegraphics[scale=0.45]{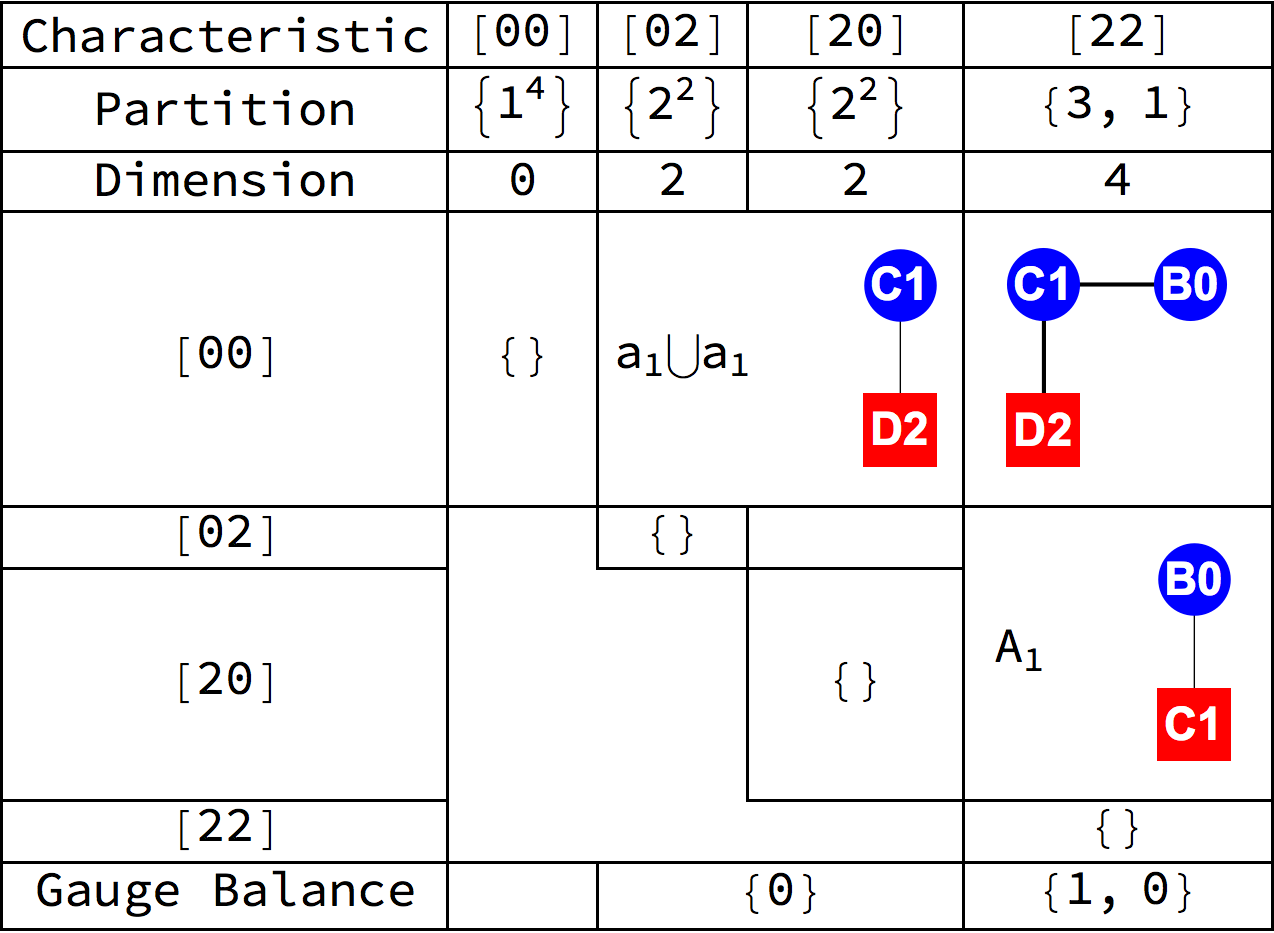}\\
\caption[O-USp Quivers for $D_2$ Slodowy Intersections.]{O-USp quivers for $D_2$ Slodowy Intersections. The Higgs branches are Slodowy intersections ${\cal S}_{\sigma ,\rho }$. Rows (columns) are labelled by Characteristics of ${ {\cal O}}_\rho$ (${ {\cal O}}_\sigma$).}
\label{fig:D2}
\end{figure}

\begin{figure}[htbp]
\centering
\includegraphics[scale=0.375]{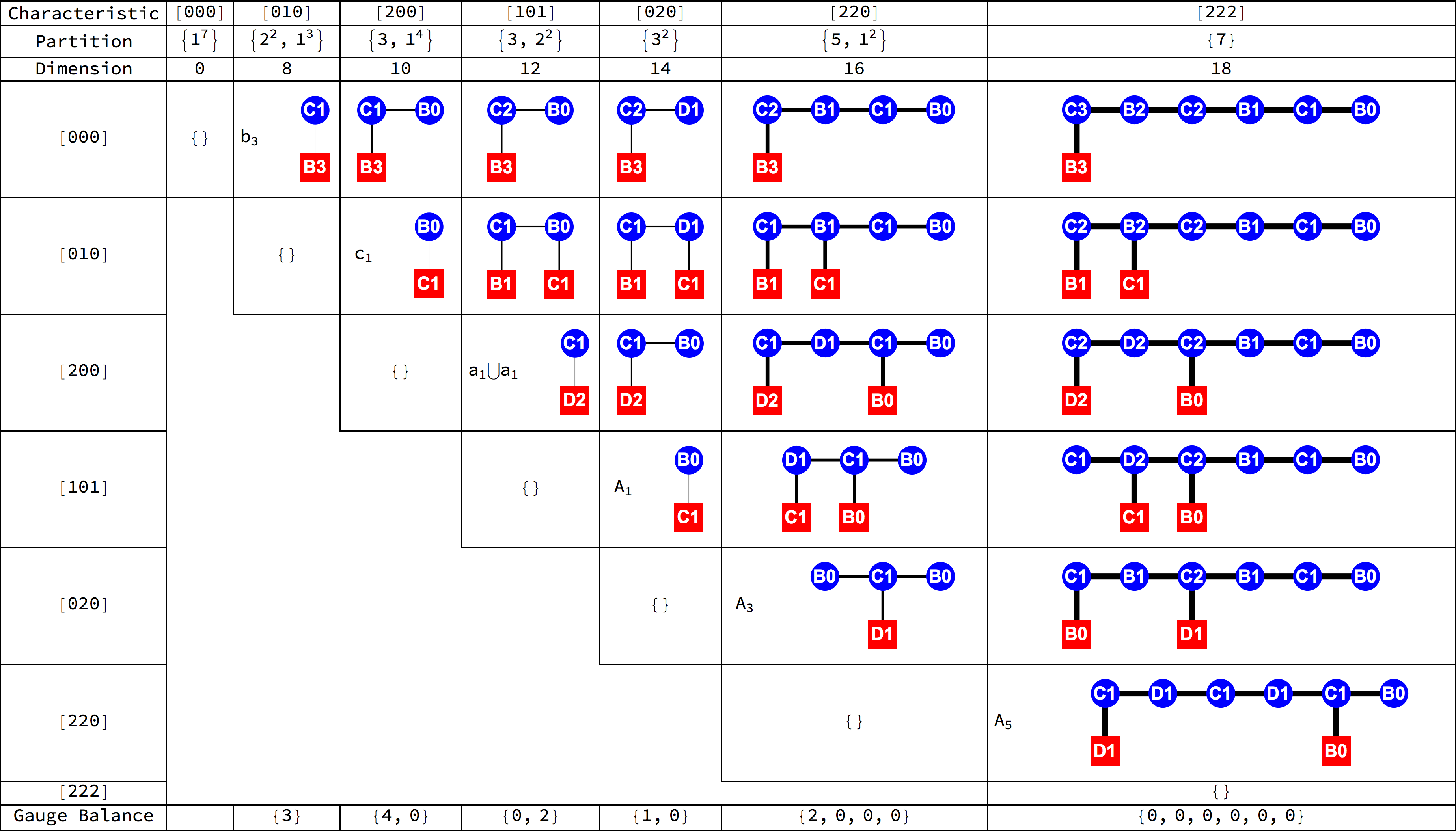}\\
\caption[O-USp Quivers for $B_3$ Slodowy Intersections.]{O-USp quivers for $B_3$ Slodowy Intersections. The Higgs branches are Slodowy intersections ${\cal S}_{\sigma ,\rho }$. Rows (columns) are labelled by Characteristics of ${ {\cal O}}_\rho$ (${ {\cal O}}_\sigma$).}
\label{fig:B3}
\end{figure}

\begin{figure}[htbp]
\centering
\includegraphics[scale=0.35]{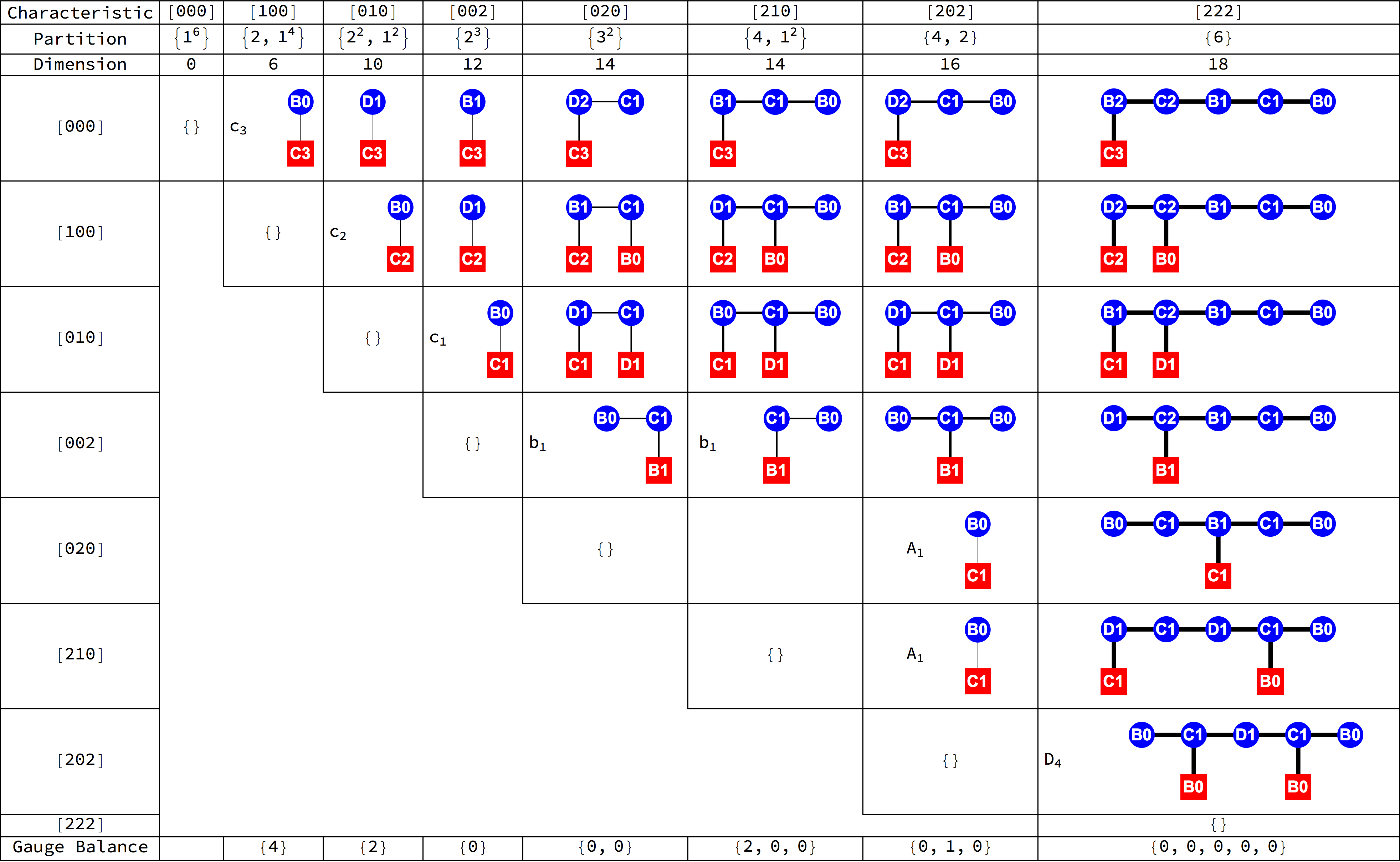}\\
\caption[O-USp Quivers for $C_3$ Slodowy Intersections.]{O-USp quivers for $C_3$ Slodowy Intersections. The Higgs branches are Slodowy intersections ${\cal S}_{\sigma ,\rho }$. Rows (columns) are labelled by Characteristics of ${ {\cal O}}_\rho$ (${ {\cal O}}_\sigma$).}
\label{fig:C3}
\end{figure}

\begin{figure}[htbp]
\centering
\includegraphics[scale=0.4]{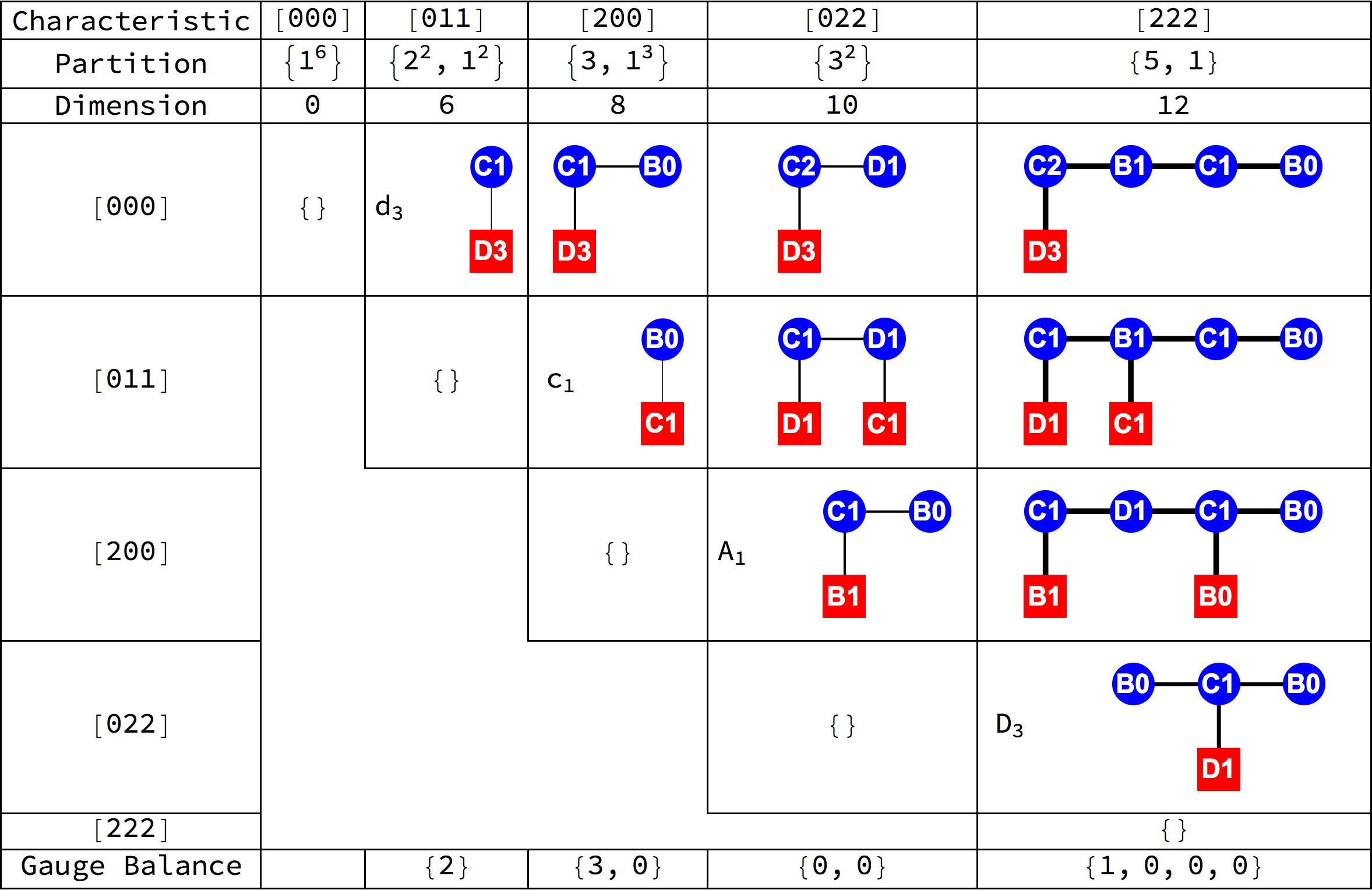}\\
\caption[O-USp Quivers for $D_3$ Slodowy Intersections.]{O-USp quivers for $D_3$ Slodowy Intersections. The Higgs branches are Slodowy intersections ${\cal S}_{\sigma ,\rho }$. Rows (columns) are labelled by Characteristics of ${ {\cal O}}_\rho$ (${ {\cal O}}_\sigma$).}
\label{fig:D3}
\end{figure}

\begin{sidewaysfigure}[htbp]
\centering
\includegraphics[scale=0.24]{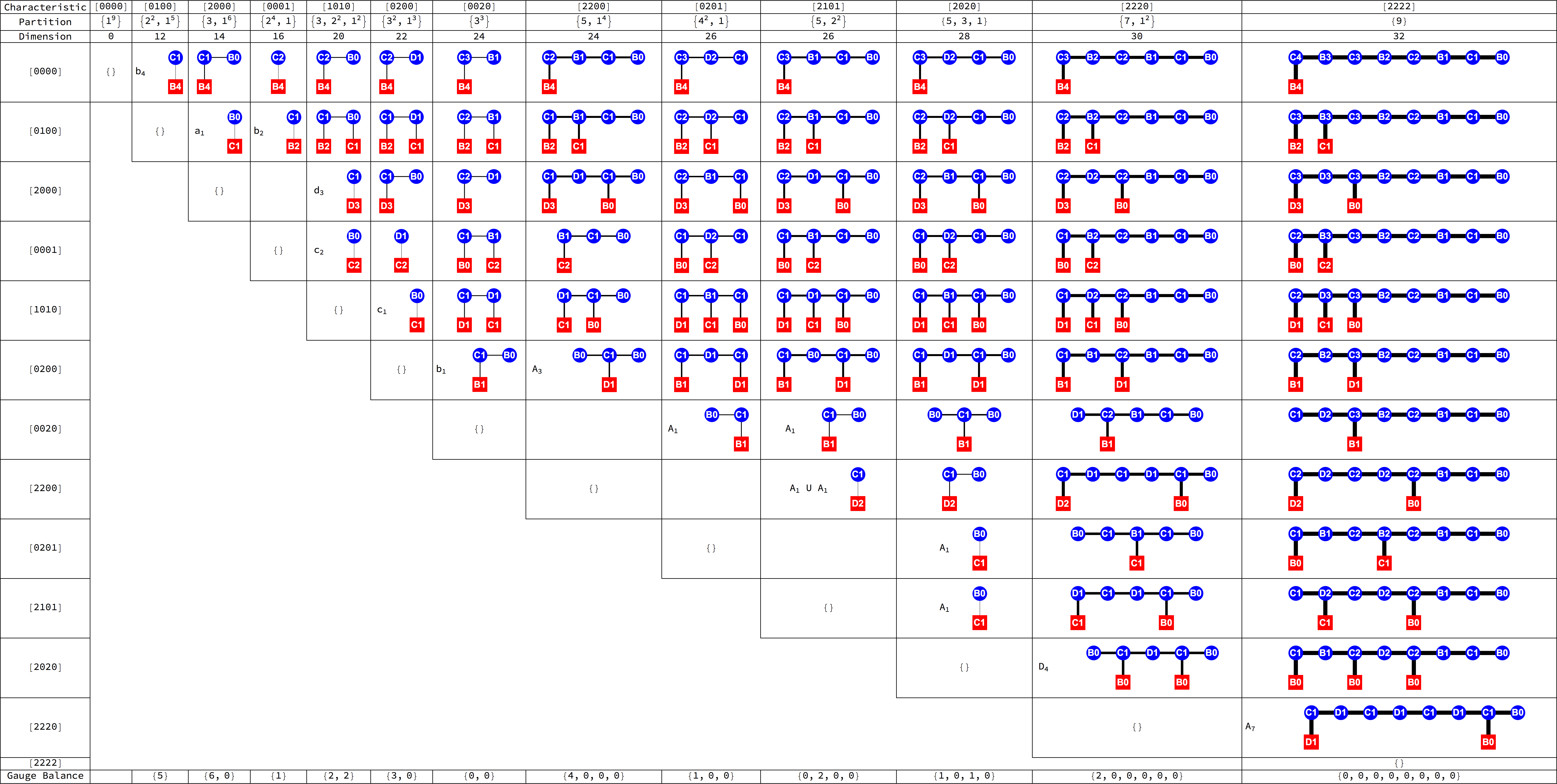}\\
\caption[O-USp Quivers for $B_4$ Slodowy Intersections.]{O-USp quivers for $B_4$ Slodowy Intersections. The Higgs branches are Slodowy intersections ${\cal S}_{\sigma ,\rho }$. Rows (columns) are labelled by Characteristics of ${ {\cal O}}_\rho$ (${ {\cal O}}_\sigma$).}
\label{fig:B4}
\end{sidewaysfigure}

\begin{sidewaysfigure}[htbp]
\centering
\includegraphics[scale=0.24]{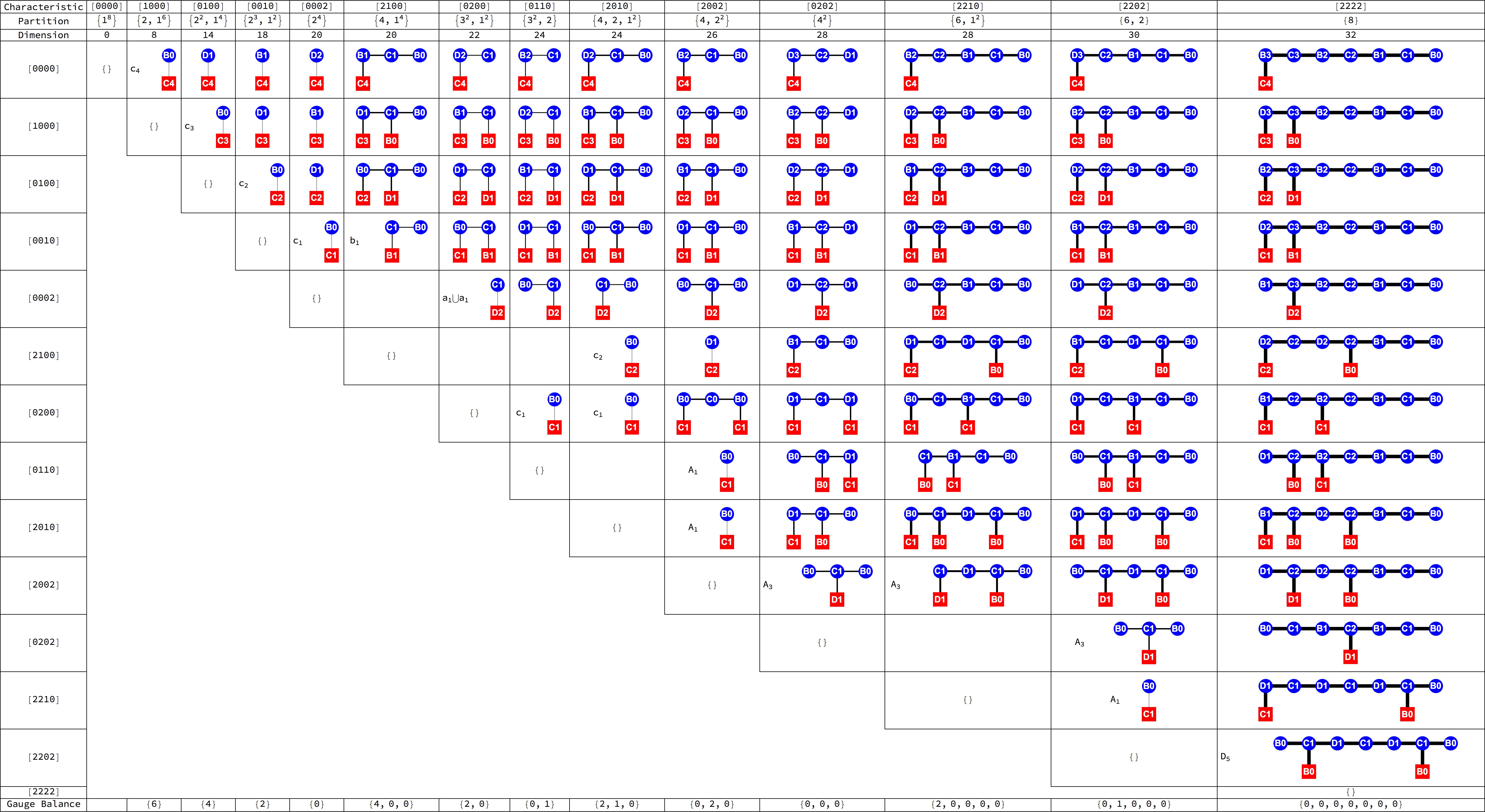}\\
\caption[O-USp Quivers for $C_4$ Slodowy Intersections.]{O-USp quivers for $C_4$ Slodowy Intersections. The Higgs branches are Slodowy intersections ${\cal S}_{\sigma ,\rho }$. Rows (columns) are labelled by Characteristics of ${ {\cal O}}_\rho$ (${ {\cal O}}_\sigma$).}
\label{fig:C4}
\end{sidewaysfigure}

\begin{sidewaysfigure}[htbp]
\centering
\includegraphics[scale=0.34]{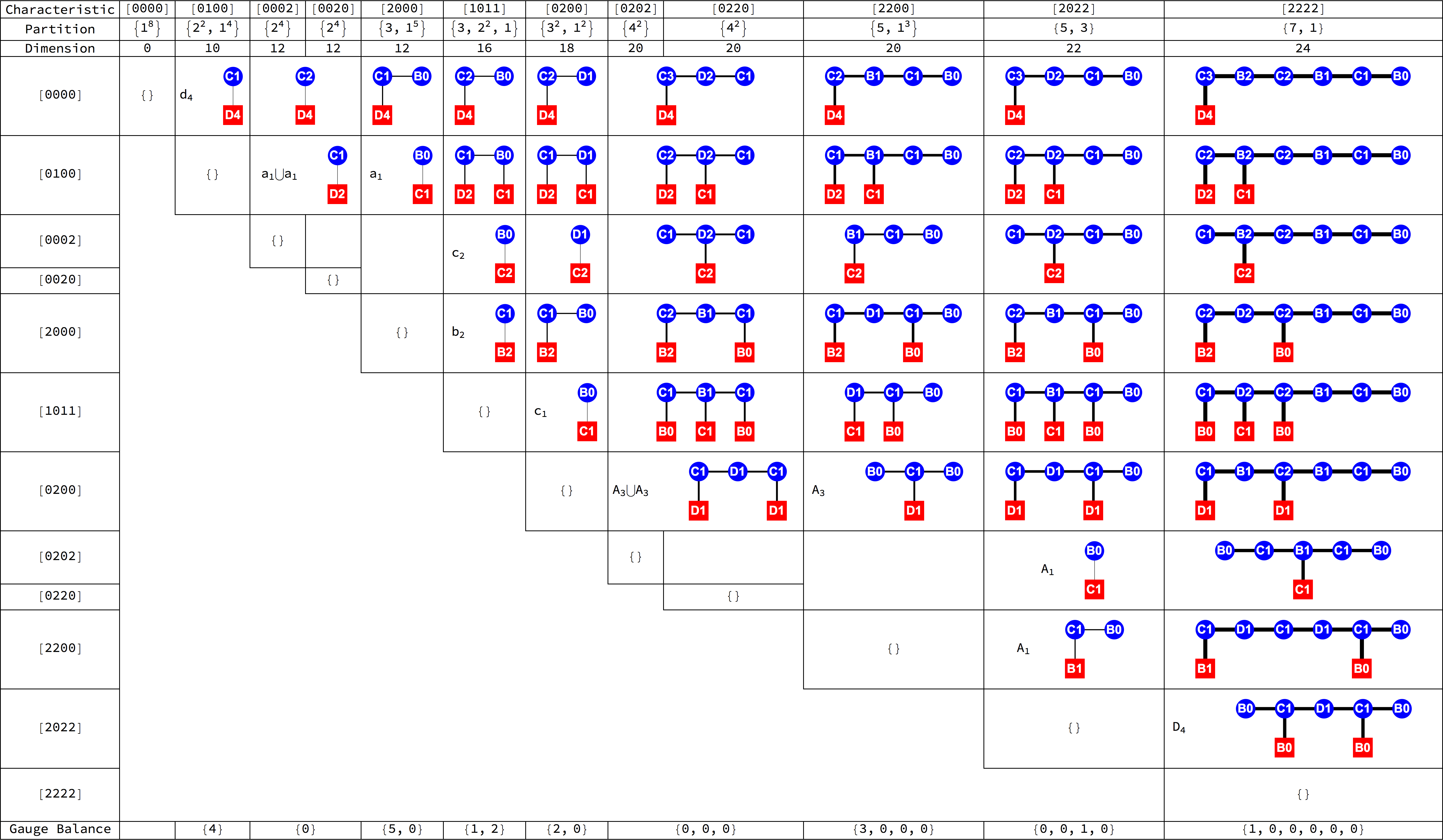}\\
\caption[O-USp Quivers for $D_4$ Slodowy Intersections.]{O-USp quivers for $D_4$ Slodowy Intersections. The Higgs branches are Slodowy intersections ${\cal S}_{\sigma ,\rho }$. Rows (columns) are labelled by Characteristics of ${ {\cal O}}_\rho$ (${ {\cal O}}_\sigma$).}
\label{fig:D4}
\end{sidewaysfigure}

These Higgs branch quivers for the ${\cal S}_{\sigma ,\rho }$ of $B$, $C$ and $D$ groups up to rank $4$ are shown in figures \ref{fig:B1} through \ref{fig:D4}. These are arranged as matrices, with rows and columns labelled by the Characteristics of ${ {\cal O}}_\rho$ and ${ {\cal O}}_\sigma$, respectively. Vector partitions and dimensions of ${ {\cal O}}_\sigma$ are also shown, as well as the balance vector $\bf B$, which by construction is constant down each column. Trivial self-intersections are denoted $\{ \}$. The Kraft-Procesi transition for each row is labelled by its minimal singularity, as described in \ref{sec:SV_KP}. Empty entries indicate the absence of an intersection. Gauge nodes of dimension zero are truncated.

The matrices are all of upper triangular form. Each top row contains quivers whose Higgs branches are closures of nilpotent orbits ${\cal S}_{\sigma ,0 }$. Each rightmost column contains quivers whose Higgs branches are Slodowy slices ${\cal S}_{{\cal N} ,\rho }$. The Higgs branches of the first non-empty entries above each diagonal are Kraft-Procesi transitions. Quivers whose Higgs branches are more general intersections ${\cal S}_{\sigma, \rho }$ appear from rank 3 upwards.

\FloatBarrier

The Slodowy intersections ${\cal S}_{\sigma ,\rho }$ in each row transform in the same group $F(\rho)$ as the slice ${\cal S}_{{\cal N} ,\rho }$, although lower dimensioned intersections (such as Kraft-Procesi transitions), may transform trivially under some component(s) of $F(\rho)$.\footnote{The non-Abelian components of the centralisers $F(\rho)$ are isomorphic with those tabulated in \cite{liebeck_seitz_2012}. } Any two quivers ${\cal M}_{BCD}(\sigma_1,\rho)$ and ${\cal M}_{BCD}(\sigma_2,\rho)$ in the same row, where $\sigma_1 > \sigma_2$, are related by quiver subtraction to a third quiver ${\cal M}_{BCD}(\sigma_1,\sigma_2)$ in a row below. All the quivers have non-negative balance, ${\bf B \ge 0}$. 

While each Slodowy intersection of $G(N_0) \in \{B_n, C_n, D_n \}$ is constructed from a pair of partitions of $N_0 \in \{2n+1,2n,2n\}$, respectively, it can also be constructed from a pair of partitions of $N_0'$, where $N_0' \equiv \sum\nolimits_{i = 1}^k {i {N_{f_i}}} $, and is thus also related to the ambient group $G(N_0')$.\footnote{The partitions can be found from a given ${\cal M}_{BCD} \left( {\bf N,{N_f}} \right)$ with $\bf B \ge 0$ by considering two linear quivers with flavour $N'_0$ and applying \ref{eq:A7} in reverse.} If an intersection transforms trivially under some component of $F(\rho)$, then $N_0' < N_0$, and it also appears amongst the intersections of $G(N_0') \subset G(N_0)$. Notably, $G(N_0')$ need not be from the same series as $G(N_0)$. A similar logic applies to sub-diagrams, which can reappear as intersection diagrams for sub-groups of $G(N_0)$.

While these matrices of ortho-symplectic quivers, whose Higgs branches are Slodowy intersections, have a similar structure to the $A$ series, being upper triangular and related by quiver subtractions, their Coulomb branches do not simply correspond to the (Special duals of) the same intersections.
\begin{enumerate}
\item As noted in section \ref{sec:SV_dual}, the Barbasch-Vogan map only generates vector partitions for \emph{special} orbits, and, furthermore, acts to interchange $B$ and $C$ partitions.
\item The dimension of a Coulomb branch equals twice the sum of the gauge node ranks, and is not proportional to $B$ series vector dimensions. Thus, rank reduces when a $D$ vector is broken into two $B$ vectors, and Coulomb branch ortho-symplectic quiver candidates for Slodowy intersections are not generally related by quiver subtractions.

\item When ortho-symplectic quivers are evaluated on the Higgs branch, $BD$ gauge nodes are taken as $O$ type, by averaging over the relevant ${\mathbb Z}_2$ finite group factors, whereas on the Coulomb branch, a careful selection needs to made between various $SO/O$ possibilities \cite{Cabrera:2017ucb}.

\end{enumerate}

For these reasons, there is no straightforward procedure for finding a complete set of ortho-symplectic quiver candidates for Coulomb branch constructions of the ${\cal S}_{\sigma ,\rho }$. Other complications include the failure of the ortho-symplectic monopole formula for quivers with zero conformal dimension, and its limitation to unrefined HS \cite{Cremonesi:2013lqa}.

Nonetheless, in \cite{Cabrera:2018ldc}, quiver candidates for Coulomb branch constructions of Slodowy slices ${\cal S}_{{\cal N} ,\rho }$, for \emph{special orbits} only, were tabulated for $BCD$ groups up to rank $4$. These are linear ortho-symplectic quivers, containing pure $BC$, $CD$ or $DC$ chains of nodes with ordered vector dimensions, but having the same Higgs branches as the quivers for ${\overline{\cal O}}_{\rho }$ (in the top rows of figures \ref{fig:B1} through \ref{fig:D4}). Their quiver subtractions yield \emph{some} Coulomb branch quiver candidates for Slodowy intersections ${\cal S}_{\sigma ,\rho }$ (of pairs of special orbits), and some other candidates can be found by $BD$ gauge node shifting to Higgs equivalent quivers, using trial and error, and with judicious choice of O/SO gauge nodes. We do not tabulate these quivers, but comment that their \emph{unrefined} Coulomb branch Hilbert series (where calculable) appear consistent with those presented herein. 


\subsubsection{Dynkin Quivers}
\label{subsec:BCDQuiverDynkin}

A set of Coulomb branch constructions, based on $ADE$ affine Dynkin diagrams, has been known since \cite{Intriligator:1996ex} for Slodowy intersections that are \emph{minimal nilpotent orbits}. This set was extended, in \cite{Cremonesi:2014xha}, to include minimal nilpotent orbits of \emph{non-simply laced} groups by modifications to the Coulomb branch unitary monopole formula, and, in \cite{Hanany:2016gbz}, to include next to (or near to) minimal orbits, by using twisted affine Dynkin diagrams.

Coulomb branch constructions are also available for many $BCD$ series intersections that are KP transitions; these are either minimal orbits (as above) or $AD$ singularities (or their unions). If the KP transitions are $A$ type singularities, they are sub-regular Slodowy slices for some $A_n$, and so have Coulomb branch constructions as in section \ref{subsec:AQuivers}. Some intersections ${\cal S}_{\sigma, \rho }$ that are adjacent to KP transitions are next to (or near to) minimal orbits of $F(\rho)$ and therefore also have Coulomb branch constructions.

It was observed in \cite{Hanany:2017ooe} that these Coulomb branch constructions for the $BCD$ series are limited to balanced Dynkin quivers which have Characteristic height $[\theta]= 2$, where $ [\theta] \equiv \bf N_f \cdot a$, with ${\bf a}$ given by Coxeter labels. This largely limits the ${{\cal D}_{BCD}}\left({\bf N, \bf N_{f}} \right)$ that are relevant to intersections to the types noted above.

Although we do not repeat tables of these Dynkin quivers here\footnote{Tables of Dynkin quivers ${\cal D}_{BCD}$ for nilpotent orbits up to rank $4$ are given in \cite{Hanany:2016gbz}.}, they generate refined Hilbert series from the Coulomb branch unitary monopole formula \cite{Cabrera:2018ldc} (section 3.3) and these are consistent with those presented herein.

\begin{figure}[htbp]
\centering
\includegraphics[scale=0.375]{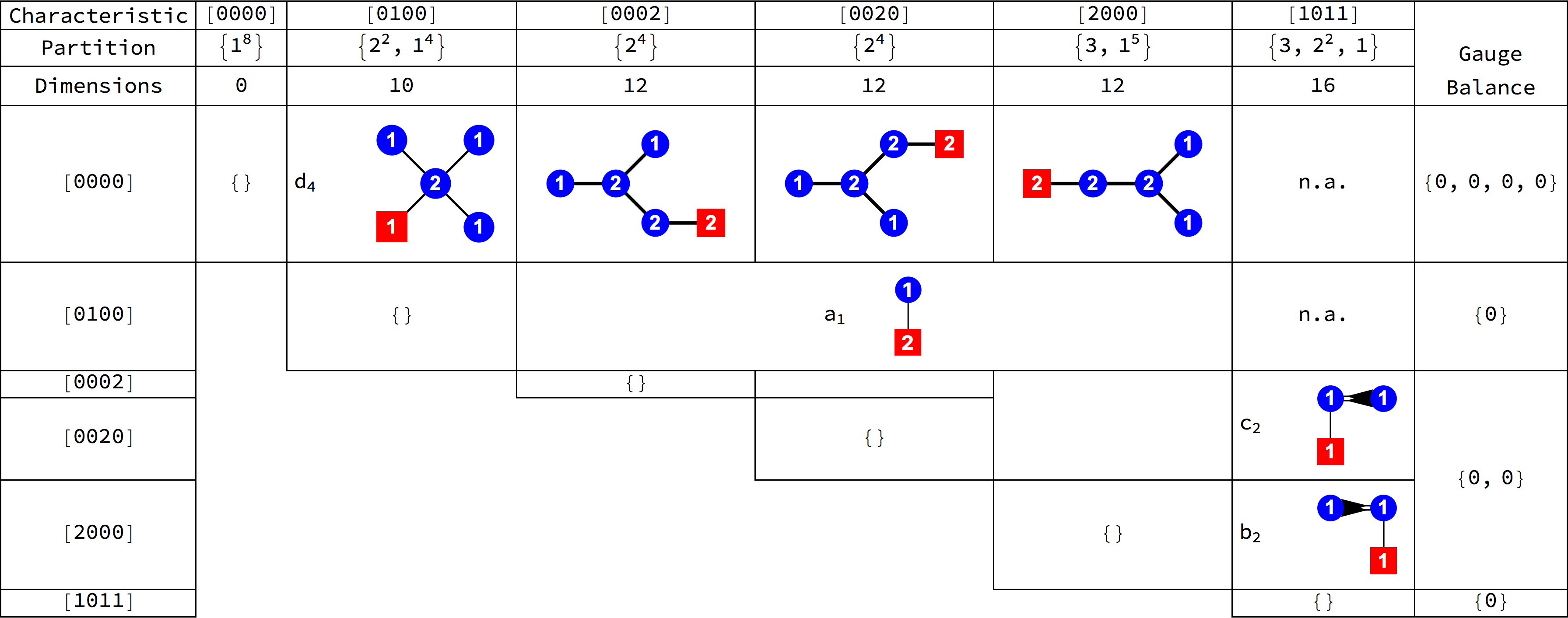}\\
\caption[$D_4$ Slodowy Intersections from Dynkin Quiver Coulomb Branches.]{$D_4$ Slodowy Intersections from Dynkin Quiver Coulomb Branches. Coulomb branches yield Slodowy intersections ${\cal S}_{\sigma ,\rho }$. Rows (columns) are labelled by Characteristics of ${ {\cal O}}_\rho$ (${ {\cal O}}_\sigma$). Only intersections between low dimensioned orbits are shown.}
\label{fig:dynkinD1}
\end{figure}

\begin{figure}[htbp]
\centering
\includegraphics[scale=0.375]{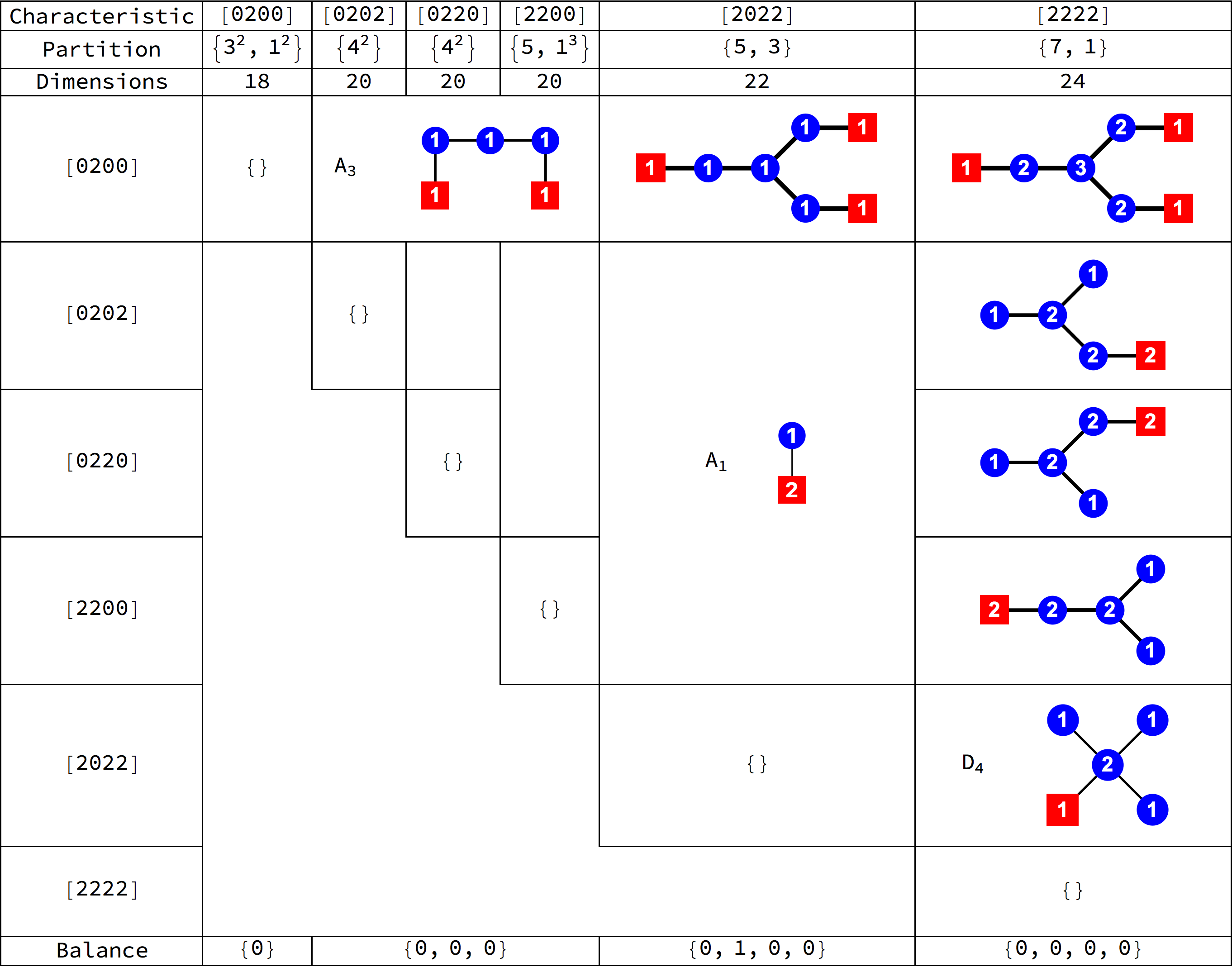}\\
\caption[$D_4$ Slodowy Intersections from Dynkin Quiver Higgs Branches.]{$D_4$ Slodowy Intersections from Dynkin Quiver Higgs Branches. Higgs branches yield Slodowy intersections ${\cal S}_{\sigma ,\rho }$. Rows (columns) are labelled by Characteristics of ${ {\cal O}}_\rho$ (${ {\cal O}}_\sigma$). Only intersections between high dimensioned orbits are shown.}
\label{fig:dynkinD2}
\end{figure}

In the case of the $D$ series, Dynkin quivers ${{\cal D}_{AD}}\left({\bf N, \bf N_{f}} \right)$, provide Higgs branch constructions for intersections close to the sub-regular Slodowy slice, as well as Coulomb branch constructions (as above) for intersections close to the minimal orbit. 
As an example, Dynkin quivers for Coulomb and Higgs branch constructions of the ${\cal S}_{\sigma, \rho }$ of $D_4$ are tabulated in figures \ref{fig:dynkinD1} and \ref{fig:dynkinD2}. The matrices are restricted to those intersections amenable to this approach.\footnote{Some of these quivers were studied in \cite{henderson_licata_2014}, where the Higgs branches of ${\cal D}_{D_n}$ quivers were shown to correspond to Slodowy slices of $C_{n-1}$ or $D_{n-1}$.}

%

There are several noteworthy features:
\begin{enumerate}
\item The quivers that appear in the Coulomb branch constructions contain Dynkin diagrams of simple subgroups of $F(\rho)$.
\item There is manifest triality between the quivers for $\left\{ {\orbit{[2000]},\orbit{[ {0002} ]},\orbit{ [ {0020} ]} } \right\}$ and likewise between their $d_{BV}$ duals $\left\{ \slice{[ {2200} ]},\slice{[ {0202} ]},\slice{ [ {0220} ] } \right\}$. This has the consequence that all these Coulomb branch constructions yield normal HS.
\item The Coulomb branch constructions are all based on Dynkin quivers with Characteristic height $[\theta]= 2$; the Higgs branch constructions contain two additional quivers with $[\theta]= 3$, including a quiver for ${{{\cal S}}_{[2022],[0200]}}$.
\item The positions of the $[\theta]= 2$ quivers in figures \ref{fig:dynkinD1} and \ref{fig:dynkinD2} can be related by Special duality, as defined in section \ref{sec:SV_dual}. Note that under the $d_{BV}$ map for $D_4$, all the orbits/partitions, with the exception of $D[1011]$, are special, with $d_{BV}^2=1$:
\begin{equation}
\small
\label{eq:D1}
\begin{aligned}
\left\{ {\left[ {0000} \right],\left[ {0100} \right],\left[ {2000} \right],\left[ {0002} \right], \ldots } \right\}  \mathop  \leftrightarrow \limits_{{d_{BV}}}& \left\{ {\left[ {2222} \right],\left[ {2022} \right],\left[ {2200} \right],\left[ {0202} \right], \ldots } \right\},\\
\left\{ {\left[ {0200} \right],\left[ {1011} \right]} \right\}\mathop  \to \limits_{{d_{BV}}}& \left\{ {\left[ {0200} \right],\left[ {0200} \right]} \right\}.
 \end{aligned}
\end{equation}
\item The quivers in figure \ref{fig:dynkinD2} can be related by quiver subtractions and the dimensions of their Higgs branches are given by a generalisation of equation \ref{eq:A3}, in which $\bf A$ is taken as the $D_4$ Cartan matrix.
\end{enumerate}
It is notable that Special duality between the Slodowy intersections constructed on the Higgs and Coulomb branches of these Dynkin quivers is limited to quivers of Characteristic height 2, for which the $d_{BV}$ map is one to one.

\FloatBarrier


\subsection{Hilbert Series}
\label{subsec:BCDHS}

Hilbert series $g_{HS}^{{\cal S}_{\sigma, \rho}}({\bf x},t)$ for $BCD$ type Slodowy intersections ${{\cal S}_{\sigma, \rho}}$ can be calculated using the Higgs branch formula, as described in \cite{Cabrera:2018ldc} (section 4.2), from ${\cal M}_{BCD} \left( {\sigma,\rho } \right)$ quivers, such as those tabulated in figures \ref{fig:B1} through \ref{fig:D4} .

The results for $BCD$ groups of rank up to 4 are summarised in tables \ref{tab:BHS12} through \ref{tab:DHS4c}. These are labelled by pairs of Characteristics $(\sigma, \rho)$ and set out, for each non-trivial Slodowy intersection, its dimension, its symmetry group $F(\rho)$, its unrefined Hilbert series $g_{HS}^{{\cal S}_{\sigma, \rho}}(t)$, and the HWG $g_{HWG}^{\svar{\sigma}{\rho}}({\bf m},t)$ (expressed as a PL) that decodes $g_{HS}^{{\cal S}_{\sigma, \rho}}({\bf x},t)$ into irreps of $F(\rho)$. Refined HS and HWGs lacking finite PLs are not tabulated (due to space constraints). Non-normal intersections are highlighted. Trivial self-intersections, with Hilbert series $g_{HS}^{{\cal S}_{\rho, \rho}}=1$, are omitted. 


\begin{sidewaystable}[h]
	\centering
\begin{tabular}{|c|c|c|c|c|c|}
\hline
	$\rho$ & $\sigma$ & $\left | {{\cal S}_{\sigma, \rho}} \right |$ & $F(\rho)$ & Unrefined HS & PL[HWG] \\
 \hline
	$[0]$ & $[2]$ & 2 & $B_1 \cong C_1$ & $\frac{1-t^4}{(1-t^2)^3}$ & $m^2 t^2$  \\
\hline
	 \hline
	 $[00]$ & $[01]$ & 4 & $B_2$ & $\frac{1+6 t^2+t^4}{(1-t^2)^4}$ &	 	$m_2^2 t^2$  \\
	 $ $ & $[20]$ & 6 & $ $ & 		$\frac{(1-t^4) (1+3 t^2+t^4)}{(1-t^2)^7}$ & 	$m_2^2 t^2 + m_1^2 t^4$  \\
	 $ $ & $[22]$ & 8 & $ $ & 		$\frac{(1-t^4) (1-t^8)}{(1-t^2)^{10}}$ & 	$m_2^2 t^2 + m_1 t^4 + m_1^2 t^4 + m_2^2 t^6 + m_1 m_2^2 t^8 - m_1^2 m_2^4 t^{16}$  \\
    	 \hline
	 $[01]$ & $[20]$ & 2 & $C_1 \otimes B_0 $ & 	$\frac {1-t^4} {(1-t^2)^3}$ & 	$m^2 t^2$  \\
	 $ $ & $[22]$ & 4 & $ $ & 		$\frac{1-t^8}{(1-t^2)^3 (1-t^3)^2}$ & 	$m^2 t^2 + m t^3 + t^4 + m t^5 - m^2 t^{10}$  \\
	  \hline
	   $[20] $ & $[22]$ & 2 & $D_1 \otimes B_0 $ & 	$\frac{1-t^8}{(1-t^2) (1-t^4)^2}$ & 	$t^2 + (q+1/q) t^4 - t^8$  \\
	     \hline
	 \hline
	 $[00]$ & $[10]$ & 4 & $C_2$ & $\frac{1+6 t^2+t^4}{(1-t^2)^4}$ &	 	$m_1^2 t^2$  \\
	 $ $ & $[02]$ & 6 & $ $ & 		$\frac{(1-t^4) (1+3 t^2+t^4)}{(1-t^2)^7}$ & 	$m_1^2 t^2 + m_2^2 t^4$  \\
	 $ $ & $[22]$ & 8 & $ $ & $\frac{(1-t^4) (1-t^8)}{(1-t^2)^{10}}$ & $m_1^2 t^2 + m_2 t^4 + m_2^2 t^4 + m_1^2 t^6 +m_1^2 m_2  t^8 -m_1^4 m_2^2  t^{16}$  \\
    	 \hline
	 $[10]$ & $[02]$ & 2 & $C_1 \otimes B_0 $ & 	$\frac {1-t^4} {(1-t^2)^3}$ & 	$m^2 t^2$  \\
	 $ $ & $[22]$ & 4 & $ $ & 		$\frac{1-t^8}{(1-t^2)^3 (1-t^3)^2}$ & 	$m^2 t^2 + m t^3 + t^4 + m t^5 - m^2 t^{10}$  \\
	  \hline
	   $[02] $ & $[22]$ & 2 & $D_1 \otimes B_0 $ & 	$\frac{1-t^8}{(1-t^2) (1-t^4)^2}$ & 	$t^2 + (q+1/q) t^4 - t^8$  \\
\hline

	\end{tabular}
	\caption[$B_1\cong C_1$ and $B_2 \cong C_2$ Slodowy Intersections]{$B_1\cong C_1$ and $B_2 \cong C_2$ Slodowy Intersections. $B_2$ and $C_2$ intersections are isomorphic, under reversal of Dynkin labels.}
	\label{tab:BHS12}
\end{sidewaystable}

\begin{sidewaystable}[h]
	\centering
	\small
	\begin{tabular}{|c|c|c|c|c|c|}
\hline
	$\rho$ & $\sigma$ & $\left | {{\cal S}_{\sigma, \rho}} \right |$ & $F(\rho)$ & Unrefined HS & PL[HWG] \\
 \hline
	$[000]$ & $[010]$ & 8 & $B_3$ & $\frac{1+13 t^2+28 t^4+13 t^6+t^8}{(1-t^2)^8}$ & $m_2 t^2$  \\
	$        $ & $[200]$ & 10 & $    $ & $\frac{(1-t^4) (1+10 t^2+20 t^4+10 t^6+t^8)}{(1-t^2)^{11}}$ & $m_2 t^2 + m_1^2 t^4$  \\
	$        $ & {\color{red} [101]} & 12 & $    $ & $\frac{(1-t^4) (1+8 t^2+36 t^4+92 t^6+t^8-6 t^{10})}{(1-t^2)^{13}}$ &
	 $m_2 t^2 + m_1^2 t^4 + m_3^2 t^4 + m_1 m_3^2 t^6 - m_1^2 m_3^4 t^{12}$  \\
	$        $ & $[020]$ & 14 & $    $ & $\frac{(1-t^4) (1+6 t^2+21 t^4+28 t^6+21 t^8+6 t^{10}+t^{12})}{(1-t^2)^{15}}$ & 
	$\begin{array}{c}
	m_2 t^2 + m_1^2 t^4 + m_3^2 t^4 + m_1 m_3^2 t^6 + m_2^2 t^8 + m_1 m_2 m_3^2 t^{10} \\- m_1^2 m_2^2 m_3^4 t^{20}
	\end{array}$  \\
	$        $ & $[220]$ & 16 & $    $ & $\frac{(1-t^4) (1-t^8) (1+3 t^2+6 t^4+3 t^6+t^8)}{(1-t^2)^{18}}$ & \nc  \\
	$        $ & $[222]$ & 18 & $    $ & $\frac{(1-t^4) (1-t^8) (1-t^{12})}{(1-t^2)^{21}}$ & \nc  \\
\hline
	$[010]$ & $[200]$ & 2 & $B_1 \otimes C_1 $ & $\frac{1-t^4}{(1-t^2)^3}$ & $m_B^2 t^2$  \\
\hline
	\end{tabular}
	\caption[$B_3$ Slodowy Intersections]{$B_3$ Slodowy Intersections. Some palindromic Hilbert series terms are abbreviated. HWGs that are not complete intersections are not shown. The orbit {\color{red} [101]} is non-normal.}
	\label{tab:BHS3}
\end{sidewaystable}

\begin{sidewaystable}[h]
	\centering
	\small
	\begin{tabular}{|c|c|c|c|c|c|}
\hline
	$\rho$ & $\sigma$ & $\left | {{\cal S}_{\sigma, \rho}} \right |$ & $F(\rho)$ & Unrefined HS & PL[HWG] \\
 \hline
	$[0000]$ & $[0100]$ & 12 & $B_4$ & $ \frac{1+24 t^2+129 t^4+220 t^6+\pal+t^{12}}{(1-t^2)^{12}}$ & $\footnotesize m_2 t^2$  \\
	$          $ & $[2000]$ & 14 & $      $ & $ \frac{1+21 t^2+105 t^4+175 t^6+\pal+t^{12}}{(1-t^4)^{-1}(1-t^2)^{15}}$ & $\footnotesize m_2 t^2 + m_1^2 t^4$  \\
	$          $ & $[0001]$ & 16 & $      $ & $ \frac{1+20 t^2+165 t^4+600 t^6+924 t^8+\pal+t^{16}}{(1-t^2)^{16}}$ & $\footnotesize m_2 t^2 + m_4^2 t^4$  \\
	$          $ & $[1010]$ & 20 & $      $ & $ \frac{1+14 t^2+106 t^4+454 t^6+788 t^8+\pal+t^{16}}{(1-t^4)^{-2}(1-t^2)^{22}}$ & 
	$\footnotesize \begin{array}{c}m_2 t^2 + m_1^2 t^4 + m_4^2 t^4 + m_1 m_3 t^6 +\\ m_3^2 t^8 - m_1^2 m_3^2 t^{12}\end{array}$  \\
	$          $ & $[0200]$ & 22 & $      $ & $ \frac{1+13 t^2+91 t^4+335 t^6+737 t^8+946 t^{10}+\pal+t^{20}}{(1-t^4)^{-1}(1-t^2)^{23}}$ & 
	$\footnotesize \begin{array}{c}m_2 t^2 + m_1^2 t^4 + m_4^2 t^4 + m_1 m_3 t^6 + m_2^2 t^8 +\\m_3^2 t^8 +m_1 m_2 m_3 t^{10} - m_1^2 m_2^2 m_3^2 t^{20}\end{array}$  \\
	$          $ & $[0020]$ & 24 & $      $ & $ \frac{1+10 t^2+56 t^4+194 t^6+438 t^8+578 t^{10}+\pal+t^{20}}{(1-t^4)^{-2}(1-t^2)^{26}}$ & \nc  \\
	$          $ & $[2200]$ & 24 & $      $ & $ \frac{1+10 t^2+55 t^4+136 t^6+190 t^8+\pal+t^{16}}{(1-t^4)^{-1}(1-t^8)^{-1}(1-t^2)^{26}}$ &  \nc  \\
\hline
 	\end{tabular}
	\caption[$B_4$ Slodowy Intersections from {[0000]} ]{$B_4$ Slodowy Intersections from [0000]. These are closures of nilpotent orbits ${\overline {\cal O}}_{\sigma}$. Some palindromic Hilbert series terms are abbreviated. HWGs that are not complete intersections are not shown.  The orbit {\color{red} [2101]} is non-normal.}
	\label{tab:BHS4a}
\end{sidewaystable}

\begin{sidewaystable}[h]
	\centering
	\small
	\begin{tabular}{|c|c|c|c|c|c|}
\hline
	$\rho$ & $\sigma$ & $\left | {{\cal S}_{\sigma, \rho}} \right |$ & $F(\rho)$ & Unrefined HS & PL[HWG] \\
 \hline
 	$[0100]$ & $[2000]$ & 2 & $B_2 \otimes C_1$ & $ \frac{1-t^4}{(1-t^2)^3}$ & $\footnotesize m^2 t^2$  \\
	
	$          $ & $[0001]$ & 4 & $      $ & $ \frac{1+6 t^2+t^4}{(1-t^2)^4}$ & $\footnotesize m_2^2 t^2$  \\
	
	$          $ & $[1010]$ & 8 & $      $ & $ \frac{(1-t^4) (1-2 t+7 t^2-2 t^3+t^4)}{(1-t)^2 (1-t^2)^7}$ & 
	$\footnotesize \begin{array}{c} m^2 t^2 + m_2^2 t^2 + m m_1 t^3 + m_1^2 t^4 -\\ m^2 m_1^2 t^6 \end{array}$  \\
	
	$          $ & $[0200]$ & 10 & $      $ & $\scriptsize \frac{ \left( \begin{array}{c} 1+3 t+11 t^2+30 t^3+69 t^4+123 t^5+204 t^6+\\287 t^7+351 t^8+366 t^9+\pal+t^{18} \end{array} \right)}{(1-t)^{-3}(1-t^2)^8 (1-t^3)^5}$ & 
	$\footnotesize \begin{array}{c}m^2 t^2 + m_2^2 t^2 + m m_1 t^3 + t^4 + m_1^2 t^4 +\\ m m_1 t^5 - m^2 m_1^2 t^{10}\end{array}$  \\
	
	$          $ & $[0020]$ & 12 & $      $ & $ \frac{1+t+5 t^2+10 t^3+20 t^4+30 t^5+52 t^6+60 t^7+66 t^8+\pal+t^{16}}{(1-t)^{-1} (1-t^4)^{-1}(1-t^2)^9 (1-t^3)^5}$ & \nc  \\
	
	$          $ & $[2200]$ & 12 & $      $ & $ \frac{ 1+2 t+7 t^2+16 t^3+30 t^4+40 t^5+55 t^6+60 t^7+\pal+t^{14}}{(1-t)^{-2} (1-t^8)^{-1}(1-t^2)^9 (1-t^3)^6}$ &  \nc  \\
	
	$          $ & $[0201]$ & 14 & $      $ & $\scriptsize \frac{ \left( \begin{array}{c} 1+t+6 t^2+9 t^3+24 t^4+39 t^5+77 t^6+110 t^7+\\174 t^8+220 t^9+270 t^{10}+266 t^{11}+\pal+t^{22}   \end{array} \right)}      {(1-t)^{-1}(1-t^2)^8 (1-t^3)^7}$ &  \nc  \\
\hline 
	\end{tabular}
	\caption[$B_4$ Slodowy Intersections from {[0100]}]{$B_4$ Slodowy Intersections from [0100]. Some palindromic Hilbert series terms are abbreviated. HWGs that are not complete intersections are not shown. The orbit {\color{red} [2101]} is non-normal.}
	\label{tab:BHS4b}
\end{sidewaystable}

\begin{sidewaystable}[h]
	\centering
	\small
	\begin{tabular}{|c|c|c|c|c|c|}
\hline
	$\rho$ & $\sigma$ & $\left | {{\cal S}_{\sigma, \rho}} \right |$ & $F(\rho)$ & Unrefined HS & PL[HWG] \\
 \hline
 	$ [2000]$ & $[1010]$ & 6 & $D_3 \otimes B_0$ & $ \frac{(1-t^4) (1+8 t^2+t^4)}{(1-t^2)^7}$ & $\footnotesize m_2 m_3 t^2$  \\
	
	$          $ & $[0200]$ & 8 & $      $ & $\frac{(1-t^4)^2 (1+5 t^2+t^4)}{(1-t^2)^{10}}$ & $\footnotesize m_2 m_3 t^2 + m_1^2 t^4$  \\
	
	$          $ & $[0020]$ & 10 & $      $ & $\frac{(1-t^4) (1+4 t^2+10 t^4+4 t^6+t^8)}{(1-t^2)^{11}}$ &
	$\footnotesize \begin{array}{c} m_2 m_3 t^2 + m_1^2 t^4 + m_2 m_3 t^4 + m_1 m_2^2 t^6 +\\  m_1 m_3^2 t^6 -m_1^2 m_2^2 m_3^2 t^{12} \end{array}$  \\
	
	$          $ & $[2200]$ & 10 & $      $ & $\frac{ 1+10 t^2+40 t^4+85 t^6+109 t^8+\pal+t^{16}}{(1-t^8)^{-1}(1-t^2)^5 (1-t^4)^6}$ &  
	$\footnotesize \begin{array}{c} m_2 m_3 t^2 + t^4 + m_1 t^4 + m_1^2 t^4 +\\ m_1 t^6 - m_1^2 t^{12} \end{array}$  \\
	
	$          $ & $[0201]$ & 12 & $      $ & $\frac{(1-t^4)^2 (1+t^2+8 t^4+t^6+t^8)}{(1-t^2)^{14}}$ &  \nc  \\
	
	$          $ & {\color{red} [2101]} & 12 & $      $ & $\scriptsize \frac{\left( \begin{array}{c} 1+8 t^2+37 t^4+127 t^6+325 t^8+573 t^{10}+\\632 t^{12}+413 t^{14}+130 t^{16}+7 t^{18}-5 t^{20} \end{array} \right)}{(1-t^2)^7 (1-t^4)^5}$ &  \nc  \\
	
	$          $ & $[2020]$ & 14 & $      $ & $ \frac{1+4 t^2+13 t^4+31 t^6+53 t^8+62 t^{10}+\pal+t^{20}}{(1-t^2)^{11} (1-t^4)^3}$ &  \nc  \\
\hline 
	\end{tabular}
	\caption[$B_4$ Slodowy Intersections from {[2000]-[0001]}] {$B_4$ Slodowy Intersections from {[2000]-[0001]}. Some palindromic Hilbert series terms are abbreviated. HWGs that are not complete intersections are not shown. The orbit {\color{red} [2101]} is non-normal.}
	\label{tab:BHS4d}
\end{sidewaystable}

\begin{sidewaystable}[h]
	\centering
	\small
	\begin{tabular}{|c|c|c|c|c|c|}
\hline
	$\rho$ & $\sigma$ & $\left | {{\cal S}_{\sigma, \rho}} \right |$ & $F(\rho)$ & Unrefined HS & PL[HWG] \\
 \hline
	$[1010]$ & $[0200]$ & 2 & $C_1 \otimes D_1 \otimes B_0$ & $\frac{1-t^4}{(1-t^2)^3}$ & $\footnotesize m^2 t^2$  \\
	
	$          $ & $[0020]$ & 4 & $     $ & $\frac{1+2 t^2+2 t^3+2 t^4+t^6}{(1-t^2)^2 (1-t^3)^2}$ & $\footnotesize t^2 + m^2 t^2 + m(q+1/q) t^3 - m^2 t^6$  \\
	
	$          $ & $[2200]$ & 4 & $      $ & $\frac{1-t^8}{(1-t^2)^3 (1-t^3)^2}$ & $\footnotesize m^2 t^2 + m t^3 + t^4 + m t^5 - m^2 t^{10}$  \\
	
	$          $ & $[0201]$ & 6 & $      $ & $\frac{(1-t^4) (1-t+t^2+2 t^3+t^4-t^5+t^6)}{(1-t) (1-t^2)^3 (1-t^3)^3}$ &  \nc  \\
	
	$          $ & {\color{red} [2101]} & 6 & $      $ & $\frac{1-t+3 t^2+5 t^4+3 t^5+7 t^6+2 t^7+4 t^8-2 t^9}{(1-t) (1-t^2) (1-t^3)^3 (1-t^4)}$ &  \nc  \\
	
	$          $ & $[2020]$ & 8 & $      $ & $\frac{1+2 t^3+3 t^4+2 t^5+2 t^6+2 t^7+3 t^8+2 t^9+t^{12}}{(1-t^2)^4 (1-t^3)^4}$ &  \nc  \\
	
	$          $ & $[2220]$ & 10 & $      $ & $\frac{ (1+t+t^2+2 t^3+3 t^4+3 t^5+\pal+t^{10}) (1-t^{12})}{(1-t)^{-1} (1-t^8)^{-1}(1-t^2)^4 (1-t^3)^5 (1-t^4)^2 (1-t^5)^2}$ &  \nc  \\
\hline 
	\end{tabular}
	\caption[$B_4$ Slodowy Intersections from {[1010]-[0200]}] {$B_4$ Slodowy Intersections from {[1010]-[0200]}. Some palindromic Hilbert series terms are abbreviated. HWGs that are not complete intersections are not shown. The orbit {\color{red} [2101]} is non-normal.}
	\label{tab:BHS4f}
\end{sidewaystable}

\begin{sidewaystable}[h]
	\centering
	\small
	\begin{tabular}{|c|c|c|c|c|c|}
\hline
	$\rho$ & $\sigma$ & $\left | {{\cal S}_{\sigma, \rho}} \right |$ & $F(\rho)$ & Unrefined HS & PL[HWG] \\
 \hline
	$[0020]$ & $[0201]$ & 2 & $  B_1  $ & $\frac{1-t^4}{(1-t^2)^3}$ & $m^2 t^2$ \\
	$          $ & {\color{red} [2101]} & 2 & $      $ & $\frac{1-t^4}{(1-t^2)^3}$ & $ m^2 t^2$  \\
	
	$    $ & $[2020]$ & 4 & $      $ & $\frac{1+t^2+4 t^4+t^6+t^8}{(1-t^2)^2 (1-t^4)^2}$ &$ m^2 t^2 + t^4 + m^4 t^4 + m^4 t^6 - m^8 t^{12}$\\
	
	$          $ & $[2220]$ & 6 & $      $ & $\frac{(1-t^8) (1-t^{12})}{(1-t^2)^3 (1-t^4)^5}$ &  \nc  \\
	
	$          $ & $[2222]$ & 8 & $      $ & $\frac{(1-t^8) (1-t^{12}) (1-t^{16})}{(1-t^2)^3 (1-t^4)^5 (1-t^6)^3}$ &  \nc  \\
 \hline
	$[2200]$ & {\color{red} [2101]} & 2 & $D_2 \otimes B_0$ & $\frac{1+4 t^2-t^4}{(1-t^2)^2}$ & $ m_1^2 t^2 + m_2^2 t^2 - m_1^2 m_2^2 t^4$  \\
	$    	    $ & $[2020]$ & 4 & $      $ & $\frac{(1-t^4)^2}{(1-t^2)^6}$ &$ m_1^2 t^2 + m_2^2 t^2$\\
\hline
	\end{tabular}
	\caption[$B_4$ Slodowy Intersections from {[0020]-[2220]}] {$B_4$ Slodowy Intersections from {[0020]-[2220]}. Some palindromic Hilbert series are abbreviated. HWGs that are not complete intersections are not shown. The orbit {\color{red} [2101]} is non-normal.}
	\label{tab:BHS4g}
\end{sidewaystable}


\begin{sidewaystable}[h]
	\centering
	\begin{tabular}{|c|c|c|c|c|c|}
\hline
	$\rho$ & $\sigma$ & $\left | {{\cal S}_{\sigma, \rho}} \right |$ & $F(\rho)$ & Unrefined HS & PL[HWG] \\
\hline
	     $[000]$ & $[100]$ & 6 & $C_3$ & $\frac{(1-t^4) (1+14 t^2+t^4)}{(1-t^2)^7}$ & $ m_1^2 t^2$  \\
	$        $ & $[010]$ & 10 & $ $ & $\frac{(1-t^4) (1+10 t^2+41 t^4+10 t^6+t^8)}{(1-t^2)^{11}}$ & $ m_1^2 t^2 + m_2^2 t^4$  \\
	$        $ & $[002]$ & 12 & $ $ & $\frac{(1-t^4)^2 (1+7 t^2+15 t^4+7 t^6+t^8)}{(1-t^2)^{14}}$ & $ m_1^2 t^2 + m_2^2 t^4 + m_3^2 t^6$  \\
	$        $ & $[020]$ & 14 & $ $ & $\frac{(1-t^4) (1+6 t^2+21 t^4+35 t^6+\pal+t^{12})}{(1-t^2)^{15}}$ & 
	$ \begin{array}{c} m_1^2 t^2 + m_2 t^4 + m_2^2 t^4 + m_1 m_3 t^6 + m_3^2 t^6 +\\ m_1 m_2 m_3 t^8 - m_1^2 m_2^2 m_3^2 t^{16}\end{array}$\\
	$        $ & $[210]$ & 14 & $ $ & $\frac{(1-t^4) (1+6 t^2+21 t^4+56 t^6+\pal+t^{12})}{(1-t^2)^{15}}$ & \nc  \\
	$        $ & $[202]$ & 16 & $ $ & $\frac{(1-t^4)^2 (1+3 t^2+7 t^4+13 t^6+\pal+t^{12})}{(1-t^2)^{18}}$ & \nc  \\
	$        $ & $[222]$ & 18 & $ $ & $\frac{(1-t^4) (1-t^8) (1-t^{12})}{(1-t^2)^{21}}$ & \nc  \\
    \hline
	\end{tabular}
	\caption[$C_3$ Slodowy Intersections from {[000]}]{$C_3$ Slodowy Intersections from {[000]}. The intersections are closures of nilpotent orbits ${\overline {\cal O}}_{\sigma}$. Some palindromic Hilbert series terms are abbreviated. HWGs that are not complete intersections are not shown.}
	\label{tab:CHS123}
\end{sidewaystable}

\begin{sidewaystable}[h]
	\centering
	\small
	\begin{tabular}{|c|c|c|c|c|c|}
\hline
	$\rho$ & $\sigma$ & $\left | {{\cal S}_{\sigma, \rho}} \right |$ & $F(\rho)$ & Unrefined HS & PL[HWG] \\
 \hline
	$[100]$ & $[010]$ & 4 & $C_2 \otimes B_0$ & $\frac{1+6 t^2+t^4}{(1-t^2)^4}$ & $\footnotesize m_1^2 t^2$  \\
	$        $ & $[002]$ & 6 & $ $ & $\frac{(1-t^4) (1+3 t^2+t^4)}{(1-t^2)^7}$ & $\footnotesize m_1^2 t^2 + m_2^2 t^4$  \\
	$        $ & $[020]$ & 8 & $ $ & $\frac{(1-t^4) (1-2 t+4 t^2-2 t^3+t^4)}{(1-t)^2 (1-t^2)^7}$ & 
		$\footnotesize \begin{array}{c} m_1^2 t^2 + m_1 t^3 + m_2 t^4 + m_2^2 t^4 + m_1 m_2 t^5 -\\ m_1^2 m_2^2 t^{10} \end{array}$  \\
	$        $ & $[210]$ & 8 & $ $ & $\frac{1+2 t+7 t^2+12 t^3+27 t^4+42 t^5+67 t^6+72 t^7+\pal +t^{14} }{(1-t)^{-2}(1-t^2)^6 (1-t^3)^4}$ & \nc  \\
	$        $ & $[202]$ & 10 & $ $ & $\frac{ 1+2 t+4 t^2+6 t^3+10 t^4+14 t^5+20 t^6+22 t^7+\pal +t^{14} }{(1-t)^{-2} (1-t^4)^{-1}(1-t^2)^9 (1-t^3)^4}$ & \nc  \\
	$        $ & $[222]$ & 12 & $ $ & $\frac{(1-t^8) (1-t^{12})}{(1-t^2)^{10} (1-t^3)^4}$ & \nc  \\
\hline
	$[010]$ & $[002]$ & 2 & $C_1 \otimes D_1 $ & $\frac{1-t^4}{(1-t^2)^3}$ & $\footnotesize m^2 t^2$  \\
\hline
	\end{tabular}
	\caption[$C_3$ Slodowy Intersections from {[100]-[202]} ]{$C_3$ Slodowy Intersections from [100]-[202]. Some palindromic Hilbert series terms are abbreviated. HWGs that are not complete intersections are not shown.}
	\label{tab:CHS3}
\end{sidewaystable}

\begin{sidewaystable}[h]
	\centering
	\small
	\begin{tabular}{|c|c|c|c|c|c|}
\hline
	$\rho$ & $\sigma$ & $\left | {{\cal S}_{\sigma, \rho}} \right |$ & $F(\rho)$ & Unrefined HS & PL[HWG] \\
 \hline
	$[0000]$ & $[1000]$ & 8 & $C_4$ & $\frac{1+28 t^2+70 t^4 +28 t^6+t^8}{(1-t^2)^8} $ & $\footnotesize m_1^2 t^2$  \\
	$          $ & $[0100]$ & 14 & $     $ & $\frac{(1-t^4) (1+21 t^2 +204 t^4+406 t^6+\pal +t^{12} )}{(1-t^2)^{15}}$ & $\footnotesize m_1^2 t^2 + m_2^2 t^4$  \\
	$          $ & $[0010]$ & 18 & $     $ & $\frac{(1-t^4) (1+17 t^2 +126 t^4+537 t^6+894 t^8+\pal+t^{16})}{(1-t^2)^{19}}$ & $\footnotesize m_1^2 t^2 + m_2^2 t^4 + m_3^2 t^6$  \\
$     $ & $[0002]$ & 20 & $ $ & $\frac{(1-t^4)^2 (1+14 t^2+79 t^4+223 t^6+317 t^8+\pal+t^{16} )}{(1-t^2)^{22}}$ & $\footnotesize m_1^2 t^2 + m_2^2 t^4 + m_3^2 t^6 + m_4^2 t^8$  \\
	$          $ & $[2100]$ & 20 & $     $ & $\frac{(1-t^4)^2 (1+14 t^2+106 t^4+574 t^6+722 t^8+\pal +t^{16})}{(1-t^2)^{22}}$ & \nc  \\
	$          $ & {\color{red} [0200]} & 22 & $ $ & $\frac{1+13 t^2+91 t^4+419 t^6+1346 t^8+2365 t^{10}+1841 t^{12}+476 t^{14}-56 t^{16}-29 t^{18}+t^{20}}{(1-t^4)^{-1}(1-t^2)^{23}}$ & \nc  \\
	$          $ & $[0110]$ & 24 & $     $ & $\frac{(1-t^4)^2 (1+10 t^2+56 t^4+194 t^6+405 t^8+512 t^{10}+\pal+t^{20}  )}{(1-t^2)^{26}}$ & \nc  \\
\hline
	\end{tabular}
	\caption[$C_4$ Slodowy Intersections from {[0000]}]{$C_4$ Slodowy Intersections from [0000]. The intersections are closures of nilpotent orbits ${\overline {\cal O}}_{\sigma}$. Some palindromic Hilbert series terms are abbreviated. HWGs that are not complete intersections are not shown. The orbit {\color{red} [0200]} is non-normal.}
	\label{tab:CHS4a}
\end{sidewaystable}

\begin{sidewaystable}[h]
	\centering
	\small
	\begin{tabular}{|c|c|c|c|c|c|}
\hline
	$\rho$ & $\sigma$ & $\left | {{\cal S}_{\sigma, \rho}} \right |$ & $F(\rho)$ & Unrefined HS & PL[HWG] \\
\hline
	$[1000]$ & $[0100]$ & 6 & $C_3 \otimes B_0$ & $\frac{(1-t^4) (1+14 t^2+t^4)}{(1-t^2)^7}$ & $\footnotesize m_1^2 t^2$  \\
	$          $ & $[0010]$ & 10 & $     $ & $\frac{(1-t^4) (1+10 t^2+41 t^4+10 t^6+t^8)}{(1-t^2)^{11}}$ & $\footnotesize m_1^2 t^2 + m_2^2 t^4 $  \\
	$     	   $ & $[0002]$ & 12 & $ $ & $\frac{(1-t^4)^2 (1+7 t^2+15 t^4+7 t^6+t^8)}{(1-t^2)^{14}}$ & $\footnotesize m_1^2 t^2 + m_2^2 t^4 + m_3^2 t^6$  \\
	$          $ & $[2100]$ & 12 & $     $ & $\scriptsize \frac{\left( \begin{array}{c}1+4 t+20 t^2+60 t^3+191 t^4+480 t^5+1049 t^6+1810 t^7+\\2641 t^8+3210 t^9+3450 t^{10}+\pal+t^{20}   \end{array}\right )}{(1-t)^{-4} (1-t^4)^{-1}(1-t^2)^{11} (1-t^3)^6}$ & \nc  \\
	$    $ & {\color{red} [0200]} & 14 & $ $ & $\frac{(1-t^4) (1-2 t+9 t^2-10 t^3+32 t^4-24 t^5+72 t^6-44 t^7+37 t^8-16 t^9+t^{10})}{(1-t)^2 (1-t^2)^{13}}$ & \nc  \\
	$          $ & $[0110]$ & 16 & $     $ & $\frac{(1-t^4) (1-2 t+7 t^2-6 t^3+15 t^4-6 t^5+17 t^6-\pal +t^{12})}{(1-t)^2 (1-t^2)^{15}}$ & \nc  \\
	$          $ & $[2010]$ & 16 & $     $ & $\scriptsize \frac{\left( \begin{array}{c}1+4 t+16 t^2+44 t^3+117 t^4+264 t^5+576 t^6+1124 t^7+2007 t^8+\\3078 t^9+4113 t^{10}+4762 t^{11}+5004 t^{12}+\pal+t^{24}   \end{array}\right)}{(1-t)^{-4}(1-t^4)^{-1}(1-t^2)^{15} (1-t^3)^6}$ & \nc  \\
\hline
	\end{tabular}
	\caption[$C_4$ Slodowy Intersections from {[1000]}]{$C_4$ Slodowy Intersections from [1000]. Some palindromic Hilbert series terms are abbreviated. HWGs that are not complete intersections are not shown. The orbit {\color{red} [0200]} is non-normal.}
	\label{tab:CHS4b}
\end{sidewaystable}

\begin{sidewaystable}[h]
	\centering
	\small
	\begin{tabular}{|c|c|c|c|c|c|}
\hline
	$\rho$ & $\sigma$ & $\left | {{\cal S}_{\sigma, \rho}} \right |$ & $F(\rho)$ & Unrefined HS & PL[HWG] \\
 \hline
	$[0100]$ & $[0010]$ & 4 & $C_2 \otimes D_1$ & $\frac{1+6 t^2+t^4}{(1-t^2)^4}$ & $\footnotesize m_1^2 t^2 $  \\
	$     	   $ & $[0002]$ & 6 & $ $ & $\frac{(1-t^4) (1+3 t^2+t^4)}{(1-t^2)^7}$ & $\footnotesize m_1^2 t^2 + m_2^2 t^4$  \\
	$          $ & $[2100]$ & 6 & $     $ & $\frac{1-2 t+9 t^2-8 t^3+9 t^4-2 t^5+t^6}{(1-t)^2 (1-t^2)^3 (1-t^4)}$ & \nc  \\
	$          $ & {\color{red} [0200]} & 8 & $ $ & $\scriptsize \frac{\left( \begin{array}{c}1+t +7 t^2+11 t^3+31 t^4+47 t^5+81 t^6+81 t^7+\\77 t^8+41 t^9+21 t^{10} -3 t^{11} -4 t^{12} -4 t^{13} \end{array} \right)}{(1-t)^{-1}(1-t^2)^5 (1-t^3)^4}$ & \nc  \\
	$          $ & $[0110]$ & 10 & $     $ & $\frac{(1-t^4)^2 (1+3 t^2+4 t^3+7 t^4+4 t^5+\pal+t^{10})}{(1-t^2)^8 (1-t^3)^4}$ & \nc  \\
	$          $ & $[2010]$ & 10 & $     $ & $\frac{1+t+7 t^2+10 t^3+32 t^4+50 t^5+102 t^6+120 t^7+162 t^8+150 t^9+\pal+t^{18}}{(1-t)^{-1}(1-t^2)^5 (1-t^3)^5 (1-t^4)}$ & \nc  \\
	$     $ & $[2002]$ & 12 & $     $ & $\frac{1+t +4 t^2+7 t^3 +15 t^4+24 t^5+40 t^6+48 t^7+58 t^8+58 t^9 +\pal+t^{18} }{(1-t)^{-1}(1-t^2)^8 (1-t^3)^5}$ & \nc  \\
\hline
	\end{tabular}
	\caption[$C_4$ Slodowy Intersections from {[0100]-[0010]}]{$C_4$ Slodowy Intersections from [0100]-[0010]. Some palindromic Hilbert series terms are abbreviated. HWGs that are not complete intersections are not shown. The orbit {\color{red} [0200]} is non-normal.}
	\label{tab:CHS4d}
\end{sidewaystable}

\begin{sidewaystable}[h]
	\centering
	\small
	\begin{tabular}{|c|c|c|c|c|c|}
\hline
	$\rho$ & $\sigma$ & $\left | {{\cal S}_{\sigma, \rho}} \right |$ & $F(\rho)$ & Unrefined HS & PL[HWG] \\
 \hline
	$[0002]$ & {\color{red} [0200]} & 2 & $D_2$ & $\frac{1+4 t^2-t^4}{(1-t^2)^2}$ & $\footnotesize m_1^2 t^2 + m_2^2 t^2 - m_1^2 m_2^2 t^4$  \\
	$          $ & $[0110]/[2010]$ & 4 & $     $ & $\frac{(1-t^4)^2}{(1-t^2)^6}$ & $\footnotesize m_1^2 t^2 + m_2^2 t^2$ \\
	$    $ & $[2002]$ & 6 & $   $ & $\frac{1+3 t^2+11 t^4+10 t^6+\pal+t^{12} }{(1-t^2)^3 (1-t^4)^3}$ & $\footnotesize  \left( \begin{array}{c} m_1^2 t^2 + m_2^2 t^2 + t^4 + m_1^2 m_2^2 t^4 + \\m_1^2 m_2^2 t^6 - m_1^4 m_2^4 t^{12}\end{array}\right)$  \\
	$          $ & $[0202]$ & 8 & $     $ & $\frac{(1-t^8) ( 1+2 t^2+7 t^4 +6 t^6+\pal+t^{12}  )}{(1-t^2)^4 (1-t^4)^5}$ & \nc  \\
	$          $ & $[2210]$ & 8 & $     $ & $\frac{(1-t^8) (1+6 t^4+t^8)}{(1-t^2)^6 (1-t^4)^3}$ & \nc  \\
	$          $ & $[2202]$ & 10 & $     $ & $\frac{(1-t^8) (1+t^2+4 t^4+4 t^6+10 t^8+\pal+t^{16})}{(1-t^2)^5 (1-t^4)^6}$ & \nc  \\
	$          $ & $[2222]$ & 12 & $     $ & $\frac{(1-t^8) (1-t^{12}) (1-t^{16})}{(1-t^2)^6 (1-t^4)^9}$ & \nc  \\
\hline
	\end{tabular}
	\caption[$C_4$ Slodowy Intersections from {[0002]-[0200]}]{$C_4$ Slodowy Intersections from [0002]-{\color{red} [0200]}. Some palindromic Hilbert series terms are abbreviated. HWGs that are not complete intersections are not shown. The orbit {\color{red} [0200]} is non-normal.}
	\label{tab:CHS4g}
\end{sidewaystable}

\begin{sidewaystable}[h]
	\centering
	\small
	\begin{tabular}{|c|c|c|c|c|c|}
\hline
	$\rho$ & $\sigma$ & $\left | {{\cal S}_{\sigma, \rho}} \right |$ & $F(\rho)$ & Unrefined HS & PL[HWG] \\
 \hline
	$[0110]$ & $[2002]$ & 2 & $C_1 \otimes B_0$ & $\frac{1-t^4}{(1-t^2)^3}$ & $\footnotesize m^2 t^2$  \\
	$          $ & $[0202]/[2210]$ & 4 & $     $ & $\frac{1-t^8}{(1-t^2)^3 (1-t^3)^2}$ & $\footnotesize m^2 t^2 + m t^3 + t^4 + m t^5 - m^2 t^{10}$  \\
	$          $ & $[2202]$ & 6 & $     $ & $\frac{(1-t^8) (1-t  +t^3+t^5 -t^7+t^8  )}{(1-t) (1-t^2)^3 (1-t^3) (1-t^6)^2}$ & \nc  \\
	$          $ & $[2222]$ & 8 & $     $ & $\frac{(1-t^8) (1-t^{12}) (1-t^{16})}{(1-t^2)^3 (1-t^3)^2 (1-t^4) (1-t^5)^2 (1-t^6)^3}$ & \nc  \\
\hline
	$[2010]$ & $[2002]$ & 2 & $C_1 \otimes B_0 \otimes B_0$ & $\frac{1-t^4}{(1-t^2)^3}$ & $\footnotesize m^2 t^2$  \\
	$          $ & $[0202]$ & 4 & $     $ & $\frac{1-t^8}{(1-t^2)^3 (1-t^3)^2}$ & $\footnotesize m^2 t^2 + m t^3 + t^4 + m t^5 - m^2 t^{10}$  \\
	$          $ & $[2210]$ & 4 & $     $ & $\frac{(1-t^6)^2}{(1-t^2)^3 (1-t^3)^2 (1-t^4)}$ &  \nc  \\
\hline
	\end{tabular}
	\caption[$C_4$ Slodowy Intersections from {[0110]-[2202]}]{$C_4$ Slodowy Intersections from [0110]-[2202]. Some palindromic Hilbert series terms are abbreviated. HWGs that are not complete intersections are not shown.}
	\label{tab:CHS4j}
\end{sidewaystable}

\begin{sidewaystable}[h]
	\centering
	\small
	\begin{tabular}{|c|c|c|c|c|c|}
\hline
	$\rho$ & $\sigma$ & $\left | {{\cal S}_{\sigma, \rho}} \right |$ & $F(\rho)$ & Unrefined HS & PL[HWG] \\
\hline
	     $[00]$ & {\color{red} [02]/[20]}& 2 & $D_2$ & $\frac{1+4 t^2-t^4}{(1-t^2)^2}$ & ${\footnotesize m_1^2 t^2 + m_2^2 t^2 - m_1^2 m_2^2 t^4}$  \\
	     $      $ & $[22]$ & 4 & $     $ & $\frac{(1-t^4)^2}{(1-t^2)^6}$ & $\footnotesize m_1^2 t^2 + m_2^2 t^2$  \\
\hline
	 {[02]/[20]} & $[22]$ & 2 & $A_1$ & $\frac{1-t^4}{(1-t^2)^3}$ & $\footnotesize m^2 t^2$  \\
\hline
\hline
	$[000]$ & $[011]$ & 6 & $D_3$ & $\frac{(1+t^2) (1+8 t^2+t^4)}{(1-t^2)^6}$ & $\footnotesize m_2 m_3 t^2$  \\
	$        $ & $[200]$ & 8 & $      $ & $\frac{(1+t^2)^2 (1+5 t^2 +t^4 )}{(1-t^2)^8}$ & $\footnotesize m_2 m_3 t^2 + m_1^2 t^4$  \\
	$        $ & $[022]$ & 10 & $      $ & $\frac{(1+t^2) (1+4 t^2+10 t^4+4 t^6+t^8)}{(1-t^2)^{10}}$ & 
$\footnotesize m_2 m_3 t^2 + m_1^2 t^4 + m_2 m_3 t^4 + m_1 m_2^2 t^6 + m_1 m_3^2 t^6 - m_1^2 m_2^2 m_3^2 t^{12} $  \\
	$        $ & $[222]$ & 12 & $      $ & $\frac{(1-t^4) (1-t^6) (1-t^8)}{(1-t^2)^{15}}$ & \nc  \\
\hline
	$[011]$ & $[200]$ & 2 & $A_1 \otimes D_1      $ & $\frac{1-t^4}{(1-t^2)^3}$ & $\footnotesize m^2 t^2$  \\
	$        $ & $[022]$ & 4 & $      $ & $\frac{1+2 t^2+2 t^3+2 t^4+t^6}{(1-t^2)^2 (1-t^3)^2}$ & $\footnotesize t^2 + m^2 t^2 + m (q+1/q) t^3 - m^2 t^6$  \\
	$        $ & $[222]$ & 6 & $      $ & $\frac{(1-t^6) (1-t^8)}{(1-t^2)^4 (1-t^3)^4}$ & \nc  \\
\hline
$[200]$ & $[022]$ & 2 & $A_1      $ & $\frac{1-t^4}{(1-t^2)^3}$ & $\footnotesize m^2 t^2$  \\
	$        $ & $[222]$ & 4 & $      $ & $\frac{(1-t^6) (1-t^8)}{(1-t^2)^3 (1-t^4)^3}$ & $\footnotesize m^2 t^2 + t^4 + m^2 t^4 + m^2 t^6 - m^4 t^{12}$  \\
\hline
$[022]$ & $[222]$ & 2 & $D_1 $ & $\frac{1-t^8}{(1-t^2) (1-t^4)^2}$ & $\footnotesize t^2 + (q+1/q) t^4 - t^8$  \\
\hline
	\end{tabular}
	\caption[$D_2$ and $D_3$ Slodowy Intersections]{$D_2$ and $D_3$ Slodowy Intersections. HWGs that are not complete intersections are not shown. The orbits {\color{red} [02]} and {\color{red} [20]} are non-normal.}
	\label{tab:DHS23}
\end{sidewaystable}

\FloatBarrier

\begin{sidewaystable}[h]
	\centering
	\small
	\begin{tabular}{|c|c|c|c|c|c|}
\hline
	$\rho$ & $\sigma$ & $\left | {{\cal S}_{\sigma, \rho}} \right |$ & $F(\rho)$ & Unrefined HS & PL[HWG] \\
\hline
	     $[0000]$ & $[0100]$ & 10 & $D_4$ & $\frac{(1-t^4) (1+17 t^2+48 t^4+17 t^6+t^8)}{(1-t^2)^{11}}$ & ${\footnotesize m_2 t^2}$  \\
	     $         $ &  {\color{red}[0002]/[0020]} & 12 & $  $ & $\frac{(1-t^4) (1+15 t^2+85 t^4+162 t^6+15 t^8-13 t^{10}-t^{12}   )}{(1-t^2)^{13}}$ & ${\footnotesize m_2 t^2 + m_3^2 t^4 + m_4^2 t^4 - m_3^2 m_4^2 t^8}$  \\
	       $         $ & $[2000]$ & 12 & $       $ & $\frac{(1-t^4)^2 (1+14 t^2+36 t^4+14 t^6+t^8)}{(1-t^2)^{14}}$ & ${\footnotesize m_2 t^2 + m_1^2 t^4}$  \\
	        $         $ & $[1011]$ & 16 & $       $ & $\frac{1+12 t^2+77 t^4+296 t^6+476 t^8+\pal+t^{16}}{(1-t^2)^{16}}$ & ${\footnotesize \left( \begin{array}{c} m_2 t^2 + m_1^2 t^4 + m_3^2 t^4 + m_4^2 t^4 +\\ m_1 m_3 m_4 t^6 - m_1^2 m_3^2 m_4^2 t^{12} \end{array} \right)}$  \\
	         $         $ & $[0200]$ & 18 & $       $ & $\frac{(1-t^4) (1+9 t^2+45 t^4+109 t^6+152 t^8+\pal+t^{16})}{(1-t^2)^{19}}$ & ${\footnotesize \left(\begin{array}{c} m_2 t^2 + m_1^2 t^4 + m_3^2 t^4 + m_4^2 t^4 + \\m_1 m_3 m_4 t^6 +m_2^2 t^8 + \\m_1 m_2 m_3 m_4 t^{10} - m_1^2 m_2^2 m_3^2 m_4^2 t^{20} \end{array}\right)}$  \\
	          $         $ &  {\color{red}[0202]/[0220]} & 20 & $       $ & $ \scriptsize \frac{\left( \begin{array}{c} 1+7 t^2+28 t^4+84 t^6+173 t^8+238 t^{10}+\\133 t^{12}+28 t^{14}-14 t^{16} -5 t^{18}-t^{20}\end{array} \right)}{(1-t^4)^{-1}(1-t^2)^{21}}$ & \nc  \\
	           $         $ & $[[2200]]$ & 20 & $       $ & $\frac{(1-t^4) (1-t^8) (1+6 t^2+21 t^4+28 t^6+\pal+t^{12}     )}{(1-t^2)^{22}}$ & \nc  \\
	            $         $ & $[2022]$ & 22 & $       $ & $\frac{(1-t^4)^3 (1+3 t^2+8 t^4+16 t^6+28 t^8+\pal+t^{16})}{(1-t^2)^{25}}$ & \nc  \\
	             $         $ & $[2222]$ & 24 & $       $ & $\frac{(1-t^4) (1-t^8)^2 (1-t^{12})}{(1-t^2)^{28}}$ & \nc  \\
%
\hline
	     $[0100]$ &  {\color{red}[0002]/[0020]} & 2 & $D_2 \otimes C_1$ & $\frac{1+4 t^2-t^4}{(1-t^2)^2}$ & ${\footnotesize m_1^2 t^2 + m_2^2 t^2 - m_1^2 m_2^2 t^4 }$  \\
	       $         $ & $[2000]$ & 2 & $       $ & $\frac{1-t^4}{(1-t^2)^3}$ & ${\footnotesize m^2 t^2 }$  \\
	        $         $ & $[1011]$ & 6 & $       $ & $\frac{1-2 t+6 t^2-2 t^3+t^4   }{(1-t)^2 (1-t^2)^4}$ & ${\footnotesize \left( \begin{array}{c} m^2 t^2 + m_1^2 t^2 + m_2^2 t^2 +\\ m m_1 m_2 t^3 - m^2 m_1^2 m_2^2 t^6 \end{array} \right)}$  \\
	         $         $ & $[0200]$ & 8 & $       $ & $\frac{(1+3 t^2+4 t^3+3 t^4+t^6) (1+2 t+3 t^2+4 t^3+6 t^4+\pal+t^8)}{(1-t)^{-2}(1-t^2)^6 (1-t^3)^4}$ & ${\footnotesize \left(\begin{array}{c} m^2 t^2 + m_1^2 t^2 + m_2^2 t^2 + m m_1 m_2 t^3 +\\ t^4 + m m_1 m_2 t^5 - m^2 m_1^2 m_2^2 t^{10} \end{array}\right)}$  \\
	  $         $ &  {\color{red}[0202]/[0220]} & 10 & $       $ & $\scriptsize \frac{\left( \begin{array}{c}1+4 t^2+3 t^3+10 t^4 +12 t^5+23 t^6+22 t^7+35 t^8+\\29 t^9 +24 t^{10}+9 t^{11}+10 t^{12}-3 t^{14}-2 t^{15}-t^{17}\end{array} \right)}{(1-t^2)^5 (1-t^3)^5}$ & \nc  \\
	           $         $ & $[[2200]]$ & 10 & $       $ & $\frac{ 1+t+3 t^2+6 t^3+8 t^4+6 t^5+\pal+t^{10}}{(1-t)^{-1} (1-t^8)^{-1}(1-t^2)^7 (1-t^3)^5}$ & \nc  \\
	            $         $ & $[2022]$ & 12 & $       $ & $\frac{ 1-t+t^2+2 t^3+6 t^6+\pal+t^{12}}{(1-t^4)^{-2}(1-t) (1-t^2)^8 (1-t^3)^5}$ & \nc  \\
	             $         $ & $[2222]$ & 14 & $       $ & $\frac{(1-t^8)^2 (1-t^{12})}{(1-t^2)^9 (1-t^3)^8}$ & \nc  \\
\hline
	\end{tabular}
	\caption[$D_4$ Slodowy Intersections from {[0000]-[0100]}]{$D_4$ Slodowy Intersections from [0000]-[0100]. Some palindromic Hilbert series terms are abbreviated. HWGs that are not complete intersections are not shown. The orbits {\color{red} [0002]}, {\color{red} [0020]}, {\color{red} [0202]} and {\color{red} [0220]} are non-normal.}
	\label{tab:DHS4a}
\end{sidewaystable}

\begin{sidewaystable}[h]
	\centering
	\small
	\begin{tabular}{|c|c|c|c|c|c|}
\hline
	$\rho$ & $\sigma$ & $\left |\svar{\sigma}{\rho} \right |$ & $F(\rho)$ & Unrefined HS & PL[HWG] \\
\hline
         {[0002]/[0020]} & $[1011]$ & 4 & $C_2$ & $\frac{1+6 t^2+t^4}{(1-t^2)^4}$ & $\footnotesize m_1^2 t^2$  \\
         
	         $         $ & $[0200]$ & 6 & $       $ & $\frac{(1-t^4) (1+3 t^2+t^4)}{(1-t^2)^7}$ & $\footnotesize m_1^2 t^2 + m_2^2 t^4$  \\
	  $         $ &  {\color{red}[0202]/[0220]} & 8 & $  $ & $\scriptsize \frac{\left( \begin{array}{c} 1+6 t^2+22 t^4+52 t^6+81 t^8+72 t^{10}+\\32 t^{12}+t^{14}-4 t^{16}-t^{18} \end{array} \right)}{(1-t^2)^4 (1-t^4)^4}$ & \nc  \\
	           $         $ & $[2200]$ & 8 & $       $ & $\frac{(1-t^4) (1-t^8)}{(1-t^2)^{10}}$ & $\footnotesize  \left( \begin{array}{c} m_1^2 t^2 + m_2 t^4 + m_2^2 t^4 + m_1^2 t^6 + \\m_1^2 m_2 t^8 - m_1^4 m_2^2 t^{16} \end{array}\right) $  \\
	            $         $ & $[2022]$ & 10 & $       $ & $\frac{1+3 t^2+8 t^4+16 t^6+22 t^8+\pal+t^{16}   }{(1-t^2)^7 (1-t^4)^3}$ & \nc  \\
	             $         $ & $[2222]$ & 12 & $       $ & $\frac{(1-t^8)^2 (1-t^{12})}{(1-t^2)^{10} (1-t^4)^5}$ & \nc  \\
%
\hline
         $[2000]$ & $[1011]$ & 4 & $B_2 \otimes B_0$ & $\frac{1+6 t^2+t^4}{(1-t^2)^4}$ & $\footnotesize m_2^2 t^2$  \\
	 $         $ & $[0200]$ & 6 & $       $ & $\frac{(1-t^4) (1+3 t^2+t^4)}{(1-t^2)^7}$ & $\footnotesize {m_2}^2 t^2+{m_1}^2 t^4$  \\
	 $         $ &  {\color{red}[0202]/[0220]} & 8 & $  $ & $\frac{(1-t^4) (1+t^2+6 t^4+t^6-t^8  )}{(1-t^2)^9}$ & \nc  \\
	 $         $ & $[2200]$ & 8 & $       $ & $\frac{1+6 t^2+16 t^4+21 t^6+\pal+t^{12}}{(1-t^8)^{-1}(1-t^2)^4 (1-t^4)^5}$ & $\footnotesize  \left( \begin{array}{c} m_2^2 t^2 + t^4 + m_1 t^4 + m_1^2 t^4 +\\ m_1 t^6 - m_1^2 t^{12} \end{array}\right) $  \\
	 $         $ & $[2022]$ & 10 & $       $ & $\frac{1+3 t^2+8 t^4+16 t^6+22 t^8+\pal+t^{16}   }{(1-t^2)^7 (1-t^4)^3}$ & \nc  \\
	 $         $ & $[2222]$ & 12 & $       $ & $\frac{(1-t^8)^2 (1-t^{12})}{(1-t^2)^{10} (1-t^4)^5}$ & \nc  \\
\hline
	\end{tabular}
	\caption[$D_4$ Slodowy Intersections from {[0002]-[2000]}]{$D_4$ Slodowy Intersections from  {[0002]/[0020]-[2000]}. Some palindromic Hilbert series terms are abbreviated. HWGs that are not complete intersections are not shown. The orbits {\color{red} [0002]}, {\color{red} [0020]}, {\color{red} [0202]} and {\color{red} [0220]} are non-normal.}
	\label{tab:DHS4b}
\end{sidewaystable}

\begin{sidewaystable}[h]
	\centering
	\small
	\begin{tabular}{|c|c|c|c|c|c|}
\hline
	$\rho$ & $\sigma$ & $\left |\svar{\sigma}{\rho} \right |$ & $F(\rho)$ & Unrefined HS & PL[HWG] \\
\hline
	 $[1011]$ & $[0200]$ & 2 & $C_1 \otimes B_0$ & $\frac{1-t^4}{(1-t^2)^3}$ & $\footnotesize m^2 t^2$  \\
	 $         $ &  {\color{red}[0202]/[0220]} & 4 & $  $ & $\frac{(1-t^4) (1-t +t^2+t^3+t^4-t^5  )}{(1-t) (1-t^2)^2 (1-t^3)^2}$ & \nc  \\
	 $         $ & $[2200]$ & 4 & $       $ & $\frac{1-t^8}{(1-t^2)^3 (1-t^3)^2}$ & $\footnotesize m^2 t^2 + m t^3 + t^4 + m t^5 - m^2 t^{10} $  \\
	 $         $ & $[2022]$ & 6 & $       $ & $\frac{1-t+t^2+2 t^4+\pal+t^8   }{(1-t) (1-t^2)^2 (1-t^3)^3}$ & \nc  \\
	 $         $ & $[2222]$ & 8 & $       $ & $\frac{(1-t^8)^2 (1-t^{12})}{(1-t^2)^3 (1-t^3)^4 (1-t^4)^2 (1-t^5)^2}$ & \nc  \\
\hline
	 $[0200]$ &  {\color{red}[0202]/[0220]} & 2 & $D_1 \otimes D_1$ & $\frac{1+t^2+3 t^4-t^6}{(1-t^2) (1-t^4)}$ & \nc  \\
	 $         $ & $[2200]$ & 2 & $       $ & $\frac{1-t^8}{(1-t^2) (1-t^4)^2}$ & $\footnotesize t^2 + (q^2+1/q^2) t^4 - t^8 $  \\
\hline
	\end{tabular}
	\caption[$D_4$ Slodowy Intersections from  {[1011]-[2022]}]{$D_4$ Slodowy Intersections from  {[1011]-[2022]}. Some palindromic Hilbert series terms are abbreviated. HWGs that are not complete intersections are not shown. The orbits {\color{red} [0002]}, {\color{red} [0020]}, {\color{red} [0202]} and {\color{red} [0220]} are non-normal.}
	\label{tab:DHS4c}
\end{sidewaystable}

\FloatBarrier

The HS $g_{HS}^{{\cal S}_{\sigma, \rho}}({\bf y},t)$ are consistent with the dimension formula \ref{eq:BCD3}, and also with results in the Literature for a variety of closures of $BCD$ series nilpotent orbits, Slodowy slices and KP transitions. As a non-trivial check, the HS for the different intersections ${\svar{\sigma}{ \rho}}$ within each slice (fixed by $\rho$) obey inclusion relations that match those of the poset of orbits $\orbit{\sigma}$ in the parent group Hasse diagram \cite{Kraft:1982fk}. Several observations can be made about the Hilbert series of $BCD$ type Slodowy intersections ${{\cal S}_{\sigma, \rho}}$ and their HWGs:

\begin{enumerate}

\item{Whenever $\orbit{\sigma}$ is normal, $g_{HS}^{{\cal S}_{\sigma, \rho}}(t)$ is palindromic, whether or not $\orbit{\rho}$ is normal. We refer to these ${{\cal S}_{\sigma, \rho}}$ as ``normal" intersections. For normal intersections, identical Hilbert series can be obtained using the SI formula \ref{eq:SV14}.}

\item{Conversely, whenever $\orbit{\sigma}$ is non-normal, then $g_{HS}^{{\cal S}_{\sigma, \rho}}(t)$ is non-palindromic, and the Kraft-Procesi transition from $\orbit{\sigma}$ to the orbit below in the Hasse diagram is always of the form ${A_{2k - 1}} \cup {A_{2k - 1}}$, for some $k$ \cite{Kraft:1982fk}. The structure of the HS of these non-normal intersections can be complicated, as discussed further below.}

\item{For a Slodowy slice, $g_{HS}^{{\cal S}_{{\cal N}, \rho}}(t)$ is always a complete intersection. Naturally, this extends to the ${{\cal S}_{{\sigma}, \rho}}$ whose quivers match those of Slodowy slices.}

\item{Several Kraft-Procesi transitions have different quivers, which generate the same $g_{HS}^{{\cal S}_{\sigma, \rho}}({\bf y},t)$; and this is also true for other intersections. For examples, see a comparison of the $B_2 \cong C_2$ quivers in figure \ref{fig:B2}, or the quivers for $\left\{ {{a_1} \cong {b_1} \cong {c_1} \cong {A_1},{b_2} \cong {c_2},{A_3}} \right\}$.}

\item{By construction, the product group $F(\rho)$ combines orthogonal and/or symplectic sub-groups. The adjoint representation of $F(\rho)$ (or its relevant subgroup) always appears at counting order $t^2$ in the HS/HWG, while other representations only appear at higher orders. The $g_{HS}^{{\cal S}_{\sigma, \rho}}({\bf y},t)$ are series of real representations, so any complex irreps are coupled with their conjugates at each counting order.}

\item{There are many identities between the ${{\cal S}_{\sigma, \rho}}$, such that the same $g_{HS}^{{\cal S}_{\sigma, \rho}}({\bf y},t)$ recur for different pairs of partitions, and for different ambient groups $G$. Such identities can only fully be identified from a comparison of Hilbert series and HWGs.}

\end{enumerate}


The situation surrounding the non-normal intersections requires further comment. These fall into one of two categories:

\begin{enumerate}

\item {Some non-normal ${{\cal S}_{\sigma, \rho}}$ (calculated on the Higgs branch) are unions of normal components. Examples include: $\sigma \in \{D[02], D[20], D[0020], D[0002], D[0202], D[0220]\}$. These arise whenever $\sigma$ is one of a spinor pair of orbits associated with a ``very even" partition of $D_{even}$. The following relations are obeyed by the intersections involved, and their HS and HWGs:

\begin{equation}
\label{eq:BCD8}
\begin{aligned}
{\cal S}_{D[ {20} ][{00} ]}^{} = {\cal S}_{D[ {02} ][ {00} ]}^{} &= {\cal S}_{D[ {20} ][ {00} ]}^{norm} + {\cal S}_{D[ {02} ][ {00} ]}^{norm} - {\cal S}_{D[{00}][{00}]}^{},\\
{\cal S}_{D[ {0020} ],\rho }^{} = {\cal S}_{D[ {0002} ],\rho }^{} &= {\cal S}_{D[ {0020} ],\rho }^{norm} + {\cal S}_{D[ {0002} ],\rho }^{norm} - {\cal S}_{D[ {0100} ],\rho }^{},\\
{\cal S}_{D[ {0220} ],\rho }^{} = {\cal S}_{D[ {0202} ],\rho }^{} &= {\cal S}_{D[ {0220} ],\rho }^{norm} + {\cal S}_{D[ {0202} ],\rho }^{norm} - {\cal S}_{D[ {0200} ],\rho }^{}.
\end{aligned}
\end{equation}

In such cases, Hilbert series for the normal components ${{\cal S}}_{D[ { \ldots 20} ],\rho }^{norm}$ and ${{\cal S}}_{D[ { \ldots 02} ],\rho }^{norm}$ can be found from the SI formula \ref{eq:SV14}, or, where available, the Coulomb branch of a Dynkin quiver. Alternatively, the $D_4$ intersections are related by triality, so the normal components ${{\cal S}_{\sigma,\rho }^{norm}}$ can also be found from normal ${{\cal S}_{\sigma,\rho }}$ by substitutions between the vector and two spinors:

\begin{equation}
\label{eq:BCD9}
\begin{aligned}
{{\cal S}}_{D[ {0002} ],\rho }^{norm}\mathop  = \limits_{{n_1} \Leftrightarrow {n_4}} {{\cal S}}_{D[ {2000} ],\rho }^{}\mathop  = \limits_{{n_1} \Leftrightarrow {n_3}} {{\cal S}}_{D[ {0020} ],\rho }^{norm},\\
{{\cal S}}_{D[ {0202} ],\rho }^{norm}\mathop  = \limits_{{n_1} \Leftrightarrow {n_4}} {{\cal S}}_{D[ {2200} ],\rho }^{}\mathop  = \limits_{{n_1} \Leftrightarrow {n_3}} {{\cal S}}_{D[ {0220} ],\rho }^{norm}.
\end{aligned}
\end{equation}
Care has to be taken over the interchange under triality of Dynkin labels within Characteristics and CSA coordinates, etc.}

\item{The remaining non-normal intersections have normal covers that are generated by the SI formula \ref{eq:SV14}. A normal cover ${{\cal S}_{\sigma, \rho}^{norm}}$ has the same dimension as ${{\cal S}_{\sigma, \rho}}$, and a palindromic Hilbert series, but contains representations (at counting degrees) that fall outside the nilcone $\cal N$ of the ambient group $G$. The cases up to rank 4 comprise $B[010]$, $B[2010]$ and $C[0200]$.}

\end{enumerate}



\subsection{Relationship to $T_{\sigma}^{\rho}$ theories}
\label{sec:BCD_Tsr}

The analysis of $3d$ mirror symmetry between $BCD$ Slodowy intersections and the relationship with $T_{\sigma}^{\rho}$ theories is not straightforward. This results from the many complications surrounding the Coulomb branches of ortho-symplectic quivers: (i) the Barbasch-Vogan map is only involutive for special orbits and exchanges $B$ and $C$ ambient groups, (ii)~Coulomb branch HS are palindromic and so do not match non-normal Slodowy intersections, (iii) quiver subtractions alone are insufficient to construct quivers with the desired Coulomb branch HS dimensions, requiring augmentation by ad-hoc shifts between $B$ and $D$ nodes, (iv) a careful choice of $O$ vs $SO$ gauge groups is required and (v)~``bad" quivers with zero conformal dimension are often encountered. As a consequence, only a subset of Slodowy intersections have Coulomb branch constructions. Furthermore, the results are limited to unrefined HS.

Most of these complications were encountered in the analysis of Slodowy slices \cite{Cabrera:2018ldc}, where it was nonetheless shown how a set of ortho-symplectic quivers, derived from the $ {\cal M}_{BCD}\left( {\bv\sigma,0} \right)$ tabulated herein, but with shifted $BD$ nodes taken as $SO$ type, yield Coulomb branch constructions of a subset of the slices $\slice{\sigma}$.

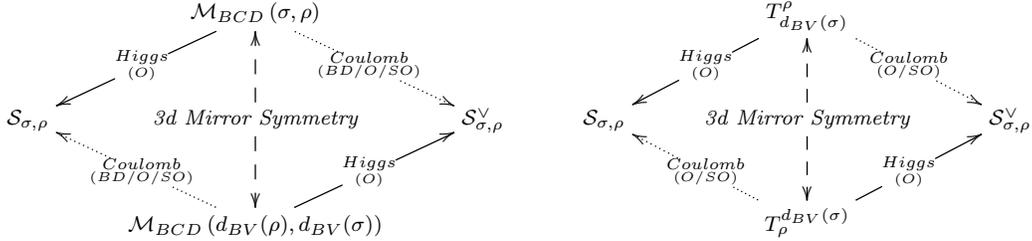
\begin{figure}[htp]
\centering
\scriptsize
\begin{displaymath}
    \xymatrix{
  &  {\cal M}_{BCD} \left( {\sigma,\rho } \right) \ar[dl]| {\mathop {Higgs}\limits_{(O)}} \ar@{.>}[dr]|{\mathop {Coulomb}\limits_{(BD/O/SO)} } & & & T_{\bv \sigma}^{\rho}   \ar[dl]|{\mathop {Higgs}\limits_{(O)}} \ar@{.>}[dr]|{\mathop {Coulomb}\limits_{(O/SO)} } & \\
 {{\cal S}_{\sigma, \rho}}  &\text{\scriptsize \it 3d~Mirror Symmetry} \ar@{-->}[d] \ar@{-->}[u] &  {{\cal S}_{\sigma, \rho}^\vee}   & {{\cal S}_{\sigma, \rho}}  & \text{\scriptsize \it 3d~Mirror Symmetry} \ar@{-->}[d] \ar@{-->}[u] &  {{\cal S}_{\sigma, \rho}^\vee}  \\
   & {\cal M}_{BCD} \left( {\bv \rho,\bv \sigma} \right) \ar@{.>}[ul]|{\mathop {Coulomb}\limits_{(BD/O/SO)} } \ar[ur]|{\mathop {Higgs}\limits_{(O)}} & & & T_{\rho}^{\bv \sigma}   \ar@{.>}[ul]|{\mathop {Coulomb}\limits_{(O/SO)} } \ar[ur]|{\mathop {Higgs}\limits_{(O)}}  & }
\end{displaymath}
\caption[$BCD$ Series 3d Mirror Symmetry]{$BCD$ series $3d$ mirror symmetry.  Under Special duality, the nilpotent orbit partitions $\rho$ and $\sigma$ of \emph{special} orbits are dualised to $\bv \rho$ and $\bv \sigma$ and then interchanged ${{{\cal S}^\vee}_{\sigma ,\rho }} \equiv {{{\cal S}}_{{d_{BV}}(\rho ),{d_{BV}}(\sigma )}}$. The ortho-symplectic Higgs branch constructions yield refined Hilbert series. Ortho-symplectic Coulomb branch constructions, where available, yield unrefined Hilbert series. $BD/O/SO$ indicates shifting between $B$ and $D$ nodes and selecting $O$ vs $SO$, as discussed in the text.}
\label{fig:BCDms}
\end{figure}

Generalising from the $A$ series, the phenomenon of (limited) $3d$ mirror symmetry for the $BCD$ series can be understood, as in figure \ref{fig:BCDms}, as a composition of the interchange of a pair of nilpotent orbits with the Barbasch-Vogan map $\bv\rho$.

Notably, Special duality, as defined in section \ref{sec:SV_dual} using $d_{BV}$, respects the accidental Lie algebra isomorphisms (e.g. $B_2 \cong C_2$ and $A_3 \cong D_3$), by assigning consistent Hilbert series to intersections defined by isomorphic pairs of Characteristics. Our account of the constructions for ${{\cal S}_{\sigma, \rho}}$ and their Special duals $ {{\cal S}_{\sigma, \rho}^\vee}$ thus sheds light on the underlying mechanism behind $3d$ mirror symmetry.

Furthermore, the approach based on the Higgs branches of ${\cal M}_{BCD} \left( \sigma ,\rho \right)$ quivers provides a complete set of constructions for $BCD$ Slodowy intersections, whereas the constructions available from $T_{\sigma}^{\rho}$ theories are limited (by the BV map) to intersections between special orbits.


\FloatBarrier


\section{Discussion and Conclusions}
\label{sec:Conclusions}

\paragraph{Classical Slodowy Intersections}

This study has outlined how unitary and ortho-symplectic multi-flavoured quivers provide constructions for the complete set of Slodowy intersections ${{{\cal S}}_{{\sigma} ,\rho }}$ of any Classical algebra. The resulting sets of quivers can be arranged as upper triangular matrices, bounded by the closures of nilpotent orbits, Slodowy slices and Kraft-Procesi transitions (modulo gaps due to the structure of Hasse diagrams of nilpotent orbits).

The key to these systematic constructions is provided by the Higgs branches of quivers of type ${\cal M}_A (\sigma, \rho)$ or ${\cal M}_{BCD}(\sigma, \rho)$. For the $A$ series, the resulting ${\cal M}_A ( \bf {N,{N_f}} )$ are the same as Dynkin quivers ${\cal D}_{A}( \bf {N,{N_f}} )$, and faithful Coulomb branch constructions for the ${{{\cal S}}_{{\sigma} ,\rho }}$ are also available via the Barbasch-Vogan map and $3d$ mirror symmetry. For the $BCD$ series, however, faithful Coulomb branch constructions are limited to those based on ${\cal D}_{BCD}( \bf {N,{N_f}} )$ of Characteristic height 2. The intersections obtained on the Coulomb branches of ortho-symplectic quivers are limited to the unrefined Hilbert series of a subset of ${{{\cal S}}_{{\sigma} ,\rho }}$ of $BCD$ algebras, as discussed in section \ref{sec:BCDSeries}. Some $D$ series Slodowy intersections can also be constructed as Higgs branches of ${\cal D}_{D}( \bf {N,{N_f}} )$ quivers.

Most of the intersections ${{{\cal S}}_{{\sigma} ,\rho }}$ are normal, and have palindromic (unrefined) Hilbert series. Whether or not an intersection ${{{\cal S}}_{{\sigma} ,\rho }}$ is normal is set by the normality of $\orbit\sigma$, and its global symmetry follows from $F(\rho) \subseteq G$.

The refined Hilbert series of intersections ${{{\cal S}}_{{\sigma} ,\rho }}$ with $\orbit\sigma$ normal, can also be constructed directly using the SI formula \ref{eq:SV14}. When $\orbit\sigma$ is non-normal, the SI formula yields either a normal component (if $\sigma$ is a very even partition of $D_{even}$), or a normal cover, of ${{{\cal S}}_{{\sigma} ,\rho }}$. This behaviour is similar to the $BCD$ Coulomb branch constructions (where these are available). In \cite{Cremonesi:2014uva} it was proposed that a localisation formula based on Hall-Littlewood polynomials be used as a proxy for the Coulomb branches of ortho-symplectic quivers, and this approach is supported by the findings herein.

\paragraph{Quiver Subtractions}

This study has used rules for quiver subtractions. These are essentially the same for both unitary and ortho-symplectic multi-flavoured linear quivers, both of which can be defined by pairs of partitions, $ {\cal M}  (\sigma, \rho) \to {\cal M} \left( \bf {N,{N_f}} \right)$, and can be summarised as:

\begin{equation} 
\label{eq:conc1}
\begin{aligned}
{{\cal M}}(\sigma ,\rho ) &= {{\cal M}}(\sigma ,\rho ') \ominus {{\cal M}}(\rho ,\rho '),
 \end{aligned}
\end{equation}
or, applying Special duality and relabelling partitions:
\begin{equation} 
\label{eq:conc2}
\begin{aligned}
{{\cal M}}(\sigma ,\rho ) &= {{\cal M}}(\sigma ',\rho ) \ominus {{\cal M}}(\sigma ',\sigma ).
 \end{aligned}
\end{equation}
Equations \ref{eq:conc1} and \ref{eq:conc2}, together with the procedures described herein, permit us, subject to certain conditions, to subtract two good quivers, that have either their first partition or their second partition in common, to obtain a quiver for a third intersection, with the partitions $(\sigma, \rho)$ tracking balance and flavour symmetries, respectively (on the Higgs branch). Quiver subtraction, as described, requires that all the quivers involved are good and that each pair of partitions obeys an inclusion relation. This study indicates these rules are consistent for Slodowy intersections calculated on the Higgs or Coulomb branches of ${{\cal M}}_A$ quivers, or on the Higgs branches of ${{\cal M}}_{BCD}$ quivers.

\paragraph{Completeness}

We have seen in sections \ref{subsec:AQuivers} and \ref{subsec:BCDQuiverOrtho} how an ambient group $G(N_0')$ of minimal dimension can be identified for any $ {\cal M} \left(\sigma, \rho \right)$ quiver from the weighted sum over flavours $N_0' \equiv \sum\nolimits_{i = 1}^k {i {N_{f_i}}} $. This implies that the Higgs branch of any good unitary or ortho-symplectic quiver with $\bf B \ge 0$ can be understood as a Slodowy intersection between a pair of Classical nilpotent orbits of such $G(N_0')$.

\paragraph{Degeneracy}

The number of distinct algebraic varieties is somewhat less than the number of (non-empty and non-trivial) Slodowy intersections $\svar{\sigma}{\rho}$ due to a combination of factors. Firstly, there are many recurrences of the quivers $ {\cal M} \left( \bf {N,{N_f}} \right)$ across different groups, as exemplified in the quiver matrices. These recurrences extend beyond Kraft Procesi transitions, to orbits and slices, indicating that all the intersections of a group G reappear in certain groups of higher dimension. Secondly, due to the accidental Classical group isomorphisms, there are several cases where different quivers $ {\cal M} \left( \bf {N,{N_f}} \right)$ generate isomorphic refined Hilbert series. The identification of such degeneracies (by a comparison of refined HS and HWGs) exposes connections between superficially different gauge theories.

\paragraph{Further Work}

While the ${\cal M} \left(\sigma, \rho \right)$ which have linear gauge nodes, only constitute a subset of quiver theories, other quivers, such as those with gauge node branches, for example, can be constructed as their combinations. Thus, Slodowy intersections provide a rich intermediate set of building blocks, whose Higgs and/or Coulomb branch quivers can be glued to construct a wide range of theories. Such approaches have been taken in \cite{Benini:2010uu, Gadde:2011uv, Cremonesi:2014vla}, together with a series of papers on class $S$ theories from \cite{Chacaltana:2010ks} through \cite{Chcaltana:2018zag}. In these studies, the building blocks have typically been (charged) Slodowy slices, glued using a combination of Coulomb and Higgs branch methods, to yield field theories with Classical or Exceptional symmetries. It may be interesting to examine and/or extend these approaches, from the perspective of the family of Slodowy intersections.

Notwithstanding the computational challenges in dealing with high dimensioned algebras, it would be interesting to explore the related matter of quiver theories whose Coulomb or Higgs branches are Slodowy intersections of Exceptional algebras. Coulomb branch constructions with Characteristic height 2 are known for (near to) minimal nilpotent orbits and it can be expected that, similar to the $D$ series, Higgs branch constructions based on ${\cal D}_{E}$ quivers will provide constructions for sub-regular slices and nearby intersections. Moreover, even when the ambient group $G$ is Exceptional, $F(\rho)$ is not generally so, and this should make several intersections of Exceptional groups accessible to constructions from Classical quivers.

It would also be interesting to extend the work in \cite{Cabrera:2017njm, Cabrera:2017ucb, Cabrera:2018ldc} to give a more systematic account of Coulomb branch constructions based on ortho-symplectic quivers for the unrefined Hilbert series of $BCD$ Slodowy intersections, for example, by providing definitive algorithms for $BD$ node shifting and the selection of $O/SO$ gauge nodes.



\paragraph{Acknowldgements}

We would like to thank Santiago Cabrera, Antoine Bourget and Marcus Sperling for helpful conversations during the development of this project. A.H. is supported by the STFC grant ST/P000762/1.

\appendix

\section{Notation and Terminology}
\label{apx:1}

\begin{enumerate}

\item We freely use the terminology and concepts of the Plethystics Program, including the Plethystic Exponential (``PE") and its inverse, the Plethystic Logarithm (``PL"). For our purposes:
\begin{equation} 
\label{eq:apxA1}
\begin{aligned}
PE\left[ {\sum\limits_{i = 1}^d {{A_i}} ,t} \right]  \equiv & \prod\limits_{i = 1}^d {\frac{1}{{\left( {1 - {A_i}t} \right)}}}\\
PE\left[ {\sum\limits_{i = 1}^d {(-{A_i})} , t} \right]  \equiv & \prod\limits_{i = 1}^d {\frac{1}{{\left( {1 + {A_i}t} \right)}}} ,\\
PE\left[ { - \sum\limits_{i = 1}^d {{A_i}} ,t} \right]  \equiv & \prod\limits_{i = 1}^d {\left( {1 - {A_i}t} \right)},\\
PE\left[ { - \sum\limits_{i = 1}^d {(-{A_i})} , t} \right]  \equiv & \prod\limits_{i = 1}^d {\left( {1 + {A_i}t} \right)},\\
PL\left[ {\frac{{\prod\limits_{j = 1}^e {\left( {1 - {B_j}} \right)} }}{{\prod\limits_{i = 1}^d {\left( {1 - {A_i}} \right)} }}} \right] = & \sum\limits_{i = 1}^d {{A_i}}  - \sum\limits_{j = 1}^e {{B_j}},\\
\end{aligned}
\end{equation}
where $A_i$ and $B_j$ are monomials in weight or root coordinates or fugacities. The reader is referred to \cite{Benvenuti:2006qr} for further detail.

\item We refer to symmetries either by Lie algebras $\mathfrak g$, or by Lie groups $G$. While such references are relatively interchangeable for $USp$ groups, with $C$ series Lie algebras, it can be important to distinguish between $O$ and $SO$ forms of orthogonal groups, which share the same $B$ or $D$ series Lie algebra, but whose representations have different characters. We highlight those areas where this distinction is important in the text.

\item We denote the characters of irreducible representations (``irreps") of a group $G$ either by ${\cal X}_{\bf [n]}^{G}({\bf x})$, or by $[irrep]_{G}$, using Dynkin labels ${\bf [n]}\equiv{[ {{n_1}, \ldots , {n_r}}]_{G}}$, where $r$ is the rank of $G$. We often represent singlets by the character $1$.

\item We typically label unimodular Cartan subalgebra (``CSA") coordinates for weights within characters by ${\bf x}\equiv (x_1 \ldots x_r)$ and simple root coordinates by ${\bf z}\equiv (z_1 \ldots z_r)$. The Cartan matrix $A_{ij}$ relates the simple root and CSA coordinates, ${z_i} = \prod\limits_j {x_j^{{A_{ij}}}}$ and ${x_i} = \prod\limits_j {z_j^{{A^{ - 1}}_{ij}}}$. We use the CSA coordinate $q$ for $U(1)$ symmetries.

\item We label highest weight (Dynkin label) fugacities within HWGs by ${\bf m}\equiv{[ {{m_1}, \ldots , {m_r}}]}$, deploying additional letter subscripts to distinguish groups, if necessary.

\item We label field (or R-charge) counting variables with $t$. Under the conventions in this paper, the fugacity $t$ corresponds to an R-charge of 1/2 and $t^2$ corresponds to an R-charge of 1.

\item We may refer to \emph{series}, such as $1 + f + {f^2} + \ldots $, by their \emph{generating functions} $1/\left( {1 - f} \right)$. Different types of generating function are indicated in table \ref{tab:apx1}; amongst these, the refined HS and HWGs faithfully encode the group theoretic information about a moduli space.\cite{Hanany:2014dia}
\begin{table}[htp]
\small
\centering
\begin{tabular}{|c|c|c|}
\hline
$\text{Generating~Function}$&$ \text{Notation}$&$ \text{Definition} $\\
\hline
$\text{Refined~HS~(Weight~coordinates)}$&$ {{g^{G}_{HS}}( {{\bf x},{t}} )}$&$\sum\limits_{n = 0}^\infty {{a_n}({\bf x})}{t^n} $\\
$\text{Refined~HS~(Simple root~coordinates)}$&$ {{g^{G}_{HS}}( {{\bf z},{t}} )}$&$\sum\limits_{n = 0}^\infty {{a'_n}({\bf z})}{t^n} $\\
$\text{Unrefined~HS}$&$ {{g^{G}_{HS}}\left( t \right)}$&$ \sum\limits_{n = 0}^\infty {{a_{ n}}} {t^n} \equiv \sum\limits_{n = 0}^\infty {{a_n}({\bf 1})}{t^n} $\\
\hline
HWG for Refined HS &$ {{g^{G}_{HWG}}({\bf m}, t)}$&$\sum\limits_{\bf [ n ]}^{} {{b_{\bf n}}(t){\bf m^n}} \cong \sum\limits_{[{\bf{n}}]}^{} {{b_{\bf{n}}}(t) \chi_{[ {\bf{n}}]}^G  }$\\
\hline
Character & $ {{g^{G}_{\chi}}({\bf x, m})}$ & $\sum\limits_{{\bf{[n]}}}^{} {\chi_{\bf [n]}^G( {\bf x})} {{\bf m }^{\bf n }} $\\
\hline
\end{tabular}
\caption{Types of Generating Function}
\label{tab:apx1}
\end{table}

\item We classify an unrefined Hilbert series $ {{g^{}_{HS}}(t)} \equiv P(t)/Q(t) $ as: (a){``freely generated", if $P(t)$ =1 and $Q(t)$ is of the canonical form $Q( t ) \equiv \prod\limits_k {{( {1 - {t^{{d_k}}}} )}^{n_k}} $ for some integers $n_k$ and $d_k$, or} (b) {a ``complete intersection", if both $P(t)$ and $Q(t)$ can  be put into canonical form, such that $g_{HS}^{}( t )$ is manifestly a quotient of geometric series, or} (c) {``(anti-)palindromic", if $P(t)$ is (anti-)palindromic. Palindromicity follows from the duality for a normal HS: $g_{HS}^{}( t ) = {t^{ - | {g_{HS}^{}} |}} g_{HS}^{}( {1/t} )$.}\cite{Gray:2008yu}.

\item We denote the Higgs and Coulomb branches of a quiver $\cal M$ as $\higgs{[\cal M}]$ and $\coulomb[{\cal M}]$, respectively.

\end{enumerate}


\section{Slodowy Intersection Formula}
\label{apx:2}

The Slodowy intersection formula \ref{eq:apxB1} is a localisation formula that yields the Hilbert series of an intersection ${\cal S}_{\sigma , \rho}$. It is related to the Hall Littlewood polynomials. As stated below, it incorporates weight space charges parameterised by Dynkin labels $\bf [n]$ of $G$, which permit the (Coulomb branch) gluing of intersections.\footnote{In \ref{eq:SV14} the charges ${\bf [n]}$ are set to $\bf [0]$, since the quivers in this paper are assumed to be free of background fluxes. A comparable formula, using somewhat different concepts and notation, appears in \cite{Cremonesi:2014uva}.}
\begin{equation}
\label{eq:apxB1}
\begin{aligned}
\hs{{\cal S}_{\sigma,\rho,{\bf [n]}}} ({\bf y},t) 
=\hs{ {\cal S}_{{\cal N}, \rho}({\bf y}, t)} {\left. {\frac{\hs{ {\overline{\cal O}}_{\sigma,{\bf [n]}} ({\bf x},t)}}{\hs{{\cal N}} ({\bf x}, t)}} \right|_{{\bf x} \to {\bf x}({\bf y},t)}}.
\end{aligned}
\end{equation}
The key ingredients of \ref{eq:apxB1} are:

\begin{enumerate}

\item the refined HS for the nilcone ${\cal N}({\bf x},t)$ of $G$, 

\item the refined HS for the \emph{charged} nilpotent orbit closure ${ {\overline{\cal O}}_{\sigma,{\bf [n]}} ({\bf x},t)}$ of $G$,

\item the fugacity map $\rho:{{\bf x} \to {\bf x}({\bf y},t)}$, from the CSA fugacities ${\bf x}$ of $G$, to the CSA fugacities ${\bf y}$ of $F(\rho)$ and the fugacity $t$ of $\mathfrak{su(2)}$.

\end{enumerate}
For a normal intersection, ${\hs{{\overline{\cal O}}_{\sigma,{\bf [n]}}^{~norm}}} {({\bf x},t)}$ can be obtained from the charged version of the Nilpotent Orbit Normalisation formula \cite{Hanany:2017ooe}:
\begin{equation}
\label{eq:apxB2}
g_{HS}^{{\overline{\cal O}}_{\sigma,{\bf [n]}}^{~norm}}\left( {{\bf x},t} \right) 
\equiv  \sum\limits_{w \in W_{G}}^{} {w \cdot \left( {\bf x}^{\bf [n]} 
\prod\limits_{\scriptsize \begin{array}{c} {\alpha  \in \tilde \Phi _G^ + (\sigma)}  \end{array}}
 {\frac{1}{{1 - {{\bf z(x)}^{\bf \alpha} }t^2}}} \prod\limits_{\beta  \in \Phi _{G}^ + }{\frac{1}{{1 - {{\bf z(x)}^{ - \bf \beta }}}}}\right)}.
\end{equation}
The summation is carried out over the Weyl group $W_{G}$ of $G$, whose elements $w$ act on the CSA fugacities, ${\bf x} \to w \cdot {\bf x} $. The subset $\tilde \Phi _{G}^ +(\sigma)$ contains those roots of $G$ that have a Characteristic height ${[\alpha] \ge 2} $, where $[\alpha] \equiv \alpha \cdot \bf q(\sigma)$, with $\bf q(\sigma)$ being the Characteristic of the nilpotent orbit $\sigma$. 

The charged formula \ref{eq:apxB1} generates some notable limiting cases or series:
\begin{enumerate}
\item In the limit $\{ \sigma \to triv., \rho \to triv. \}$, \ref{eq:apxB1} reduces to the Weyl Character formula:
\begin{equation}
\label{eq:apxB3}
\begin{aligned}
\hs{{\cal S}_{triv. , triv. ,{\bf [n]} }}  ({\bf y},t) =
 {\hs{ {\overline{\cal O}}_{triv. ,{\bf [n]}}} ({\bf x},t)}  =\hs{ \overline {\cal O}_{{{triv.}},[{\bf{n}}]}}({\bf{x}},0) = \chi _{{\bf{[n]}}}^G({\bf{x}}).
 \end{aligned}
\end{equation}
\item In the limit $\{ \sigma \to {\cal N},\rho \to triv. \}$, \ref{eq:apxB1} reduces to the \emph{modified} Hall Littlewood formula:
\begin{equation}
\label{eq:apxB7}
\begin{aligned}
\hs{{\cal S}_{ {\cal N}, triv. ,{\bf [n]} }}   ({\bf y},t) = 
\hs{\overline {\cal O}_{{{\cal N}},[{\bf{n}}]}}   ({\bf{x}},t) = mHL _{{\bf{[n]}}}^G({\bf{x}},t).
\end{aligned}
\end{equation}
The $mHL$ functions obey the orthogonality, 
\begin{equation}
\label{eq:apxB8}
\begin{aligned}
\oint \limits_G {}  {d{\mu _{mHL}^G}({\bf x},t)} ~ mHL_{{\bf{[m]}}}^G({\bf{x}}^*,t) ~ mHL_{[{\bf{n}}]}^G({\bf{x}},t) & = {\delta _{{\bf{[m]}},[{\bf{n}}]}} N_{[ \bf n ]}^G (t),
\end{aligned}
\end{equation}
where ${d{\mu _{mHL}^G}({\bf x},t)} $ is the Haar measure for the $mHL$, and the $N_{[ \bf n ]}^G$ are normalisation factors:
\begin{equation}
\label{eq:apxB6}
\begin{aligned}
N_{[ \bf n ]}^G (t) & \equiv PE \left[rank[G]{~}t^2 - \sum\limits_{i = 1}^r t^{2 {d_i} ({\bf [n]})} \right].\\
\end{aligned}
\end{equation}
In \ref{eq:apxB6} the $d_i ({\bf [n]})$ are the degrees of the symmetric Casimirs of the subgroup(s) of $G$ identified by the Dynkin diagram formed by the zeros of $\bf [n]$; non-zero Dynkin labels contribute U(1) Casimirs with $d_i=1$.
\item In the limit $\{ \sigma \to {\cal N},\rho \to triv., {\bf [n] \to [0]} \}$, \ref{eq:apxB1} reduces to the nilcone $\cal N$ of $G$: 
\begin{equation}
\label{eq:apxB9}
\begin{aligned}
\hs{{\cal S}_{ {\cal N}, triv. ,{\bf [0]} }}({\bf y},t) = 
 mHL_{\bf [0]}^G({\bf x},t) & =\hs{\cal N}({\bf x},t).
\end{aligned}
\end{equation}
\item In the limit $\{ \sigma \to \rho, {\bf [n] \to [0]} \}$, \ref{eq:apxB1} evaluates to a trivial self-intersection:
\begin{equation}
\label{eq:apxB9a}
\begin{aligned}
\hs{{\cal S}_{\rho, \rho ,{\bf [0]} }}  ({\bf y},t) = 
\hs{{{\cal S}}_{{\cal N},\rho }}({\bf{y}},t){\left. {\frac{{\hs{{\overline {{\cal O}} }_{\rho ,[0]}}({\bf{x}},t)}}{{{\hs{\cal N}}({\bf{x}},t)}}} \right|_{{\bf{x}} \to {\bf{x}}({\bf{y}},t)}} 
 =1.
\end{aligned}
\end{equation}
\end{enumerate}

As discussed in \cite{Hanany:2016gbz, Hanany:2017ooe}, any (charged) nilpotent orbit of $G$ can be expanded as a finite sum over the basis functions provided by the $mHL$ of $G$:
\begin{equation}
\label{eq:apxB10}
\begin{aligned}
\hs{{\overline {\cal O}_{\sigma,{\bf [m]}}}}  ({\bf{x}},t)  = \sum\limits_{{\bf [n]}} {{a_{{\bf [m]},{\bf{[n]}}}}(t)}{~} mHL_{{\bf{[n]}}}^G({\bf{x}},t).
\end{aligned}
\end{equation}
Inserting \ref{eq:apxB7} and \ref{eq:apxB10} into \ref{eq:apxB1}, it follows that this decomposition extends to any (charged) Slodowy intersection. Indeed, ${{\cal S}_{\sigma ,\rho ,{\bf [m] }}}({\bf{y}},t)$ can be expanded as a sum of charged Slodowy slices:
\begin{equation}
\label{eq:apxB11}
\begin{aligned}
\hs{{\cal S}_{\sigma ,\rho ,{\bf [m] }}}({\bf{y}},t) = \sum\limits_{{\bf{[n]}}} {{a_{{\bf [m]},{\bf{[n]}}}}(t)} \hs{{\cal S}_{{{\cal N}},\rho ,{\bf{[n]}}}}({\bf{y}},t),
\end{aligned}
\end{equation}
where the polynomial coefficients ${{a_{{\bf [m]},{\bf{[n]}}}}(t)}$ are inherited from \ref{eq:apxB10}. Decompositions such as \ref{eq:apxB11} provide a further set of relationships that can be used to cross-check the Hilbert series for Slodowy intersections.


\paragraph {Example: ${\cal S}_{A[202], A[101]}$.}
Consider the $A_3$ intersection ${\cal S}_{A[202], A[101]}$. We start by evaluating the expressions \ref{eq:apxB9} and \ref{eq:apxB2}, for the nilcone ${\cal N \equiv} \orbit{A[222]}$ and the orbit $ \orbit{A[202]}$, respectively, and take their quotient:
\begin{equation}
\label{eq:apxBx1}
\begin{aligned}
g_{HS}^{\cal N} &= PE \left[ [1, 0, 1]_A t^2 - t^4 - t^6 - t^8 \right],\\
g_{HS}^{\orbit{A[202]}}  &=PE \left[ [1, 0, 1]_A t^2 \right]  \left(PE \left[ - t^4 - t^6 - t^{10} \right] -PE \left[ -2 t^2 - t^4  \right] [1, 0, 1]_A  t^6  \right),\\
g_{HS}^{\orbit{A[202]}/{\cal N}} & = PE \left[ t^8- t^{10} \right] - PE \left[-2 t^2+t^6 + t^8 \right] [1, 0, 1]_A t^6.
\end{aligned}
\end{equation}
We now identify the CSA fugacity map for the $SU(2)$ embedding induced by the orbit with Characteristic $\rho =A [101]$:
\begin{equation}
\label{eq:apxBx2}
\begin{aligned}
\rho: \{x_1,x_2,x_3\}  \to \{ q^{1/2} t, y t, t /q^{1/2} \},
 \end{aligned}
\end{equation}
where the CSA fugacities for $SU(2)$, $A_1$ and $U(1)$ are $t$, $y$ and $q$, respectively. As can readily be verified, this maps $A_3$ characters to $SU(2) \otimes  A_1 \otimes U(1)$:
\begin{equation}
\label{eq:apxBx2a}
\begin{aligned}
\rho: [1,0,0]_A  & \to  q^{1/2}[1]_{SU}+q^{-1/2} [1]_{A},\\
\rho: [1,0,1]_A  & \to  [2]_{SU}+ [1]_{SU}[1]_{A}(q+1/q) +[0]_{SU}( [2]_{A}+1).
 \end{aligned}
\end{equation}
Thus, the $A_3$ adjoint decomposes to the $SU(2)$ partition $(3,2^4,1^4)$, consistent with the tables in \cite{Hanany:2016gbz}. Applying the map \ref{eq:apxBx2} to the nilcone $\cal N$, using \ref{eq:SV13}, we find the slice:
\begin{equation}
\label{eq:apxBx3}
\begin{aligned}
g_{HS}^{\slice{A[101]}} &= PE \left[ ([2]_A +1)  t^2 + [1]_A (q + 1/q) t^3 -t^6 - t^8 \right].
\end{aligned}
\end{equation}
We also apply the map \ref{eq:apxBx2} to the quotient $\orbit{A[202]}/{\cal N}$:
\begin{equation}
\label{eq:apxBx4}
\begin{aligned}
g_{HS}^{\left. \orbit{A[202]}/{\cal N} \right |_{{\bf x} \to {\bf x}(y,q,t)}} &= PE \left[t^6 + t^8 - 2 t^2 \right]\\
&\times { \left( \begin{array}{c} 1 + 2 t^2 + 2 t^4 + t^6 + 2 t^8 + 2 t^{10} + t^{12}-\\
 \ldots [1]_A (q+1/q )(t^5 + t^7)  - [2]_A t^6 \end{array} \right)},
\end{aligned}
\end{equation}
Combining \ref{eq:apxBx3} and \ref{eq:apxBx4} gives the refined Hilbert series for ${{\cal S}_{A[202], A[101]}}$:
\begin{equation}
\label{eq:apxBx5}
\begin{aligned}
g_{HS}^{{\cal S}_{A[202], A[101]}}({y},q,t) & =
PE \left[[2]_A t^2 - t^2+[1] (q + 1/q)t^3 \right]\\
& \times \left( \begin{array}{c}  1 + 2 t^2 + 2 t^4 + t^6 + 2 t^8 + 2 t^{10} + t^{12} -\\ \ldots [1]_A  (q+1/q)  (t^5 + t^7)- [2]_A t^6  \end{array} \right),
\end{aligned}
\end{equation}
which simplifies to the unrefined form consistent with table \ref{tab:AHS1}:
\begin{equation}
\label{eq:apxBx5a}
\begin{aligned}
g_{HS}^{{\cal S}_{A[202], A[101]}}({1},1,t) & =
PE \left[2 t^2 +2 t^3 \right] \times \left( 1 + 2 t^2 + 2 t^3+ 2 t^4 + t^6 \right).
\end{aligned}
\end{equation}

Alternatively, we can find the refined Hilbert series for ${{\cal S}_{A[202], A[101]}}$, using \ref{eq:apxB11} and the tables in \cite{Hanany:2016gbz}. Table 4 of \cite{Hanany:2016gbz} shows that:
\begin{equation}
\label{eq:apxBx6}
\begin{aligned}
\hs{\overline {\cal O}_{A[202]}} = mH{L_{[0,0,0]_A}} - mH{L_{[1,0,1]_A}} {~}{t^6},
\end{aligned}
\end{equation}
and so, applying \ref{eq:apxB11}, we obtain:
\begin{equation}
\label{eq:apxBx7}
\begin{aligned}
\hs{{\cal S}_{A[202], A[101]}}({\bf{y}},t) = \hs{{\cal S}_{{\cal N}, A[101],[0,0,0]_A}}({\bf{y}},t) -\hs{{\cal S}_{{\cal N}, A[101],[1,0,1]_A}}({\bf{y}},t) {~}{t^6},
\end{aligned}
\end{equation}
The first RHS slice in \ref{eq:apxBx7} was calculated in \ref{eq:apxBx3}. We use \ref{eq:apxB1} to find the second (charged) slice as:
\begin{equation}
\label{eq:apxBx8}
\begin{aligned}
g_{HS}^{{\cal S}_{{\cal N}, A[101],[1,0,1]_A}}({\bf{y}},t) =& PE\left[ {[2]_A{t^2} + [1]_A(2){t^3} - {t^2}} \right]\\
& \times \left( {2 + 1/{t^2} - {t^4} - {t^6} + [1]_A ({q} + 1/{q})(t + 1/t) + [2]_A} \right).
\end{aligned}
\end{equation}
Combining \ref{eq:apxBx7}, \ref{eq:apxBx3} and \ref{eq:apxBx8} recovers \ref{eq:apxBx5}.

The HS can be simplified by transforming to an HWG with respect to the $A_1$ fugacities, using \ref{eq:apxB12}. Inserting necessary terms from the Haar measure for $A_1$ and an $A_1$ character generating function, and carrying out the contour integration, we have:
\begin{equation}
\label{eq:apxBx9}
\begin{aligned}
g_{HWG}^{{\cal S}_{A[202], A[101]}}(m,q,t) &= \oint\limits_{A_1} {d{\mu^{A_1}}} {g_\chi ^{A_1} \left( y^*, m \right)}{~}g_{HS}^{{\cal S}_{A[202], A[101]}} \left( y,q,t \right),\\
&=\oint\limits_{A_1} \frac{{dy}}{y}\left( {1 - {y^2}} \right) \frac{1}{{1 - m/y}}{~}g_{HS}^{{\cal S}_{A[202], A[101]}} \left( y,q,t \right),\\
\ldots\\
&=\frac{1-m^2 t^6}{(1-t^2) (1-m^2 t^2)  (1-m q t^3)(1-{m/q t^3})},\\
&= PE \left[t^2 + m^2 t^2 + m (q+1/q) t^3 -m^2 t^6 \right],
\end{aligned}
\end{equation}
as shown in table \ref{tab:AHS1}.


\clearpage


\bibliography{RJKBibLib}


\end{document}